\documentclass[]{emulateapj}

\usepackage{graphicx,amssymb,verbatim,natbib}

\newcommand\hii{\hbox{H$\rm II$~}}

\newcommand\ha{H$\alpha$}

\newcommand\ip{$i_{775}$}
\newcommand\inp{$I_{814}$}
\newcommand\zp{$z_{850}$}
\newcommand\bp{$B_{435}$}
\newcommand\up{$U_{330}$}
\newcommand\vp{$V_{606}$}

\shorttitle{New evidence for a merger-driven formation of LBGs}
\shortauthors{Overzier et al.}

\begin{document}      
                      
 \title{HST morphologies of local Lyman break galaxy analogs I:\\
Evidence for starbursts triggered by merging}

\author{Roderik A. Overzier\altaffilmark{1,2}, Timothy
  M. Heckman\altaffilmark{1}, Guinevere Kauffmann\altaffilmark{2},
  Mark Seibert\altaffilmark{3}, R. Michael Rich\altaffilmark{4},
  Antara Basu-Zych\altaffilmark{5}, Jennifer Lotz\altaffilmark{6},
  Alessandra Aloisi\altaffilmark{7}, St\'ephane
  Charlot\altaffilmark{8}, C. Hoopes\altaffilmark{1}, D. Christopher
  Martin\altaffilmark{9}, David Schiminovich\altaffilmark{5}, Barry Madore\altaffilmark{3}}
\email{overzier@mpa-garching.mpg.de}

\altaffiltext{1}{Department of Physics and Astronomy, The Johns Hopkins University, 3400 North Charles Street, Baltimore, MD 21218.}
\altaffiltext{2}{Max-Planck-Institut f\"ur Astrophysik, D-85748 Garching, Germany.}
\altaffiltext{3}{Observatories of the Carnegie Institution of Washington, 813 Santa Barbara Street, Pasadena, California 91101, USA.}
\altaffiltext{4}{Deptartment of Physics and Astronomy, Division of Astronomy and Astrophysics, University of California, Los Angeles, CA 90095-1562, USA.}
\altaffiltext{5}{Department of Astronomy, Columbia University, MC 2457, 550 West 120th Street, New York, NY 10027.}
\altaffiltext{6}{Department of Astronomy and Astrophysics, University of California, Santa Cruz, CA 95064.}
\altaffiltext{7}{Space Telescope Science Institute, 3700 San Martin Drive, Baltimore, MD 21218.}
\altaffiltext{8}{Institut d'Astrophysique du CNRS, 98 bis Boulevard Arago, F-75014 Paris, France.}
\altaffiltext{9}{California Institute of Technology, MC 405-47, 1200 East California Boulevard, Pasadena, CA 91125.}

\begin{abstract}
\noindent
Heckman et al. (2005) used the Galaxy Evolution Explorer (GALEX) UV
imaging survey to show that there exists a rare population of nearby
compact UV-luminous galaxies (UVLGs) that closely resembles high
redshift Lyman break galaxies (LBGs).  We present HST images in the
UV, optical, and H$\alpha$, and resimulate them at the depth and
resolution of the GOODS/UDF fields to show that the morphologies of
UVLGs are also similar to those of LBGs. Our sample of 8 LBG analogs
thus provides detailed insight into the connection between star
formation and LBG morphology. Faint tidal features or companions can be seen
in all of the rest-frame optical images, suggesting that
the starbursts are the result of a merger or interaction.  The
UV/optical light is dominated by unresolved ($\sim$100-300 pc) super
starburst regions (SSBs). A detailed comparison with the galaxies Haro 11 and VV 114 at $z=0.02$ indicates that 
the SSBs themselves consist of diffuse stars and (super) star clusters. 
The structural features revealed by the new
HST images occur on very small physical scales and are thus not
detectable in images of high redshift LBGs, except in a few cases
where they are magnified by gravitational lensing. We propose,
therefore, that LBGs are mergers of gas-rich, relatively low-mass
($M_*\sim10^{10}$ $M_\odot$) systems, and that the mergers trigger the
formation of SSBs. If galaxies at high redshifts are dominated by
SSBs, then the faint end slope of the luminosity function is predicted
to have slope $\alpha\sim2$. Our results are the most direct
confirmation to date of models that predict that the main mode of star
formation in the early universe was highly collisional.
\end{abstract}



\section{Introduction}
\label{sec:intro}

\noindent
How did galaxies form? Ultimately, this simple question captures most,
if not all of the most widely pursued research topics in
cosmology. Surveys of galaxies at different redshifts are being used
to study which galaxies contain most of the stellar mass at a given
epoch, and which galaxies are undergoing the strongest
evolution. These measurements serve to constrain models of structure
formation.  The main formative processes associated with typical
galaxies (e.g. star formation, merging, and feedback) were largely
completed within the first half of the Hubble time, making their study
a very challenging one.  One of the best probes for studying star
formation in the early universe is provided by galaxies that are
luminous in the rest-frame UV due to intense star formation. The most
luminous of these, the Lyman break galaxies (LBGs), are easily
detected at $z=2-6$ in deep pencil beam surveys from the ground and
with the {\it Hubble Space Telescope} (HST)
\citep[e.g.][]{steidel93,steidel99,shapley01,giavalisco04,adelberger04,bouwens06,yoshida06}. The
cosmic star formation rate (SFR) reached a maximum at $z\sim1-3$ and
decreased dramatically toward $z=0$
\citep[e.g.][]{lilly96,schiminovich05}. There is only a modest
decrease in the UV luminosity density from $z\sim3$ to $z\sim6$
\citep[e.g.][]{ouchi04,bouwens06}, indicating that LBGs represent a
major phase in the early stages of galaxy formation and
evolution. Based on their clustering and number statistics, LBGs are
the precursors of present day massive galaxies undergoing a phase of
intense star formation, and a large fraction of LBGs may merge to form
elliptical galaxies that are situated in present-day groups and
clusters
\citep[][]{governato01,giavalisco01,moustakas02,ouchi04b,adelberger05}.

Studies of the structure and sizes of LBGs indicate that high redshift
galaxies are compact \citep[$\sim$0.1\arcsec--0.3\arcsec, or
$\sim$1--2.5 kpc;][]{giavalisco96a,lowenthal97,ferguson04,bouwens04},
and that large ($\gtrsim$0.4\arcsec) low surface brightness galaxies
are rare. The morphologies are best characterized as being dominated
by one to several UV-bright knots embedded in diffuse emission, with
little difference between the rest-frame UV and optical light
distributions \citep{papovich01,papovich05}.  The morphologies of LBGs
are unlike those of nearby galaxies on the Hubble sequence. This is
true even if the images of the nearby objects are degraded so that
they have similar resolution and depth to those of LBGs.  LBG
morphologies are more similar to those of ultraluminous infrared
galaxies (ULIRGs), which occupy a region of the
concentration/asymmetry plane that lies between double-nucleated
mergers, spheroids and disks \citep[e.g.][and references
therein]{abraham96,conselice04,lotz04,lotz06,law07a}.  However, the
interpretation of the UV morphologies of high redshift populations
remains problematic \citep[see][]{law07a}, because it is unclear
whether the irregular morphologies are a consequence of the merging of
gas-rich galaxies, or whether one is just seeing patchily distributed
star formation within a single system.  The answer to this question
and how it evolves with redshift is essential for understanding how
galaxies formed.

A detailed study of LBGs is limited by two factors. First, their
distances render them small and faint, and, second, star forming
galaxies at high redshift are systematically different from starburst
galaxies in the nearby universe.  Local starbursts (SBs) show strong
correlations among their basic properties that are different from
LBGs.  In particular, there is a systematic increase in the amount of
dust obscuration ($L_{FIR}/L_{FUV}$) with increasing SFR,
corresponding to a systematic increase in metallicity with mass.  LBGs
are substantially less obscured than local SBs of similar SFR or mass
\citep[e.g.][]{reddy06}.  The relatively low extinction of LBGs is
also a characteristic of blue compact dwarfs (BCDs, alsoor \hii\
galaxies), but the latter have SFRs that are typically two orders of
magnitude smaller than LBGs and they have lower masses
\citep[e.g.][]{fanelli88,telles97,hunter99,gildepaz03}. Although BCDs
may resemble the faintest LBGs \citep[e.g.][]{meurer95,noeske06}, and
LIRGS and ULIRGS may resemble the dustiest LBGs
\citep[e.g.][]{adelberger00,goldader02,daddi05,chapman05,huang05,vandokkum06},
typical local SBs are different from typical LBGs.  LBGs have
systematically lower metallicities and higher gas-mass fractions than
local SBs with the same SFRs.  The typical mechanisms that trigger
starbursts might also be very different at high redshift compared to
low redshift.

Clearly, if we could find relatively nearby starburst galaxies whose
properties are the best match to LBGs, we could study for the first
time at a much higher physical resolution how the vigorous star
formation and morphologies of typical high redshift galaxies are
related.  However, local LBG analogs are currently very rare; as
indicated by the Galaxy Evolution Explorer \citep[GALEX;][]{martin05}
imaging survey in the near (NUV, $\lambda$$\sim$2250\AA) and far (FUV,
$\lambda$$\sim$1550\AA) UV, the co-moving number density of galaxies
with FUV luminosities similar to LBGs has declined by a factor of
several hundred between $z=3$ and $z=0$. \citet{heckman05} and
\citet{hoopes07} (henceforward called ``H05'' \& ``H07'') identified a
rare population of low-redshift ($z<0.3$) galaxies with properties
remarkably similar to those of LBGs by matching sources in the GALEX
all-sky survey with the Sloan Digital Sky Survey
\citep[SDSS;][]{york00} spectroscopic sample, and selecting according
to 2 criteria specifically designed to match the typical UV properties
of LBGs \citep[e.g.][]{steidel96,shapley01}: 
$L_{FUV} \geq 10^{10.3} L_{\odot}$\ $\wedge$\ $I_{FUV} \geq 10^9
L_{\odot} \mathrm{kpc}^{-2}$, where $L_{FUV}$ is the FUV luminosity,
$\lambda P_\lambda$, and $I_{FUV}$ is the mean FUV surface brightness
interior to the SDSS {\it u}-band half-light radius ($I_{FUV} =
\frac{1}{2} L_{FUV}/\pi r_{e,u}^2$).  By further limiting the sample
to $z<0.3$ and excluding broad line active galactic nuclei (AGN), this
resulted in a sample of 31 ``supercompact UV-luminous galaxies''
(UVLGs).  Interestingly, once selected according to these criteria,
the UVLG sample proved to be similar to typical LBGs in all other
measurable properties (see H05 and H07).  Further details on the
sample are given by \citet{basuzych07}, who investigated the radio to
UV continuum properties of the sample and found that they follow the
radio-far infrared correlation of normal star-forming galaxies, but
generally have less dust than other star forming galaxies with such
high specific SFRs.  In this respect, they are similar to LBGs. This
suggests that the process that causes star formation in the
supercompact UVLGs differs from other local star forming galaxies, but
may be similar to LBGs.  Our sample of UVLGs (``local LBG analogs'')
is therefore highly valuable for understanding the nature of Lyman
break galaxies at high redshifts\footnote{See the Appendix for a
  detailed analysis that demonstrates that the interpretation and
  conclusions presented in the recent paper by Scarpa et al. (2007)
  are in fact in error.}.

In this paper (Paper I) we present the first set of high resolution
HST images of 8 UVLGs and discuss their morphologies. In a subsequent
paper (Paper II) we will carry out a more detailed study using a large
HST data set to be observed in Cycle 16. The structure of this paper
is as follows. In Sect. 2 we describe the new observations with
HST. In Sections 3 and 4 we present the data, investigate UVLG
morphologies both qualitatively and quantitatively, resimulate our
data at higher redshift, and compare with morphologies of star-forming
galaxies and LBGs as well as two of the most nearby LBG-like galaxies
known (Haro 11 and VV 114). We discuss the implications of our results
for understanding the nature of LBGs at high redshifts in Section 5,
followed by a summary of the results (Section 6). We use a cosmology
[$\Omega_M$, $\Omega_\Lambda$, $h$]$=$[0.27,0.73,0.73] with $H_0=100h$
km s$^{-1}$ Mpc$^{-1}$.

\section{Data}

\subsection{HST observations}

\noindent
We have observed 8 of the nearest ($0.091<z<0.204$) and brightest
local LBG analogs of H05 \& H07 with HST in Cycle 15.  To date, 7
UVLGs have been observed with the Advanced Camera for Surveys (ACS)
High Resolution Camera (HRC) in the filter F330W, and with the Wide
Field Channel (WFC) through the filter F850LP. In addition, one UVLG
has been observed with the Solar Blind Channel (SBC) in the filter
F150LP, and with the Wide Field and Planetary Camera 2 (WFPC2) through
F606W, given the new constraints following the failure of ACS during
the course of 2007. Ramp filter images with a central wavelength equal
to that of redshifted H$\alpha$ were also obtained for all but the
latter source.  Each filter probes star formation and morphology at a
different characteristic timescale.  The F330W filter
(central wavelength $\lambda_c$$\approx$3334\AA\ with an effective width $\Delta\lambda\sim548\AA$) is the most sensitive to star formation
over the past $\sim$100 Myr \citep{leitherer95} and provide the
closest match to the GALEX NUV filter ($\lambda_c$$\approx$2315\AA, $\Delta\lambda\sim730\AA$).
The F150LP filter ($\lambda_c$$\approx$1614\AA, $\Delta\lambda\sim$234\AA), available for one of
the UVLGs, provides an excellent match to the GALEX FUV filter
($\lambda_c$$\approx$1530\AA, $\Delta\lambda\sim255\AA$).  The F606W
($\lambda_c$$\approx$6001\AA) and F850LP ($\lambda_c$$\approx$9170\AA)
images are well-matched to the SDSS images in the {\it r-} and {\it
  i}-band and probe older ($>$Gyr) stellar populations. The ACS ramp
filters (FR647M with a width of 207\AA\ or FR782N with a width of
52\AA, depending on the redshift) probe redshifted H$\alpha$ from
\hii\ regions tracing the youngest stellar population (O stars with
lifetimes $<$10 Myr), and allow us to search for evidence of
galaxy-wide outflows of ionized gas.

The targets were observed for one orbit per filter with ACS, and two
orbits per filter with WFPC2.  The F330W image of
SDSS J092600.41+442736.1 failed to execute due to a guide star problem.
The F150LP, F330W, ramp filter, F606W, and F850LP images consisted of
3, 3, 2, 6 and 3 exposures, respectively, to facilitate the removal of
cosmic rays. The images were combined using {\it Multidrizzle}
\citep{koekemoer02}, producing registered, cosmic-ray free,
geometrically corrected images. The SBC and HRC images have a plate
scale of 0\farcs025 pixel$^{-1}$ and a resolution of $\sim$$0\farcs07$
(FWHM). The WFPC2/PC images have a plate scale of 0\farcs046
pixel$^{-1}$ with a resolution of $\approx$$0\farcs11$ (FWHM). The WFC
images have a scale of 0\farcs05 pixel$^{-1}$ with a resolution of
$\approx$$0\farcs12$ (FWHM). Magnitudes in the AB system were measured
using SExtractor \citep{bertin96} from
$m_{AB}=-2.5\mathrm{log}_{10}(\mathrm{counts}/T_{exp})+\mathrm{ZPT}$,
where the zeropoints (ZPT) are 22.448, 24.085, 23.004, and 24.862 mag
for F150LP, F330W, F606W and F850LP, resp.  All magnitudes were
corrected for Galactic extinction using the dust maps of
\citet{schlegel98}.  The reader is referred to Table \ref{tab:log} for
a log of the observations, Table \ref{tab:sfrs} for SFRs and stellar
masses of the objects in our sample, and Table \ref{tab:phot} for the
main photometric data from GALEX and ACS.

\subsection{Continuum subtracted \ha\ images}

\noindent
The \ha\ ramp filter images were continuum-subtracted as follows.
First we scaled the ACS \zp\ image to an artificial \ip\ continuum
image, by determining the ratio of the flux of the \ip\ continuum at
the wavelength of redshifted \ha\ (measured from the SDSS fiber
spectrum on opposite sides of the \ha\ line) to the total flux in the
\zp\ image.  The total flux was measured within an aperture similar in
size to the 3\arcsec\ diameter SDSS fiber aperture.  We then created a
``continuum-free'' \ha\ image by subtracting the artificial \ip\
continuum image from the ACS ramp filter image. The resulting \ha\
image should be a very good approximation, under the (very reasonable)
assumption that the morphology of the continuum does not change
between \ip\ and \zp.  In the case of object SDSS J080844.26+394852.4 the
subtraction was problematic due to the fact that it relied on
subtraction of two bright point sources in \zp\ and \ha.

\subsection{Spitzer observations}

\noindent
Infrared photometry was obtained with the Infrared Array Camera (IRAC)
and the Multi-band Imaging Photometer for Spitzer (MIPS) aboard the
{\it Spitzer Space Telescope} (PI: C. Hoopes, \#20390).  The total
integration time with IRAC at 3.6, 4.5, 5.8 and 8.0 $\mu$m was 60 s
for each channel. The total integration time with MIPS at 24 and 70
$\mu$m was 42 and 60 s, resp.  We used a minimum 5 step dither pattern
for removal of cosmic rays. The post-basic calibrated data (BCD)
delivered by the pipeline were used to measure the integrated flux
densities. The infrared photometry is given in Table \ref{tab:ir}.

\subsection{Comparison data}

\noindent
The UVLGs in our sample, which have a median redshift of $z\sim0.15$,
lie at redshifts that are intermediate between local starburst
galaxies ($\sim0.02$) and LBGs at high redshift ($z>1.5$).  Throughout
the paper, we therefore find it instructive to present our results
with respect to the following two sets of comparison data:

\subsubsection{Archival imaging data of {\it Haro 11} and {\it VV 114}}
\label{sec:archival}

\noindent
As shown in H07 (see their Fig. 12), a number of very nearby blue
compact starburst galaxies and (U)LIRGS fall very near to the boundary
of the UVLG selection window as defined in the $L_{FUV}$ vs. $I_{FUV}$
plane. Because there is quite some leverage in the definition of a
typical `LBG', we will compare our data to two of these objects, which
have been found to possess some properties that are similar to high
redshift LBGs. The two local comparison objects are Haro 11 and VV
114, both of which lie at $z=0.02$ \citep[for details see][and
references therein]{knop94,scoville00,goldader02,bergvall02,kunth03,grimes06,grimes07}.
Haro 11 is a blue compact dwarf galaxy consisting of a number of UV
bright knots believed to be in the process of merging. VV 114 is a
merging system consisting of a UV-luminous Western component (VV
114W), and an IR-luminous Eastern component (VV 114E) with very little
associated UV emission.

In this paper, we make use of an ACS/HRC image of Haro 11 taken
through the filter F220W ($\lambda_c$$\approx$2255\AA) obtained from
the HST archive (Program 10575, PI: G\"oran \"Ostlin). For VV 114(W), we
use a Space Telescope Imaging Spectrograph (STIS) image in the NUV
($\lambda_c$$\approx$2365\AA) from Program 8201 (PI: Gerhardt Meurer).

\subsubsection{High redshift samples from \citet{lotz06}}

\noindent
Our high redshift comparison samples consist of 55 starburst galaxies
at $z\sim1.5$ selected from GOODS, and two (largely non-overlapping)
samples of in total 82 $z\sim4$ LBGs in GOODS and the UDF from
\citet{lotz06}.  The galaxy sample at $z\sim1.5$ is based on a large
spectroscopic sample of strong emission line galaxies. We used the
publicly available GOODS and UDF data, and extracted postage stamps of
the objects in \bp\ for the $z\sim1.5$ sample, and in \vp$+$\ip\ for
the $z\sim4$ sample.  A detailed description of the data and sample
selection can be found in \citet{lotz06}.

\section{Results}

\subsection{Images and notes on individual galaxies}

\begin{figure*}[t]
\begin{center}
\includegraphics[width=0.75\textwidth]{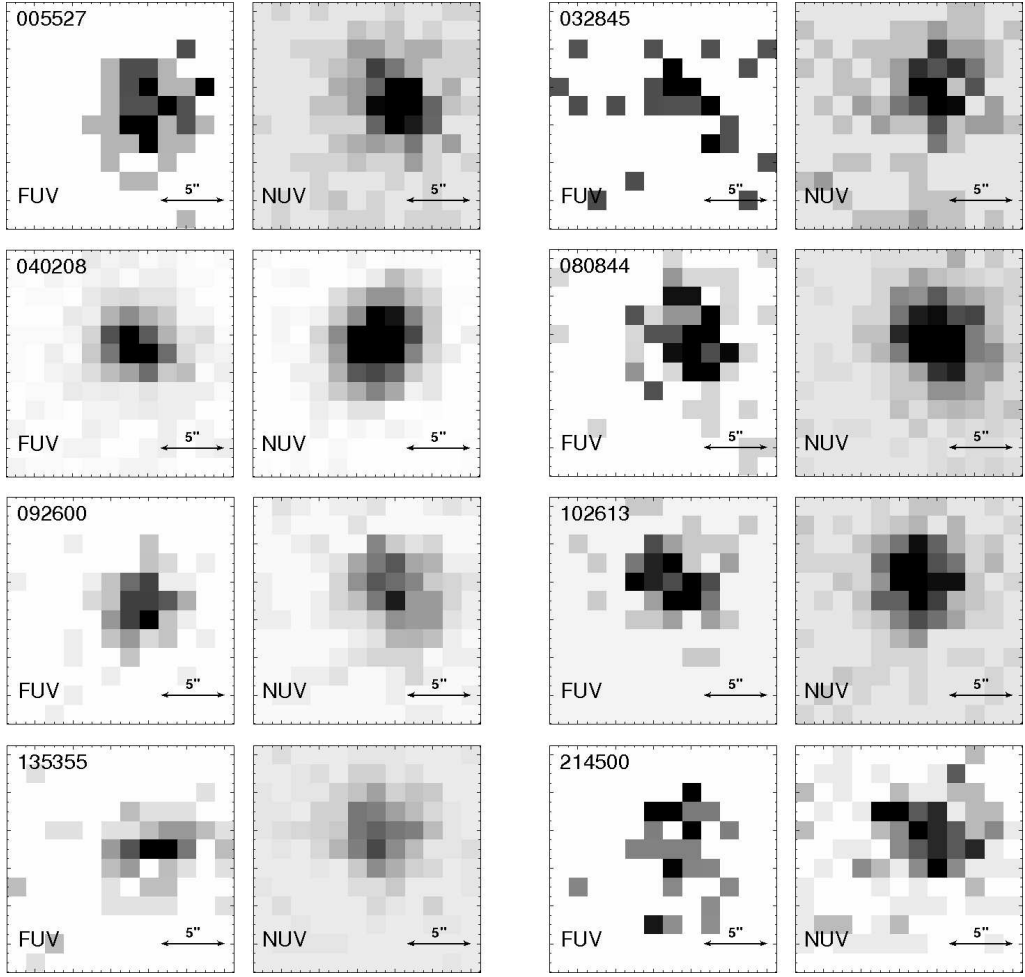}
\end{center}
\caption{\label{fig:uv}Panels show Galaxy Evolution Explorer (GALEX) in the far-UV (FUV) and near-UV (NUV) of the 8 objects in our sample. Bars in the 
lower right of each panel measure 5\arcsec, and are comparable to the full-width-at-half-maximum of the GALEX point spread function. North is up, East is to the left.}  
\end{figure*}

\begin{figure*}[ht]
\includegraphics[width=\textwidth]{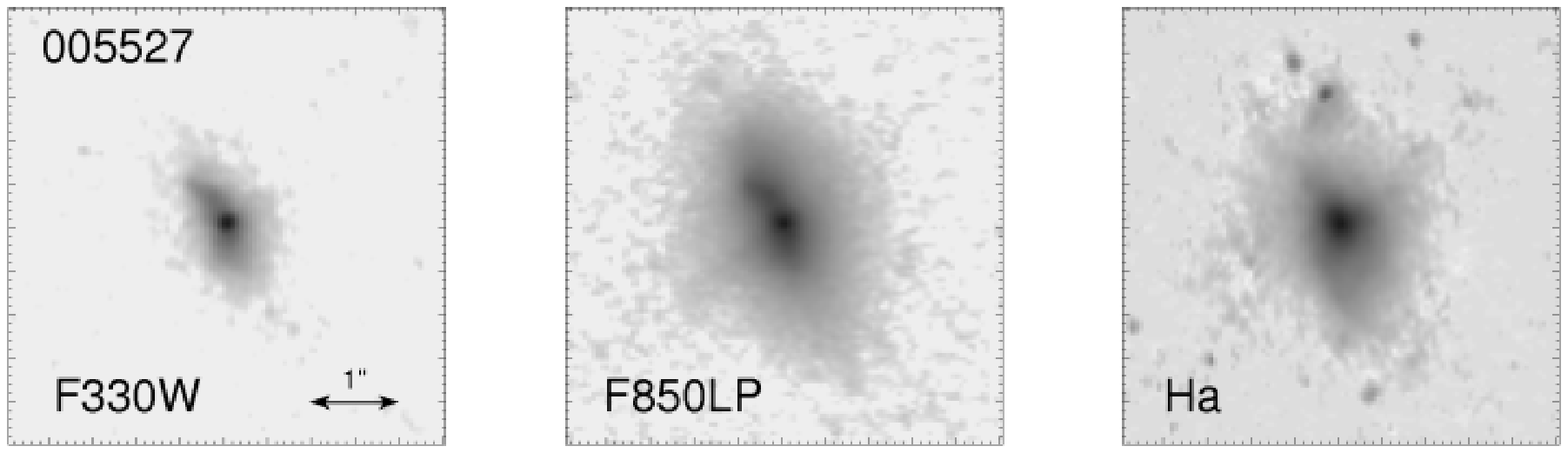}
\includegraphics[width=\textwidth]{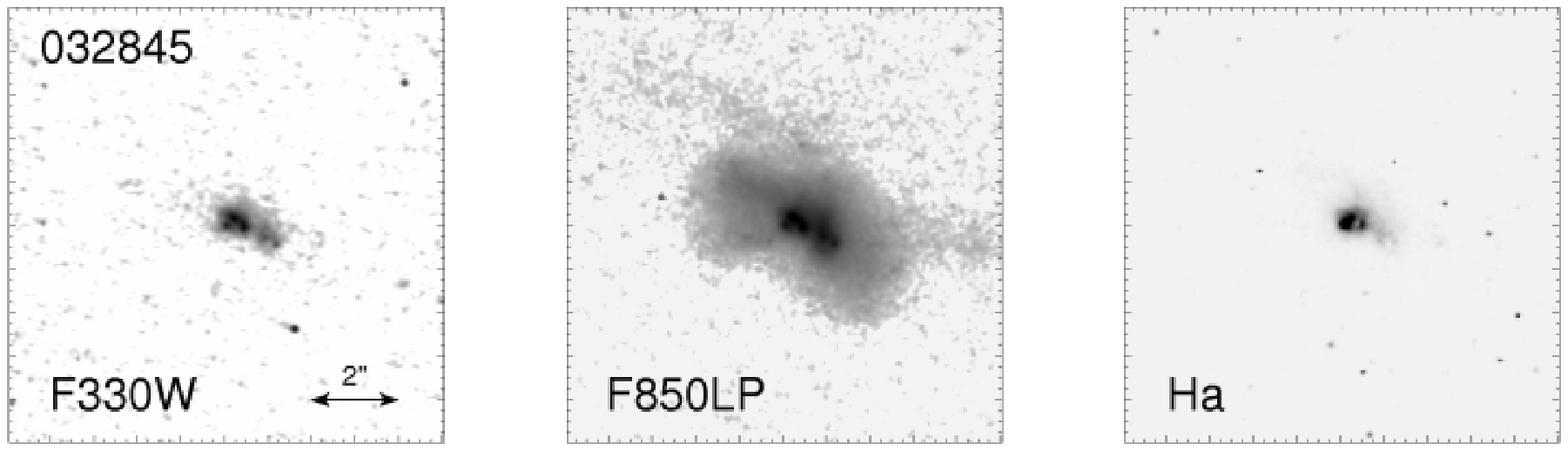}
\includegraphics[width=\textwidth]{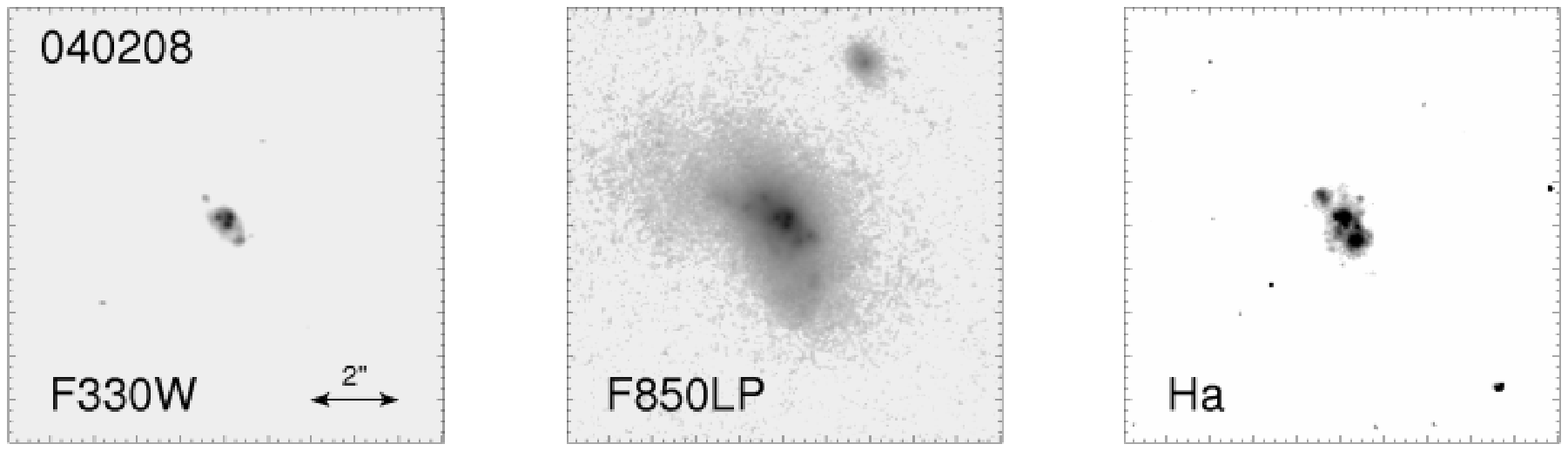}
\includegraphics[width=\textwidth]{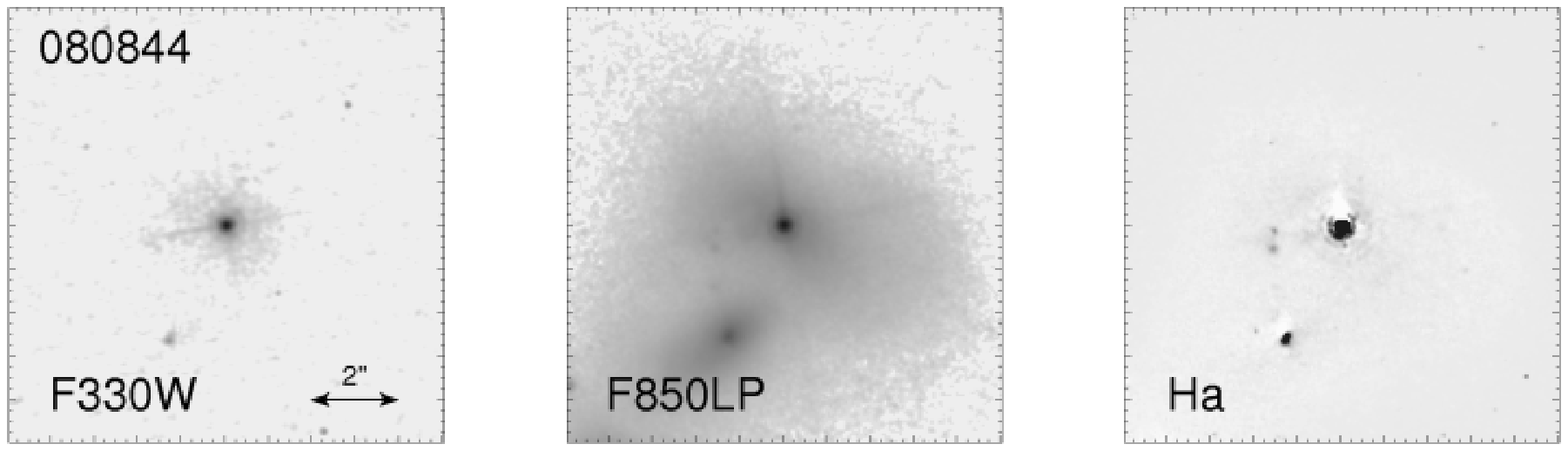}
\caption{\label{fig:stamps}Panels indicate the HST morphologies in the UV ({\it left}), optical ({\it middle}), and H$\alpha$ ({\it right}). 
The images were smoothed using a Gaussian filter of 1.5 pixels (FWHM) to reduce some of the noise at the sub-PSF level, 
and shown using a logarithmic stretch. The physical scales range from 1.6 kpc per arcsec at $z=0.09$ to 3.2 kpc per arcsec at $z=0.20$. 
The F330W image of 092600 and H$\alpha$ image of 214500 are not available. 
The H$\alpha$ image of 080844 suffers from poor PSF subtraction. North is up, East is to the left.}
\end{figure*}

\noindent
In Fig. \ref{fig:uv} we show the GALEX FUV and NUV images.  The UV,
H$\alpha$, and optical images taken with HST are shown in
Fig. \ref{fig:stamps}, which have a resolution of $\sim50$ times that
of the GALEX images. Qualitative remarks on the morphologies of the UVLGs in Fig. \ref{fig:stamps} are as follows.\\

\noindent {\bf SDSS J005527.46--002148.7}. This object consists of a strong
point source surrounded by diffuse emission that can be seen in \up,
\zp\ and \ha.  The diffuse emission has a high surface brightness
region extending to the NE with respect to the nucleus, which suggests
either the infall of a small diffuse galaxy or an additional
off-center star forming region.  The continuum-subtracted \ha\
emission is peculiar with an `arm' of emission extending to about
1\arcsec\ South of the nucleus. The morphology of the \ha\ is very
different from the \zp\ morphology, and at its outer extremity, the
\ha\ is not associated with any UV emission. We suspect that the
extended \ha\ is due to an outflow over the larger region probed by
the \zp\ continuum.

\noindent {\bf SDSS J032845.99+011150.8}. The \zp\ image suggests a merger
of two low surface brightness galaxies, as evidenced by the tidal
tails extending symmetrically from the center to the Northwest and
Northeast. The starburst dominates the nuclear region, in which five
isolated point sources can be seen.  The easternmost knot dominates in
\ha.

\noindent {\bf SDSS J040208.86--050642.0}. The \zp\ image suggests a merger
of two diffuse systems as the contours of faint \zp\ continuum
emission show a sharp bend in position angle. The image center is
dominated by three point sources, and several fainter ones lie further
out on either side of the center. One of these faint, compact regions
(about 1\arcsec\ South of the nucleus) is more extended in \ha.

\noindent {\bf SDSS J080844.26+394852.4}. There is a bright, unresolved
source seen in the \up\ image that is located in the centre of an
extended, highly diffuse galaxy seen in \zp.  The unresolved source is
remarkably bright as both \up\ and \ha\ show the ACS diffraction
spikes. The \ha\ image suffers from poor point source subtraction.
The object seen $\sim2\arcsec$ to the South lies at a similar
redshift, and may be interacting. The nucleus of this companion is
also detected in \up, but its flux is $\sim75\times$ lower.

\noindent {\bf SDSS J092600.41+442736.1}. The faint contours of the diffuse
emission in \zp\ suggest that at least two diffuse galaxies may be
merging.  A bright, compact component dominates in \zp\ and \ha. The
faint extension about 1\arcsec\ East of the nucleus in \zp\ has little
associated \ha\ emission. The \up\ image failed to execute due to a
guide star problem.

\noindent {\bf SDSS J102613.97+484458.9}. This galaxy has a ring-shaped
morphology with most of the starburst occurring along the Eastern rim.
There is also an isolated knot to the Southwest. The star-forming
region contains several bright, unresolved components and diffuse
emission. A very faint diffuse companion can be seen in \zp, about
2\arcsec\ to the South, and very faint extended emission, possibly
tidal debris, is seen directly to the East along the entire extent of
the galaxy. The overall morphology is very similar to that of the
``drop-through'' ring galaxy NGC922 studied by \citet{wong06}.

\noindent {\bf SDSS J135355.90+664800.5}. This UVLG is highly irregular in
all bands, showing many star-forming knots and diffuse emission.  The
object is interacting or merging with a warped, edge-on or filamentary
galaxy located just to the East, which is much redder and only just
detected in \up. A longslit spectrum along the major axis of the
system shows that the two objects are likely to be a counter-rotating
merger (Overzier et al., in prep.).

\noindent {\bf SDSS J214500.25+011157.6}. This is the only object in our
sample for which the ACS image provides a direct match at the
wavelength of the GALEX FUV image. The F150LP image shows a very
compact, but elongated starburst region in which we can discern three
different knots. Fainter FUV emission comes from a slightly more
extended region of $\sim$1\arcsec.  The optical morphology in \vp\ is
strikingly different. The image is still dominated by the emission
surrounding the starburst region, and a faint spiral structure may be
present. A companion object appears to have gone straight through the
star forming nucleus as evidenced by its tadpole-like morphology with
a faint tail pointing towards the main galaxy and a trail of tidal
debris that can be traced back to a location that is about equidistant
from the nucleus on the opposite side of the main galaxy.

\subsection{Emission lines}

\noindent
Although the sample of UVLGs does not contain any broad line AGN, we
want to make sure that we are studying the morphologies of starburst
galaxies rather than those of narrow line AGN. The main optical
emission line diagnostic diagrams involving the lines
[OIII]$\lambda5007$\AA, H$\beta$, [OI]$\lambda6300$\AA, H$\alpha$,
[NII]$\lambda6584$\AA, and the [SII]$\lambda\lambda$6713,6731\AA\
doublet (measured from the SDSS spectra) are shown in
Fig. \ref{fig:bpt} (see also H07). The 8 objects observed by HST are
indicated by the large filled circles and their IDs, while small
filled circles indicate objects in our supercompact UVLG sample
without HST data.  Most of the UVLGs lie along the main star forming
sequence (points) in log([OIII]/H$\beta$) vs. log([NII]/H$\alpha$)
(left panel of Fig. \ref{fig:bpt}), albeit offset towards higher
values of log([OIII]/H$\beta$ probably due to a more intense ionizing
radiation field. It is interesting to note that a similar offset from
the main star-forming sequence has been observed for high-redshift LBGs as
well \citep{shapley05,erb06a}. 

About one quarter of the sample falls in the region between the
relations of \citet{kauffmann03} (solid line) and \citet{kewley06}
(dashed line) that is typically populated by objects having a
composite spectrum consisting of a metal rich stellar population and
an AGN.  When we plot log([OI]/H$\alpha$) vs. log([OIII]/H$\beta$)
(middle panel) or log([SII]/H$\alpha$) vs. log([OIII]/H$\beta$) (right
panel), the objects that were in the AGN/starburst composite region
move to the far, opposite side of the star forming track (see object
SDSS J080844.26+394852.4 that is most relevant to this paper). It is not exactly clear
what causes this behavior of the line ratios, but it is consistent
with starburst model predictions having very high ionization
parameters \citep{kewley01} as might be expected for our sample.  A
more detailed analysis of the ionization parameters and the possibly 
remaining contribution from AGN will be given elsewhere (Overzier et al., in prep.). 
In any case, \citet{kauffmann03} have shown that the continuum emission from local
narrow-line AGN is almost always dominated by its stellar population;
this assures us that the HST images accurately reflect the morphology
of the stellar light.

\subsection{HST-based size measurements} 
\label{sec:sizes}
\subsubsection{UV sizes}

\noindent
Our sample of UVLGs is selected to have high surface brightness, which
is estimated from the FUV flux and a half-light radius measured from
fitting a seeing-convolved exponential profile to the SDSS {\it
  u}-band images (see Section 1, H05 \& H07). The resulting
seeing-deconvolved sizes were on the order of 1--2 kpc, but were
somewhat uncertain because most of the UVLGs are only barely resolved
in the {\it u}-band SDSS images.  The F330W images are a factor 10
higher in resolution and thus provide a reliable and independent test
for the true sizes of these UVLGs.  We have used SExtractor's circular
aperture method to estimate the radii; the 50 and 90\% radii were
computed using the light enclosed within 2.5 times the elliptical Kron
radius also determined using SExtractor (Table \ref{tab:sizes}). The
sizes measured from the HST images are somewhat smaller than our
previous estimates based on SDSS. In the case of SDSS J080844,
approximately half the light in F330W is coming from a region that is
still unresolved even at the resolution of the HRC ($\sim0\farcs07$,
or $\sim$110 pc at $z=0.091$).  Excluding this extreme case, the radii
range from 0.4 to 1.9 kpc.

One could question (see Appendix) whether the sizes measured in \up\
or {\it u}-band (used to calculate the UV surface brightness) are
representative for the sizes of the regions that emit in the FUV and
NUV as seen by GALEX at the much lower resolution of $\sim4\arcsec$
(FWHM). We can test this simply by measuring the slope, $\beta$, of
the UV continuum (with $f_\lambda\propto\lambda^\beta$) across the
GALEX bands, and comparing the flux measured in the HST image to the
extrapolated value at \up. The UV slopes\footnote{Calculated from
  $\beta_{GALEX}=2.32(m_{FUV}-m_{NUV})_{AB}-2$.} are given in Table
\ref{tab:phot}, and we note that they are very similar to that of
high-redshift LBGs \citep[e.g.][]{ouchi07}.  The measured flux in \up\
is lower than the predicted values by a factor of $\sim1.3$ on
average.  The measured flux in {\it u}-band is $\sim1.2\times$ lower
than the value extrapolated from the UV.  The generally good
correspondence between the predicted and measured continuum flux at
1500--3500\AA\ indicates that the emission structure seen in the HST
and SDSS images is representative of the light distribution in the
FUV.  A further test is provided by the case of SDSS J214500 
for which we have an ACS/SBC image taken through F150LP, a filter that
is almost identical to the GALEX FUV filter \citep[see][]{teplitz06}.
The corresponding FUV surface brightness calculated from the ACS data
is $\textrm{log}_{10}I_{FUV,ACS}=9.53$ $L_\odot$ kpc$^{-2}$, in
agreement with H07.

\subsubsection{H$\alpha$ and Optical sizes}

\noindent
The distribution of the optical and \ha\ light need not be the same.
We find, however, that the sources are compact at all observed
wavelengths (Table \ref{tab:sizes}), and the starbursts dominate the
structures.  The half-light radii measured from the continuum
subtracted \ha\ images are slightly larger than those measured from
\up\ ($\sim$0.5--2 kpc).  In some cases, the \ha\ morphology suggests
an outflow (SDSS J005527, and perhaps J032845 and J092600; see
Fig. \ref{fig:stamps}), but this does not affect the measured sizes.
The half-light radii in F850LP are typically 1--2 kpc, with the
exception of SDSS J135355 which has $r_{50,850}\approx4$ kpc
(note, however, that it has a nearby companion). Both the 50 and 90\%
radii are typically twice as large as the UV size ($\sim$2--5 kpc),
indicating that these systems have relatively faint, underlying
structures that were already in place prior to the current episode of
star formation.

\subsection{UV-optical Colors}

\noindent
We have used SExtractor to derive (circular) \up--\zp\ radial color
profiles out to the 90\% flux radius.  The results are shown in
Fig. \ref{fig:radial}.  Most objects have large gradients in their
UV-optical color profile, with objects typically being bluer near the
center, and redder further out.  Because the profiles were calculated
with respect to the centroid of the objects in the optical image, for
some objects (032845, 040208 \& 102613) the bluest regions are
sometimes slightly offset from the nominal center of the galaxy (see
Fig. \ref{fig:stamps}). We have defined an ``inner'' color measured
within the \up\ half-lightradius and an ``outer'' color measured over
the region between the 50 and 90\% radii in \zp\ (see Table
\ref{tab:sizes}).  In all cases the inner color is bluer than the
outer color, and the steepest color gradients are seen for the most
compact objects (in \up). This indicates that these objects are
composites of recent, central starbursts within older (or more dusty)
extended structures.

\subsection{Extinction}
\label{sec:slopes}

\noindent
For starburst galaxies there exists a good correlation between the
amount of UV flux that is absorbed by dust (usually taken to be a
foreground ``screen''), and the amount of flux (re-)emitted in the
far-IR \citep{meurer97,calzetti01,kong04,seibert05}. To estimate the
internal extinction for our sample we have calculated the attenuation
in the FUV, $A_{FUV}$, using the ratio of the bolometric dust
luminosity, $L_{TIR}$, to bolometric FUV luminosity, $L_{FUV}$, and
the fitting formula of \citet{burgarella05}:
\begin{equation}
A_{FUV}=a_1x^3+a_2x^2+a_3x+a_4,
\end{equation}
with $[a_1,a_2,a_3,a_4]=[-0.028,0.392,1.094,0.546]$ and $x=\mathrm{log_{10}}(L_{TIR}/L_{FUV})$. The bolometric dust luminosity is given by  
\citep{dale02}:\\
\begin{equation}
\L_{TIR}=\xi_1\nu_{24}L_{24}+\xi_2\nu_{70}L_{70}+\xi_3\nu_{160}L_{160},
\end{equation}
with $[\xi_1,\xi_2,\xi_3]=[1.559,0.7686,1.347]$. Although we do not
have observed flux densities at 160 $\mu$m, we can estimate 160 $\mu$m
using the strong empirical correlation between
$(f_8/f_{24})_{\mathrm{dust}}$ and $(f_{70}/f_{160})$ by employing the
synthetic models of \citet{dale02}. The correlation arises because
more intense radiation fields have larger $(f_{70}/f_{160})$ due to
hotter large grains, and smaller $(f_8/f_{24})_{\mathrm{dust}}$ due to
larger emission by small grains at 24 $\mu$m.  The dust emission at
8.0 and 24 $\mu$m was calculated from the observed flux densities
using $(f_8)_{\mathrm{dust}}=f_8-0.232f_{3.6}$ and
$(f_{24})_{\mathrm{dust}}=f_{24}-0.032f_{3.6}$, which includes a
correction for the contribution due to stars based on the 3.6 $\mu$m
flux \citep{dale05}. Finally, we calculate the extinction using the
formula $E(B-V)_{\mathrm{stars}}= A_{FUV}/k(\lambda)$ with
$k(\lambda)\approx10.77$ \citep{calzetti01}. The extinction ranges
from $E(B-V)\approx0.01$ to 0.14.  The IR flux densities, bolometric
luminosities, $A_{FUV}$ and extinction are given in Table
\ref{tab:ir}. The extinction will be needed in Section 4 when we
estimate masses and ages from the data using a set of model star
formation histories. We refer the reader to \citet{basuzych07} for a
more detailed discussion and analysis of the extinction properties of
this sample using alternative estimators.

\subsection{Morphologies}

\subsubsection{Gini coefficient, $M_{20}$, and Concentration}
\label{sec:morphologies}

\noindent
In order to compare the morphologies of the UVLGs to those of nearby
galaxies and of LBGs at high redshift, we calculated the Gini
coefficient ($G$; a measure of the equality with which the flux is
distributed across a galaxy), $M_{20}$ (the log of the ratio of the
second order moment of the pixels containing the 20\% brightest flux
to the total second order moment), and concentration ($C$; five times
the log of the ratio of the circular radii containing 80 and 20\% of
the flux). To calculate the structural parameters we follow the
procedures described in full detail in \citet{lotz04,lotz06}. First,
we use SExtractor to make an object segmentation map and mask out
neighboring objects. The image is background subtracted, and we
calculate an initial Petrosian radius ($r_P$ with $\eta\equiv0.2$)
using the object center and (elliptical) shape information from
SExtractor. We then smooth the image by $\sigma=r_P/5$ and create a
new segmentation map by selecting those pixels that have a surface
brightness higher than the mean surface brightness at the Petrosian
radius. We recalculate the object center by minimizing the second
order moment of the flux, and then recalculate the Petrosian radius in
the original image using this center. The total flux is defined as the
flux within a radius of $1.5\times r_P$ and $C$ is then calculated in
circular apertures. The individual measurements are listed in Table
\ref{tab:morphologies}, and the results are shown in
Fig. \ref{fig:morphologies}.  For each object, the \up\ measurements
are plotted as stars and the \zp\ as circles, and are connected by a
dotted line.  For 080844, $G$ and $C$ values in \up\ must be
considered lower limits, and its $M_{20}$ value is an upper limit
because the central pixel of the unresolved component contains
$\approx$20\% of the total flux.

Although our sample is small, we will attempt to make a general
comparison with the morphologies of nearby galaxies.  As shown in
Fig. \ref{fig:morphologies}, the UVLGs populate the region roughly
defined by $M_{20}\lesssim-1.3$ and $0.53\lesssim G\lesssim0.7$ (left
panel), and $3\lesssim C\lesssim5$ (right panel). We can compare these
measurements to the various divisions of morphological parameter space
discussed by \citet{lotz06}. Our UVLGs have smaller $M_{20}$ than
merging galaxies with two clearly separated nuclei
($M_{20}\gtrsim-1.1$, left hatched region), larger $G$ than the
expectations for exponential disks (dotted box), $M_{20}$ similar or
larger than those expected for bulge-dominated objects (right hatched
region), and $G$ and $C$ that are either similar or smaller than those
expected for bulge-dominated galaxies seen face-on (solid box). We do
not find any systematic differences between the morphologies measured
in \up\ and \zp\ for the sample as a whole. The morphological
parameters in the UV and optical are thus dominated by the same
(bright) regions, even though some of the objects possess very faint,
extended structures in \zp.

Fig. \ref{fig:morphologies} shows that two UVLGs qualify as
bulge-dominated in \up, while four objects qualify as bulge-dominated
in \zp.  None of the objects qualifies as a pronounced merger, and
only one object lies just on the border of the exponential disk region
in \zp, with the remainder populating the regions in between. The
overall distribution of morphologies is very similar to that of star
forming galaxies at $z\sim$1.5 and LBGs at $z\sim4$ selected from
GOODS and the UDF. \citet{lotz06} find median values of $G\sim0.55$,
$M_{20}\sim-1.5$ and $C\sim3.3$ at $z\sim1.5$, and $G\sim0.58$,
$M_{20}\sim-1.6$ and $C\sim3.8$ at $z\sim4$ (encircled crosses in
Fig. \ref{fig:morphologies}). However, a fair comparison of the
morphologies of UVLGs with the higher redshift comparison samples
should be carried out at the same $S/N$ and resolution. In the next
section, we will carry out such a comparison.

\subsubsection{Redshift simulations}
\label{sec:sims}

\noindent
We now investigate whether the conclusions of the previous subsection
are still upheld if the analysis is carried out using images that are
simulated to have the same depth and resolution as LBGs observed at
high redshift.

We follow common practice \citep[e.g.,
see][]{giavalisco96b,hibbard97,bouwens98,papovich03,conselice03,lotz04},
and apply corrections for cosmological surface brightness dimming and
for changes in physical resolutions to the \up\ images, in order to
match the instrumental conditions under which our objects would be
observed in typical surveys when placed at $z=1-4$.

The first step of the procedure is to rebin the \up\ images by a
factor $b=(\theta_1/\theta_2)(s_2/s_1)$, where $\theta_i$ is the angle
on the sky of an object of fixed size $d$ at $z=z_i$, and $s_i$ is the
instrumental pixel scale (in arcsec pixel$^{-1}$).  The rebinning
factor can be expressed in terms of either the angular diameter
distance, $D_{A_i}=d/\theta_i$, or the luminosity distance,
$D_{L_i}=(1+z_i)^2D_{A_i}$, at low and at high redshift ($z_2>z_1$):
\begin{equation}
b=\frac{D_{A_2}}{D_{A_1}}\frac{s_2}{s_1}=\left(\frac{1+z_1}{1+z_2}\right)^2\frac{D_{L_2}}{D_{L_1}}\frac{s_2}{s_1},
\end{equation}  
The second step is to reduce the surface brightness of each (rebinned) pixel according to the relative 
amount of cosmological dimming of a galaxy at $z_2$ 
with respect to that at $z_1$. We calculate the scaling by making use of the fact that the absolute rest-frame magnitude 
(or luminosity) of the object before and after redshifting will be conserved 
($M_{\lambda_2/(1+z_2)}=M_{\lambda_1/(1+z_1)}$ with matched filters so that $\lambda_2=\lambda_1(1+z_2)/(1+z_1)$). 

In order to compare our results to the high redshift samples of
\citet{lotz06}, objects were simulated at $z=1.5$ in \bp, and at
$z=3.0$ and 4.0 in both \vp\ and \ip. Objects were simulated at the
depths of GOODS (3, 2.5 and 2.5 orbits in \bp, \vp\ and \ip, resp.)
and the UDF (56, 56, 150 orbits in \bp, \vp\ and \ip, resp.).  For
completeness and future reference, we also simulated the sample at the
shallower COSMOS survey in the \inp\ filter (1 orbit).  For each
filter and for each redshift, we used the HST images that correspond
most closely in terms of rest-frame central wavelength. Where
necessary, we rescaled the flux of the input images assuming the UV
slopes determined in Section \ref{sec:slopes}.  The final steps of the
simulation consist of applying Poissonian noise to the simulated
profiles based on the typical exposure times of COSMOS/GOODS/UDF,
convolving the images with a Gaussian to match the desired output PSF
size ($\sim0\farcs12$), and placing the simulated object inside an
empty region in a COSMOS, GOODS or UDF image to obtain a realistic
background. The simulated pixel scale was $0\farcs03$ pixel$^{-1}$ for
GOODS and the UDF, and $0\farcs05$ pixel$^{-1}$ for COSMOS.

The rest-frame UV images of Haro 11 and VV 114 (see top panels of
Fig. \ref{fig:haro11vv114sims}) were artificially redshifted in an
identical manner using a two-step process.  First, we simulated their
F330W images at a redshift ($z=0.15$), depth (2500 s), plate scale
($0\farcs025$ pixel$^{-1}$), and seeing ($\sim0\farcs075$) comparable
to our observations of the UVLGs. These images are shown in the bottom
panels of Fig. \ref{fig:haro11vv114sims}.  Both Haro 11 and VV 114
possess multiple bright nuclei ($\sim$1\arcsec\ apart) when observed
in \up\ at $z=0.15$. VV 114 also shows a significant amount of diffuse
emission in between the knots.  Next, these images were used to
simulate how these objects would appear at high redshift, analogous to
the UVLG simulations described above.  To match the desired rest-frame
wavelengths of the output filters, small flux extrapolations were
performed using $\beta=-1.4$ for both objects as measured by
\citet{goldader02} and \citet{bergvall06}.

\subsubsection{Results}

\noindent
Figs. \ref{fig:sims_goods_uvlgs}, \ref{fig:sims_udf_uvlgs}, and
\ref{fig:sims_cosmos_uvlgs} compare postage stamps of the local LBG
analogs (including Haro 11 and VV 114) that are simulated at the depths of the GOODS, UDF, and COSMOS
surveys.  Results are shown at the three different redshifts discussed
above ($z=1.5$, 3.0 and 4.0). The images demonstrate that as redshift
increases, components that are well separated at low redshift blend
due to the lower spatial resolution, and that low surface brightness
features are lost due to surface brightness dimming. The latter effect
is most severe in the shallow COSMOS data, and least severe in the
deep UDF data.

Fig. \ref{fig:morphologies_redshift} shows how the measured
morphological parameters $G$ (top panel), $M_{20}$ (middle panel) and
$C$ (bottom panel) change as a function of redshift and survey depth.
There is a systematic drop of $\sim$0.05--0.10 in $G$ from $z\sim0.15$
to $z\sim1.5$, followed by a further less significant decrease of a
few hundredths out to $z=4$.  The first decrease is caused by the
strong drop in resolution, as well as the loss of faint extended
features that cause the flux to be more evenly distributed over the
galaxy profile. Simulations by \citet{lotz06} show that the second
decrease is mainly caused by the loss of low surface brightness pixels
lowering $G$ somewhat further at $z>1.5$. The loss of low surface
brightness features with redshift also tends to increase $M_{20}$ (the
total second order moment is lowered while the second order moment of
the 20\% brightest flux stays roughly constant), and tends to lower
the concentration, because the 80\% flux radii are systematically
underestimated. The measurements at the COSMOS depth show quite a
large scatter with respect to GOODS and the UDF, indicating that the
lower $S/N$ and its larger pixel scale result in relatively unstable
morphology measurements for LBGs.

The main results of our morphological comparison are presented in
Fig. \ref{fig:morphologies_lotz}. We compare the UVLG morphologies
(filled stars) with those of the low redshift LBG analogs (open stars)
and the high redshift comparison sample (plusses) of
\citet{lotz06}. In order to make sure that the morphologies are
measured in a consistent manner, we used our own code to recalculate
the morphologies of the 55 $z\sim1.5$ starburst galaxies in GOODS
(bottom panels), and the 82 $z\sim4$ LBGs in GOODS (middle panels) and
the UDF (top panels) analyzed by \citet{lotz06}. \citet{lotz06} showed
that the morphologies of the $z\sim1.5$ sample are very similar to
that of the $z\sim4$ sample. The distribution of the UVLGs is very
similar to that of the high redshift objects as well, although the
latter has a larger scatter due to the much larger sample size.  

\subsubsection{Comparison with Haro 11 and VV 114}

\noindent
Haro 11 and VV 114 (open stars in Fig. \ref{fig:morphologies_lotz})
have higher $M_{20}$ and lower $G$ and $C$ when compared to the
UVLGs. We can understand this given that both objects possess several
bright nuclei with a separation that enables them to be significantly
resolved even when simulated at $z=0.15-4.0$. Their qualitative
morphology perhaps most closely resembles that of the high redshift
``clump-cluster galaxies'' \citep[e.g., see][]{elmegreen04}. Although
objects with similar morphological characteristics are certainly
present in high redshift samples, they do not make up the majority of
LBGs as can be seen from the distribution in morphological types of
LBGs \citep[e.g.][and Fig. \ref{fig:morphologies_lotz} in this
paper]{lotz06,ravindranath06,elmegreen07}. 

It is important to note that our GALEX/SDSS selection of local LBG
analogs at $z\approx0.1-0.3$ is likely biased against finding objects
having such widely separated nuclei as Haro 11 and VV 114. Because we
selected objects principally on having a high FUV luminosity, as well
as a high UV surface brightness based on the half-light radius
measured in the {\it u}-band, we are most sensitive to objects that
are not or barely resolved in SDSS. The surface brightness requirement
is necessary in order to obtain a relatively clean separation between
the LBG analogs on one hand, and large, UV-luminous (predominantly
spiral) galaxies on the other. However, we will further investigate
whether we can find more LBG analogs of the type of Haro 11/VV 114 by
looking closer at the objects that straddle the boundary of the LBG
analog selection criteria in a future paper (Overzier et al., in prep.).

\subsubsection{Summary}

\noindent
We conclude that the morphologies of the UVLGs cannot be distinguished
from the morphologies of high redshift LBGs when measured from the
redshifted images simulated at the same depth as GOODS and the
UDF. This further strengthens our conclusion that the relatively
nearby supercompact UVLGs and high redshift LBGs are very similar.

In the following sections, we will turn our attention again to the
undegraded HST images of Fig. \ref{fig:stamps}. In
Sect. \ref{sec:sscs} we will study in detail the nature of the
starburst regions, and in Sect. \ref{sec:disc} we discuss the
connection between morphology and the mechanisms responsible for
triggering the vigorous star formation observed in these local LBG
analogs.

\section{Super starburst regions}
\label{sec:sscs}

\noindent
As discussed in Section 3, in all our sources the UV light is
distributed in a series of bright, unresolved knots embedded in a
region of more diffuse emission.  In this section we will show that
these regions can be interpreted as being (super) starburst regions
(SSBs).  An understanding of the physical nature of these regions is
crucial for interpreting the UV emission from LBGs.

\subsection{Identification and photometry}

\noindent
In Fig. \ref{fig:ssc_apertures} we indicate all the starburst regions
identifiable by eye from the \up-band images (small circles).
Although our identification by eye is somewhat subjective, most
regions are bright and isolated and thus easily recognised.  We
identify a total of 41 regions in our sample of 8 galaxies. Some
galaxies contain only one knot, while others contain as many as ten.
Next, we measured the fluxes and colors of each region using circular
aperture photometry in matched \up\ and \zp\ images (resolution of
$\sim0\farcs12$, FWHM).  We were able to isolate all of the starburst
regions identified in Fig. \ref{fig:ssc_apertures} using circular
apertures of 0\farcs3 in diameter (corresponding to physical radii of
$\sim$240-480 pc).
The encircled energy (EE) measurements of \citep{sirianni05} show that
circular apertures of this diameter enclose $\sim80$\% of the total
light of a point source observed with ACS. We did not apply the EE
correction.  Magnitudes were corrected for Galactic extinction, and
the magnitude distribution of the 41 starburst regions is shown in
Fig. \ref{fig:ssc_maghist}. Absolute magnitudes were calculated from
$M=m-5\mathrm{log}_{10}(D_L\mathrm{(pc)})+5+2.5\mathrm{log}_{10}(1+z)$. We
did not apply a $K$-correction as the central rest-frame wavelengths
at which the absolute magnitudes are determined differ only by a small
amount ($\lesssim300$\AA).  Fig. \ref{fig:ssc_maghist} shows that our
`by eye' selection is relatively complete at $m_{330}\lesssim23$ mag
($M_{330/(1+z)}\lesssim-16.5$).
We also determined the flux and color of the large, diffuse region
that generally surrounds the compact starburst knots (large circles in
Fig. \ref{fig:ssc_apertures}). These regions measure $\sim$2-5 kpc in
radius.  The fluxes measured in the small apertures were subtracted
from the larger aperture.

We find that the UV light is typically dominated by emission from the
compact, luminous regions: the combined flux of the unresolved regions
contributes an average of $\sim$50\% to the total flux in \up. Object
102613 has the lowest contribution from SSBs ($\sim34$\%) and object
080844 the highest ($\sim80$\%).  Although we do not have a \up\ image
for 092600, we can infer that its structure is very similar to the
other objects given that the unresolved knots contribute $\sim$42\% to
the total flux in \zp, and its morphology in \ha\ (assumed to be a
good proxy for its \up\ morphology, see Fig. \ref{fig:stamps}), shows
a similar knotty structure.

\subsection{Color-magnitude diagram}

\noindent
In Fig. \ref{fig:ssc_cm} we show the color-magnitude diagram of all
regions identified in Fig. \ref{fig:ssc_apertures}. Objects 092600 and
214500 were omitted.  Starburst regions from each galaxy are plotted
using a different filled symbol, with their corresponding ID number
from Fig. \ref{fig:ssc_apertures} indicated for reference. Open
symbols of corresponding shape indicate the diffuse regions that
surround the compact regions. In order to compare the measured colors
and magnitudes to model starburst tracks, we `dereddened' the
measurements using the global reddening value for the stellar
continuum that was derived from the bolometric dust to FUV luminosity
ratio (see Sect. \ref{sec:slopes}). We note that this correction is in
general quite small.
The compact regions span quite a large range both in color and in
absolute magnitude (filled symbols in Fig. \ref{fig:ssc_cm}). The
diffuse regions, defined by the larger areas in
Fig. \ref{fig:ssc_apertures} but with the circular regions around all
the identified point sources removed, lie at $M_U\sim-20$ and
(\up--\zp)$\sim0.5$ (open symbols in Fig. \ref{fig:ssc_cm}).

To interpret these colors, we have used STARBURST99
\citep{leitherer95,leitherer99} to predict the color and magnitude
evolution of a starburst as observed through our filters.  The tracks
in the left panel of Fig. \ref{fig:ssc_cm} represent an {\it
  instantaneous burst model} with burst masses of $M=10^7$ $M_\odot$
and $M=10^8$ $M_\odot$, while tracks in the right panel indicate a
{\it continuous star formation model} having SFRs of 0.1, 1 and 5
$M_\odot$ yr$^{-1}$.  We use a Kroupa initial mass function (IMF) with
a slope of $\alpha=1.3$ between 0.1 and 0.5 $M_\odot$ and $\alpha=2.3$
between 0.5 and 100 $M_\odot$.
We also use the Padova 1994 models with thermally pulsating AGB stars
added \citep{vazquez05}.  The tracks shown were obtained after
redshifting the rest-frame spectra to $z=0.15$, the average redshift
of our sample, and measuring the magnitudes in \up\ and \zp.  The blue
lines correspond to a metallicity equal to that of the Large
Magellanic Cloud ($Z=0.008$), while the red lines are for solar
metallicity.  Ages (in Myr) have been indicated along the
tracks. Dashed lines indicate the contribution from the nebular
continuum (assuming no reddening).  The general behaviour of all
tracks is to become redder with age. After about 10 Myr, the colors
become steadily redder, because the most massive O stars have become
supernovae and lower mass O stars have become red supergiants. The red
supergiant dominated phase at 10--20 Myr is highly dependent on
metallicity, as their number is lower and their temperature is warmer
in lower metallicity starbursts \citep{vazquez05}. After 20 Myr, O
stars have disappeared leading to a gradual reddening with age.
Continuous models become brighter, while instantaneous models fade
with time.

\subsection{Results}

\noindent
The instantaneous burst and continuous star formation models have been
chosen to bracket the range of possible star formation histories of
our compact starburst regions. Most of the regions are clearly
detected in \ha\ indicating that they must at least have had very
recent star formation.  In Fig. \ref{fig:ew} we plot the SSBs in the
plane of color vs. the rest-frame equivalent width (EW) of \ha\
measured for each SSB.  The measured equivalent widths are 30--500\AA,
and are seen to correlate with the \up-\zp\ color, suggesting that the
\ha\ EW is a reasonably good age indicator.  We have overplotted the
predicted EW of \ha\ based on the STARBURST99 models shown in Fig
\ref{fig:ssc_cm}.  For instantaneous models, the \ha\ EW is less than
$\sim$100\AA\ after 6 Myr, and less than 10\AA\ after only 10 Myr
(dashed lines). For continuous star forming models, a high \ha\ EW is
maintained over a much longer period of time, with
$EW_{H\alpha}\gtrsim$100\AA\ for ages $\lesssim$1 Gyr.

Comparison of the instantaneous burst and continuous star formation
tracks with the observed compact and diffuse regions lead us to the
following conclusions:

\noindent {\it (i)} The $M\simeq10^{7-8}$ $M_\odot$ instantaneous
starburst tracks shown in the left panel of Fig. \ref{fig:ssc_cm}
provide a generally good match to the majority of the compact regions,
but the ages of $\sim$10-100 Myr inferred are inconsistent with their
large \ha\ EWs, as shown in Fig. \ref{fig:ew}.  A better match to both
the color-magnitude diagram and the \ha\ EWs is provided by the
continuous star formation models shown in the right panel of
Fig. \ref{fig:ssc_cm}.  The exact ages are uncertain, because of
small-scale differences between the \up\ and \ha\ morphologies (note
that the apertures were defined on the \up\ image), uncertainties in
the small-scale dust distribution (now assumed to be a single
foreground screen), and because of the unknown fraction of the light
that is due to the background population inside each of the small
apertures (not subtracted).  Our best estimates for the ages of the
starburst regions range from a few tens of Myr to a few hundreds of
Myr based on a comparison with the tracks in Fig
\ref{fig:ssc_cm}. These ages should be regarded as upper limits,
because we may have underestimated the contribution from dust and
older stars.  The scatter in $M_U$ magnitude can be interpreted as a
sequence in SFR ranging from $\sim$0.1 to $\sim$5 $M_\odot$ yr$^{-1}$
(or total burst mass ranging from $\sim10^7$ $M_\odot$ to several
$10^8$ $M_\odot$). The scatter in (\up--\zp) color can be interpreted
as a sequence in age (or burst age in the case of instantaneous
bursts).  Note that our results are almost independent of metallicity.

\noindent {\it (ii)} For the brightest starburst regions, there is a
strong tendency for neighbouring regions to have a very similar color
and magnitude, e.g. regions (2,4,7) in object 135355, regions (1,4,5)
in object 040208, and regions (4,5) in object 032845. This suggests
that the starburst in these objects tends to occur in the form of
several co-eval and equally massive bursts.

\noindent {\it (iii)} The two compact regions that are outliers (in
the sense that they appear to be both extremely young, $\sim$6 Myr, as
well as very massive, $M_*\sim10^8$ $M_\odot$) each correspond to
objects where the UV emission is dominated by a {\it single}, luminous
knot (objects 005527 \& 080844).  Both objects have a very large \ha\
EW, indicating that they are massive, young starbursts.  The SDSS
spectrum\footnote{http://cas.sdss.org/dr5/en/tools/explore/obj.asp?id=588015508738539596}
of 005527 shows a broad feature around a bright and narrow HeII
4686\AA\ line that is characteristic of Wolf-Rayet (WR) galaxies
\citep[e.g.][]{conti91,sargent91,izotov97}.  The WR phase is a
relatively short phase in the life of the most massive stars during
which they experience a large mass loss.  The presence of WR stars is
a unique indicator of young starbursts, as WR stars disappear after
$\sim6-10$ Myr depending on the metallicity of the burst, with less
mass loss occurring at lower metallicity \citep{leitherer95}. Further
confirmation of this young age comes from the large equivalent width
of H$\beta$ ($\sim$60\AA), consistent with an age of $\sim$6 Myr
\citep{leitherer95}.

\noindent {\it (iv)} Some other regions with very blue colors and high
EW in \ha\ (e.g. region 2 in 032845 and region 1 in 102613) appear to
be as young as the previous two, but are $\sim$2-3 magnitudes fainter.

\noindent {\it (v)} The more diffuse, inner regions in which the SSBs
are embedded (open symbols in Fig. \ref{fig:ssc_cm}) are typically a
few tenths of a magnitude redder than the average color of the compact
SSBs they surround, and their total magnitudes are very similar to the
combined flux of their SSBs.  These regions contain an older stellar
population. Given their large physical extent, the instantaneous burst
model is unlikely to be appropriate.  We find that the colors and
luminosities of these regions are in good agreement with continuous
star formation models with SFRs of $\simeq1-5$ $M_\odot$ yr$^{-1}$ and ages 
of $\simeq0.2-1.0$ Gyr.

\noindent {\it (vi)} The outer annuli (defined as the region between
$r_{50,850}$ and $r_{90,850}$, see Table \ref{tab:sizes}) are
consistent with older populations having ages between a few Gyr to a
Hubble time, depending on the star formation history.

\subsection{Summary}

\noindent
The starburst regions observed in the local LBG analogs are highly
compact (radii of 100-300 pc), and characterized by bright, unresolved
knots of emission within a larger region of diffuse star formation
that extends out to a radius of a few kpc.  The total UV emission from
these compact starburst regions is substantial, indicating that
30--80\% of the total SFR is generated in these knots. Comparison with
STARBURST99 evolutionary tracks indicates that some of the regions are
due to very young ($\sim$6 Myr) bursts, while other regions may have
been forming stars in a more continuous manner for several tens to
several hundreds of Myr as indicated by their relatively red colors
and large \ha\ equivalent widths.  There is a tendency for neighboring
knots to have similar colors and luminosities. This implies that they
may be co-eval and of similar mass. Most of the SSBs are still
unresolved in the unbinned \up\ images ($\approx0\farcs075$,
FWHM). The stellar mass densities implied are $\sim10^{2-3}$ $M_\odot$
pc$^{-2}$ estimated from their masses of $\sim10^{7-8}$ $M_\odot$ and
assuming an effective radius of $\sim$100 pc.

The masses of the super starburst regions are one to two orders of
magnitude larger than the masses of the most massive clusters found in
the local Universe. We may therefore conclude that the (unresolved)
super starburst regions seen in Fig. \ref{fig:ssc_apertures} are
likely composed of smaller units. \citet{meurer95} found that typical
starburst regions in local starburst galaxies consist of diffuse stars
as well as clusters in the ratio of respectively 80 to 20\%. The most
prominent of the clusters, the so-called super star clusters (SSCs),
have masses of $\approx10^{5-6}$ $M_\odot$ and sizes of a few to
$\sim$10 pc and likely correspond to globular clusters in the process
of formation (provided the stellar IMF extends down to 0.1 $M_\odot$).

We can get an even better idea of the sub-resolution structure of the
super starburst regions by comparing the starburst regions in Haro 11
and VV 114 shown in the top panels of Fig. \ref{fig:haro11vv114sims}
with the same regions simulated at $z=0.15$ shown in the bottom panels
of the same figure. The majority of the bright nuclei seen in the
images simulated at $z=0.15$ are a blend of numerous smaller super
starclusters that can clearly be seen in their unredshifted images.
However, the most Northern and brightest knot in the redshifted image
of Haro 11 is still largely unresolved in its unredshifted image. This
indicates that single, highly massive star clusters are capable of
dominating the rest-frame UV/optical morphology of LBG-like galaxies
at virtually any redshift. A detailed determination of the sizes,
masses and ages of the star clusters in Haro 11 and VV 114 is beyond
the scope of this paper, and can be found elsewhere \citep[][Adamo et
al., in prep.]{scoville00,hayes07}.

\section{Discussion}
\label{sec:disc}

\subsection{Triggering mechanism}

\noindent
Although the number of objects in our HST sample is currently quite
small, the results presented in this paper can nevertheless guide us
in the interpretation of the morphologies of these local starburst
galaxies, and by extension, those of LBGs at higher redshifts.\\

\noindent
{\it (i) Are the morphologies evidence for merging?}\\

\noindent
Although some of the SDSS images showed evidence for close companions
or faint extended emission, the galaxies were extremely compact and it
was not at all clear from the SDSS images that they were highly
disturbed. The biggest surprise from the HST images was that most of
the UVLGs show disturbed morphologies at scales well below the SDSS
seeing or sensitivity.  Each UVLG shows at least one of the typical
signs of merging or interaction, such as multiplicity or twists in the
faint, outer isophotes (e.g., 005527, 032845, 040208, 092600), tidal
debris or tails (e.g., 032845, 102613, 214500), and close companions
(e.g., 080844, 102613, 135355, 214500).

While none of the UVLGs appear as fully-formed Hubble sequence
galaxies, in one case, 214500, a small spiral structure seems to be
present both in the FUV and in \vp.  Nonetheless, even in this object
there is circumstantial evidence that the starburst was triggered by a
merging event, as evidenced by its companion that appears to have gone
straight through its nucleus leaving behind a trail of tidal
debris. The projected distance between the centroid of the starburst
region and the companion is about 5 kpc. The companion galaxy would
traverse this distance in $\sim$10-100 Myr assuming a velocity of a
few hundred km s$^{-1}$ and a straight trajectory in the plane of the
sky. If we asume that the starburst is triggered by the interaction
(as suggested by the \vp\ morphology), this timescale constrains the
maximum possible age of the burst. The results are consistent with
30-100 Myr ages derived from the colors and magnitudes of these
systems.  In many of the other cases, the region that dominates the UV
morphology often lies at the interface of two diffuse merging
structures, as identified by the contours of their outer isophotes in
\zp. In the remaining cases, a companion is always seen at a distance
of less than $\lesssim$5 kpc from the UVLG.  It is well-known that
close neighbours can affect the SFR in galaxies, with close pairs
having separations of $\lesssim50$ kpc showing enhanced star formation
\citep[e.g.][Li et al. in
prep.]{larson78,lambas03,nikolic04,woods06,owers07} This supports the
idea that the starbursts are linked to mergers. The merging galaxies
must be very gas-rich in order to trigger the very high level of
enhanced star formation and the luminous super starburst regions that
we observe.

It has been proposed that starburst galaxies at high redshift, such as
Lyman break galaxies, are triggered by mergers of relatively gas-rich
disk galaxies \citep{somerville01}.
Simulations indicate that mergers of gas-rich galaxies can trigger
massive starbursts. The conditions of the starburst depends on a
number of key parameters of the merging system \citep[e.g. see][and
references therein]{mihos94a,mihos94b,springel05,dimatteo07}.  During
a merger of disk galaxies, bar-like patterns develop that efficiently
funnel gas into the central region.  The gas flow to the center is
generally not disrupted or depleted due to star formation during the
early stages of the merger, although the inflow of gas can be limited
due to resonances if a bulge component is present, reducing the
central gas densities and overall SFR and duration of the central
starburst.  The presence of gas is not always a sufficient condition
for generating a starburst. The star formation efficiency of a merger
depends largely on the galaxy spin direction (retrogade encounters
have higher efficiency), and on the amount of gas that is being
expelled due to the tidal interaction at first passage
\citep{springel05,dimatteo07}.

We remind the reader that typical ($L$$\sim$$L^*_{z=3}$) LBGs are not
dwarf galaxies, but have masses of $\sim$10$^{10}$ $M\odot$ and SFRs
that are much higher than those of typical (unobscured) local
starbursts \citep{papovich01,shapley01}. They do not show evidence for
spiral structure \citep{papovich05}.  As such, they are very similar
to our local LBG analogs. We conclude that the main mechanism
responsible for triggering star formation in both high redshift LBGs
and their local analogs is likely to be mergers.\\

\noindent
{\it (ii) Are the compact super-starburst regions triggered by the mergers?}\\

\noindent
Luminous star clusters (e.g. SSCs) occur in many irregular and dwarf
galaxies (e.g. 30 Dor in the LMC). The triggering of compact SSBs as
massive as the ones in our sample of UVLGs probably requires a highly
efficient inflow and compression of the gas that is typically only
seen in galaxy mergers or interactions. The inflow ensures that the
pressure in the interstellar medium becomes larger than the internal
pressure of giant molecular clouds, and that the clouds can collapse
and form SSCs before they are disrupted by supernova explosions \citep{bekki04}.
SPH simulations of galaxy mergers suggest that the properties of the
starburst are determined by the magnitude, timescale and geometry of
this inflow
\citep[see][]{mihos94a,bekki01,bekki04,dimatteo07}. \citet{mihos94a}
found that the central starbursts resulting from gas-rich mergers can
have total sizes of $\sim350$ pc, and contain most of the total mass
in new stars. \citet{bekki01} predict that (i) SSCs have a narrow age distribution
($\sim200$ Myr) because they form most efficiently during the stage
when the gas density is the highest, (ii) the total number and
intrinsic masses of SSCs are larger in major mergers due to larger
tidal effects and higher peak SFRs, and (iii) SSC production is more
efficient in merging galaxies \citep[e.g. ULIRGS,][]{sanders96} than
in tidally interacting galaxies \citep[e.g. M82,][]{kennicutt98}.

Observations show that starbursts indeed occur in regions where the
gas densities are the highest, typically in the cores \citep[e.g. Arp
220,][]{shaya94} and in circum-nuclear starburst regions
\citep[e.g. VV114,][]{scoville00}. Stars may form in large complexes
having total stellar masses of $\sim10^{7-9}$ $M_\odot$, and
containing several tens to hundreds of SSCs as well as smaller star
clusters. This is true both in ULIRGS and in compact blue galaxies
\citep[e.g.][and
therein]{meurer95,ostlin98,whitmore99,zepf99,goldader02}.

We conclude that the characteristic morphologies, ages, sizes and
masses of the starburst regions we have observed are consistent with
those predicted by simulations of gas-rich mergers. They are also
analogous to the large star-forming complexes observed in very nearby
merging galaxies.

\subsection{Low redshift lessons for high redshift LBGs}

\noindent
Because of their relative proximity, studies of local star-forming
galaxies can help us understand the nature of starburst galaxies at
high redshifts.  As explained in Section 1, up to now these
comparisons have been of limited usefulness, because typical local
starburst galaxies are systematically different from LBGs at high
redshift. The sample of local LBG analogs first studied by H05 was
specifically designed to bridge this gap. In this paper, we have
studied the detailed morphologies of these systems and we now discuss
how our results impact on the study of star-forming galaxies at high
redshifts.

\subsubsection{Morphologies and sizes}

\noindent
Our analysis of the HST data has established that UVLGs have a light
distribution dominated by very compact starbursting regions (ranging
in number from 1 to more than 10 per galaxy) that are triggered by
mergers or interactions.  In most cases, an older, diffuse, stellar
component is also present. Our redshift simulations showed that the sizes and morphologies at
$z=1.5-4$ are similar to that of LBGs
\citep[e.g.][]{giavalisco96a,conselice06,lotz06,overzier07,younger07}.
The morphological parameters measured at rest-frame UV and optical
wavelengths are very similar. This is unlike typical local galaxies,
but very similar to LBGs as both are dominated by young stars
\citep[e.g.][]{papovich05}.

The redshift simulations presented in Section 3 yield a number of
important conclusions: (i) at high redshift, the compact starburst
regions can blend giving rise to a galaxy image that appears to have
the one or two bright `knots' with a total half-light radius of 1--2
kpc, (ii) the diffuse emission is partly lost in the background noise
or may blend with the knots, and (iii) the {\it undegraded} optical
images were in most cases required to establish a definite connection
between the (UV) starbursts and merging (Section 5.1).

If `knots' are present in LBGs they are typically not resolved in HST
observations, ground-based spectroscopy with adaptive optics
\citep[resolution $\sim0\farcs1$ or $\sim$1 kpc;][]{genzel06,law07b},
or standard spectroscopy \citep[resolution $\sim0\farcs5$ or $\sim$4
kpc;][]{forster06,erb06b}. Because imaging observations of high
redshift LBGs lack the sensitivity and resolution to see the (often
subtle) features associated with merging, 2D or 3D spectroscopy is the
most promising method of studying the connection between star
formation and morphology in these objects. Small samples studied to
date show complex kinematics and high nuclear gas fractions that are
consistent with the merger hypothesis
\citep{erb03,erb06b,forster06,law07b}. However, some star-forming
objects at $z\sim2$ may have more ordered underlying structures
\citep{forster06,genzel06,wright07}.

Coming back to the important, long-standing question of whether the
irregular morphologies of LBGs are a sign of merging or of patchy star
formation within a single (forming) disk \citep[e.g.][]{law07a}, it is
safe to say that in {\it virtually every single case} presented in
this paper we do not see any direct evidence for merging based on the
HST UV images alone. However, in {\it virtually every single case}
evidence for merging is readily apparent from the rest-frame optical
image, and we reiterate that most of the features suggestive of
merging would be too faint or too small to be seen at $z$$\sim$2--4.
In some of the local LBG analogs the burst occurs predominantly in
only one of the members of a merging or interacting pair. In others
the merger seems in such an advanced state that one cannot speak about
separate systems any more, making the distinction between `merging'
and `patchy disk' essentially irrelevant. We conclude that the UV
morphologies of LBGs are likely {\it patchy as the result of merging}.

Although both the UV and optical morphologies are dominated by compact
starbursts, some of our objects show evidence for an older, extended
component, which is generally not seen in high redshift LBGs
\citep[e.g.][]{papovich05}.  In this respect, the UVLGs may be more
evolved than LBGs at high redshift. However, it is important to
remember that the age of the universe at $z\sim3$ is only $\sim2$
Gyr. The spectral energy distributions of LBGs are consistent with the
presence of such a ``maximally old'' stellar component
\citet{papovich01}.  The limiting surface brightness of rest-frame
optical images of LBGs in the NICMOS HDFN indicates that these
features are likely often too faint to be detected.  Also, the
rest-frame wavelength of the reddest NICMOS images of LBGs
($H_{160}$-band) corresponds to only about half that of our {\it
  z-}band observations of the UVLGs and so are less sensitive to
detection of an older population. Resolution effects could play a role
as well.  In paper II we will carry out redshift simulations of a
large set of rest-frame optical images of UVLGs to investigate these
optical structures in detail.

Because high surface brightness regions in LBGs are very likely to be
blends of super starburst regions,
we stress that one must be cautious in interpreting size/morphology
measurements in terms of disk- or bulge-like components.  The
half-light or effective radius may be more directly related to the
typical radius over which the compact, super starburst regions are
distributed, rather than to the scale size of a (forming) bulge or
disk \citep[see also][]{noguchi99,immeli04,law07b}, although the two
are likely closely related if stars predominantly form inside these
burst regions and relaxation or tidal disruption spreads them out over
time.  Although we do not rule out the possibility that some fraction
of LBGs has bulges at redshifts as high as $z\sim4$, the high physical
resolution data presented here suggests that this is possibly far less
common than currently believed based on morphological parameters
(e.g. Lotz et al. 2006 estimate $\sim$30\%).

Evidence that the small-scale structure of LBG knots is indeed of the
nature advocated in this paper is found in a few rare cases where LBGs
are observed at high magnification due to gravitational lensing. By
studying a lensed galaxy at $z=4.92$, \citet{franx97} were the first
to show that high surface brightness super starburst regions having
sizes of a few hundred pc form an essential contribution ($\sim$75\%)
to the UV flux in LBGs. Similar features can be seen in other lensed
systems that have been discovered at high redshift
\citep{bunker00,ellis01,smail07,swinbank07}.  We note that some
fraction of LBGs with very knotty structures may be discarded in
current large surveys, because their light profiles are similar to
those of stars.  Our sample indicates that this number could be on the
order of 13-25\% (1--2 out of 8 objects). If we compare the
substructure of the super starburst regions in Haro 11 and VV 114 at
$z=0.02$ with the same regions simulated at $z=0.15$
(Fig. \ref{fig:haro11vv114sims}), we see that majority of the bright
nuclei seen in the images simulated at $z=0.15$ are a blend of smaller
super star clusters identified in their unredshifted images.  However,
the most Northern and brightest knot in the redshifted image of Haro
11 is still largely unresolved in its unredshifted image, indicating
that single, luminous star clusters are capable of dominating the
rest-frame UV/optical morphology of LBG-like galaxies at virtually any
redshift.

As shown by \citet{dahlen07}, the fraction of star-forming galaxies
having a bulge-like morphology at rest-frame 2800\AA\ decreases from
$\sim$30\% at $z\approx2$ to $\sim$10\% at $z\approx0.5$, illustrating
the relative importance of concentrated star formation at high
redshift compared to low redshift.  \citet{zirm07} have pointed out
the existence of a population of compact, massive galaxies at
$z\sim2.5$ having effective radii and high stellar mass surface
densities. It is possible that these
are the result of massive and concentrated starbursts in highly dissipative, gas-rich mergers at high redshift ($z\sim3-6$). 
Extrapolation of our results to higher redshifts suggests that very dense stellar `cores' 
are indeed actively being formed in the LBG population.

\subsubsection{Prediction for the faint end slope of the LBG luminosity function}

\noindent
The UV luminosity function (LF) has now been evaluated over the entire
redshift range from $z=0-6$. Interestingly, the faint end slope of the
LF has been found to flatten with decreasing redshift from
$\alpha\sim$1.74 at $z=6$ to $\sim1.2$ at $z=0$
\citep{yan04a,yan04b,wyder05,bouwens07,ryan07}.  Various simulations
and models have been tried to reproduce the LF and explain its
flattening at faint magnitudes
\citep[e.g.][]{night06,finlator07,khochfar07}.  As we will show, the
existence of compact starbursts in LBGs may offer a natural
explanation for the steep slopes that are observed at high
redshifts. The LF of star-forming regions is well-known to have a
slope of $\alpha\approx2$ over a wide range in physical scales from
\hii\ regions to SSCs and beyond \citep[e.g.][and references
therein]{kennicutt89,meurer95,elmegreen97,zepf99,alonso02,bradley06}.

We consider the following, very simple, toy model. Let $N(L,z)$ be the
total number density of galaxies with UV luminosity between $L$ and
$L+dL$ and redshift between $z$ and $z+dz$. We will also assume that
the stellar IMF, as well as the star formation efficiency in galaxies,
does not change with redshift.  The UV luminosity of an individual
galaxy at a given time $t$ is the sum over the luminosity of all its
previous star-formation events, $L_i$, at that time:
\begin{equation}
\label{eq:lf}
L_{\mathrm{LBG}}^{UV}(t)=\sum_i L_i (t).
\end{equation}
The luminosity arising from an `event' can take any arbitrary form
(e.g. burst, constant, declining, etc.), and it is not important what
caused the event.  For LBGs, it is instructive to make a distinction
between the contribution to the total luminosity from a series of $N$
(semi-)discrete starbursts, $L_{\mathrm{burst}}^i$, and that due to a
diffuse star-forming component that is slowly evolving with time,
$L_{\mathrm{diffuse}}$, so that
\begin{equation}
L_{\mathrm{LBG}}^{UV}(t)  =  L_{\mathrm{diffuse}}(t) + \sum_{i=1}^N L_{\mathrm{burst}}^i(t).
\end{equation}
We have learned that the UV luminosity from an LBG is typically
dominated by a series of $N_{\mathrm{knot}}$ `knots' ($\sim$1--2 kpc
in size), and that each knot can be further resolved into a series of
$N_{\mathrm{SSB}}$ super starburst regions ($\sim$100-300 pc in size),
and thus
\begin{equation}
L_{\mathrm{LBG}}^{UV}(t)=L_{\mathrm{diffuse}}(t) + \sum_i^{N_{\mathrm{knot}}} \sum_j^{N_{\mathrm{SSB}}} L_{\mathrm{SSB}}^{i,j} (t),
\end{equation}
where $L_{\mathrm{SSB}}^{i,j}(t)$ is the luminosity of a SSB region
$j$ in knot $i$ at time $t$. For (proto-)galaxies that are just
undergoing their first episode of massive star formation the above
expression will simplify enormously, as we can neglect the diffuse
component, $L_{\mathrm{diffuse}}(t)$, and both $N_{\mathrm{knot}}$ and
$N_{\mathrm{SSB}}$ will be near unity.  The most simple case is
represented by $N_{\mathrm{knot}}=N_{\mathrm{SSB}}=1$, as observed in
some of our local LBG analogs, and (at least in this extreme case) the
total UV LF will be purely determined by a random sampling of the LF
of the SSBs over the galaxy population in a given redshift
interval. As shown by \citet{meurer95}, the `anatomy of starbursts'
typically consists of irregularly shaped cloud(s) (SSBs) with embedded
compact sources (SSCs), the brightest of which are massive forming
(globular) clusters. Star formation thus occurs in hierarchically
clustered systems ranging from sub-pc to multi-kpc scales, and is
believed to be generated at all scales by self-gravity and turbulence
within an underlying hierarchical gas mass distribution $n(M)dM\propto
M^{-\alpha} dM$ \citep[][and references
therein]{elmegreen96,elmegreen97}.  Because the UV luminosity of
starbursts is dominated by its OB stars at all scales, on average
$M\propto L$ with $\alpha\sim1.7-2$ for the references given above.
{\it Therefore, in the regime where the LF is dominated by young
  galaxies generating their first significant amount of stellar mass,
  i.e. at relatively faint magnitudes and high redshift, our simple
  model predicts that the galaxy LF observed will essentially be a
  random subsampling of the LF of starburst regions thereby
  maintaining its intrinsic slope.} It is interesting to note that,
for example, the faintest \bp, \vp, or \ip-dropout galaxies observed
at $z\sim4-6$ have stellar masses very similar to those of our
starburst regions ($\sim10^{8-9}$ $M_\odot$).

Generally, it will be impossible to model the full UV LF
quantitatively in this way (e.g. Eq. \ref{eq:lf}) without resorting to
semi-analytic models or simulations because it involves the
convolution of large numbers of galaxies with a range of star
formation histories including effects such as merging, feedback and
dust.  However, qualitatively one expects that, when all processes are
considered, the slope will flatten from its initial value.  Thus, we
expect that the faint end slope of the LF at least steepens with
redshift and will deviate less and less from the `pure' star formation
LF with a slope of $\alpha\approx2$, exactly as observed.

\section{Conclusions}

\noindent
The study of galaxies in the early Universe can benefit enormously
from the study of objects that are relatively nearby, provided that
suitable samples of local analogs can be found. H05 and H07 selected
``supercompact'', UV-luminous galaxies in a large GALEX-SDSS
cross-matched sample (see \S1).  Because these objects match LBGs in
terms of size, SFR, surface brightness, mass, metallicity, kinematics,
dust, and UV--optical color, this sample of local LBG analogs is well
suited for investigating the connection between star formation and
morphology at a level of detail and precision that is impossible for
high redshift galaxies.  Emission line diagnostics indicate that they
are dominated by starbursts, and that the line ratios show small
offsets with respect to the local star-forming population analogous to
those observed for high redshift LBGs (\S3.2).

In this paper we present HST imaging data in \up, \zp\ and \ha\ of a
subsample of 8 local LBG analogs in order to investigate their
morphologies. The effective radii estimated from the \up\ images range from
unresolved ($\lesssim$100 pc) to $\sim2$ kpc, confirming that the
objects are highly compact. The objects are slightly more extended in
\zp\ ($\sim$1--2 kpc), and they generally have color gradients with
relatively blue inner colors and redder outer colors. These gradients
are due to the fact that the starburst regions are very compact and
are embedded in an older, more extended population (\S\S3.3--3.4). The
internal extinctions amount to $E(B-V)\approx0.01-0.14$ mag based on
the bolometric dust to FUV luminosity ratios observed with Spitzer and
GALEX (\S3.5).  We have calculated the morphological parameters $G$,
$M_{20}$ and $C$ and find that local LBG analogs have irregular
morphologies that, on average, lie in between those expected for major
mergers, disks and bulges. We have simulated our data at the depth and
resolution of LBGs at $z=[1.5,3.0,4.0]$ in the COSMOS, GOODS, and UDF
surveys, and we have carried out direct comparisons with the
morphologies of $z\sim1.5$ star-forming galaxies and $z\sim4$ LBGs
extracted from these surveys.  The morphological parameters measured
for the artificially redshifted sample are very similar to those
obtained for the distant LBGs.  This establishes that local LBG
analogs and high-redshift LBGs are likely to be in a similar phase of
their evolution (\S3.6). The UVLGs are more typical of LBGs in
comparison with two very nearby LBG analogs, Haro 11 and VV 114 at
$z=0.02$.  The latter two are more similar to the ``clump cluster
galaxies'' due to their multiple nuclei at relatively large separation
\citep{elmegreen07}. Although more local LBG analogs of this type may
be found as well, we note that our current sample is biased against
finding such objects (\S3.6.4). 

Although a large fraction of LBGs at $z\sim3-4$ is consistent with
having exponential or $r^{1/4}$ light profiles, their visual
morphologies include multiple, double and perturbed systems.
Morphological parameters indicate that the majority of LBGs are likely
to be in a range of stages associated with minor and major galaxy
merging \citep[e.g.][]{giavalisco96a,lotz06,ravindranath06,elmegreen07}. In
contrast, \citet{burgarella06} find that the majority of luminous,
UV-selected star-forming galaxies at $z\sim1$ has a disk or spiral
morphology.  At a similar redshift, but at much lower UV luminosities,
\citet{demello06} find a mix of morphological types that include
early-type galaxies as well as starbursts. We note that this is a
direct consequence of the selection of galaxies purely based on their
UV emission. As discussed here and by H05, at relatively low redshifts ($z\sim0-1$) 
the construction of a sample of galaxies that most closely resembles
high redshift LBGs in terms of their physical properties requires
additional selection criteria (e.g. UV surface brightness) rather than
selection on UV luminosity or the presence of a `Lyman break' alone.

We have carried out a detailed investigation of the small-scale
structure of the local LBG analogs in terms of their star forming
regions (\S4).  The starburst regions observed in the local LBG
analogs are highly compact (radii of 100-300 pc), and characterized by
bright, unresolved knots of emission within a larger region of diffuse
star formation that extend up to a few kpc in size.  The total UV
emission from these compact super starburst regions (SSBs) is
substantial, indicating that 30--80\% of the total SFR is generated in
these knots. Comparison with STARBURST99 evolutionary tracks indicates
that some of the regions are due to very young ($\sim$6 Myr) and
massive ($\sim10^{8}$ $M_\odot$) bursts, while other regions may have
been forming stars in a more continuous manner for several tens to
several hundreds of Myr as indicated by their relatively red colors
and high \ha\ equivalent widths. The super starburst regions are
likely to be a blend of diffuse star forming regions, stellar
associations and super or globular star clusters, as demonstrated by
comparing the images of Haro 11 and VV 114 at $z=0.02$ to the same
images redshifted to $z=0.15-4.0$.
The SSBs are generally embedded in a
diffuse, older component having ages ranging from a few Gyr to a
Hubble time, depending on the star formation history. In this respect,
the precursors of the UVLGs may have been more evolved than those of
LBGs at high redshift (although the presence of a ``maximally old''
stellar component in LBGs is consistent with the observations).  

Although some of the SDSS images showed limited evidence for close
companions or faint extended emission, the optical HST images clearly
reveal disturbed morphologies at scales well below the SDSS seeing or
sensitivity.  Each object shows evidence of merging or interactions,
such as multiplicity of position angles, tidal debris or tails, or
close companions. Most morphological information about high redshift
galaxies has been derived from rest-frame UV images.  In our sample,
the \up-band does not establish a definite case for a merger in any
single one of our galaxies. The disturbed optical ({\it z}-band) morphologies, together with the
luminous, compact super starburt regions, suggests that mergers of
relatively gas-rich objects trigger vigorous episodes of star
formation, with general properties reminiscent of simulations of
collisional starbursts (\S5.1). Our results on local LBGs and their similarity to their high-$z$ LBGs
constitute the most direct evidence to date that the onset of star
formation and morphological structures of high redshift LBGs are
largely driven by highly dissipational merging.

We discuss several implications of our results for the interpretation
of lower resolution data on high redshift LBGs.
Resolution effects will cause the compact starburst regions to blend
into typically one or two bright LBG-like `knots'. Although these
knots will appear relatively featureless with total half-light radii
of 1--2 kpc,  
the substructure of our LBG analogs indicates that they should generally
not be interpreted as being evidence of bulges or disks. Evidence that the
substructure of high redshift LBGs is indeed dominated by SSBs is
found in a few cases where LBGs are highly magnified by gravitational
lensing (\S5.2.1). However, the strong relation we find between
merging and the triggering of very compact, massive super starburst
regions indicates that these events may be closely linked to the
formation of stellar bulges. Furthermore, we suggest that the
prominence of luminous, unresolved (super) starburst regions in
forming galaxies may provide a natural explanation for the value of
the faint end slope of the UV luminosity function at high redshift by
an extension of the local star-forming region luminosity function
which is well-known to have a power law slope of $\alpha\approx2$
(\S5.2.2).
 
{\it Future work.--} In Paper II we will study a significantly larger
sample of local LBG analogs to be observed with HST in Cycle 16 in the
FUV/optical using the ACS/SBC and WFPC2.  Follow-up spectroscopy is
being used to study emission line ratios and kinematics, measure
outflows, and identify companion objects.  The ongoing surveys with
GALEX will provide larger samples of local LBGs that will be used to
derive better sample statistics.

\begin{acknowledgments}
  This paper has benefited from discussions and helpful comments from
  numerous friends and colleagues.  We thank Casey Papovich, Masami
  Ouchi and Isa Oliveira for carefully reading through the
  manuscript. We further thank Rychard Bouwens, Nick Cross, Ricardo Demarco,
  Marijn Franx, Lisa Kewley, Cheng Li, Crystal Martin, Alessandro Rettura, Samir
  Salim, Christi Tremonti, Arjen van der Wel and Andrew Zirm for
  discussion of various parts of this paper. RAO thanks Gabrelle Saurage for her 
excellent support during observations at APO. 

  Based on observations made with the NASA/ESA Hubble Space Telescope,
  which is operated by the Association of Universities for Research in
  Astronomy, Inc., under NASA contract NAS 5-26555.  These
  observations are associated with program \# 10920.  Based on
  observations obtained with the Apache Point Observatory (APO)
  3.5-meter telescope, which is owned and operated by the
  Astrophysical Research
  Consortium. 
\end{acknowledgments}

\clearpage

\begin{table*}[t]
\scriptsize
\caption{\label{tab:log}Sample and log of HST/ACS observations.} 
\begin{tabular}{lllrrcccccc}
\hline
\hline
\multicolumn{1}{c}{Galaxy} & \multicolumn{1}{c}{$\alpha$} & \multicolumn{1}{c}{$\delta$} & \multicolumn{1}{c}{$z$\tablenotemark{a}} & \multicolumn{1}{c}{UT date} & \multicolumn{5}{c}{$T_{exp}$ ($s$)} & $E(B-V)_{gal}$\tablenotemark{b}\\
 & (J2000) & (J2000) &  &  &  \multicolumn{1}{c}{$FUV_{150}$} &  \multicolumn{1}{c}{\up}  & \multicolumn{1}{c}{H$\alpha$} & \multicolumn{1}{c}{\vp} & \multicolumn{1}{c}{\zp} & \multicolumn{1}{c}{(mag)}\\
\hline 
SDSS J005527.46--002148.7 & 00$^h$55$^m$27.46$^s$ & --00$\degr$21$\arcmin$48.7$\arcsec$ & 0.167 & 11/01/06 & --   & 2514 & 2302 & --   & 2238 & 0.03\\
SDSS J032845.99+011150.8  & 03$^h$28$^m$45.99$^s$ &  +01$\degr$11$\arcmin$50.8$\arcsec$ & 0.142 & 10/07/06 & --   & 2514 & 2302 & --   & 2238 & 0.11\\
SDSS J040208.86--050642.0 & 04$^h$02$^m$08.86$^s$ & --05$\degr$06$\arcmin$42.0$\arcsec$ & 0.139 & 10/31/06 & --   & 2514 & 2302 & --   & 2238 & 0.10\\
SDSS J080844.26+394852.4  & 08$^h$08$^m$44.26$^s$ &  +39$\degr$48$\arcmin$52.3$\arcsec$ & 0.091 & 10/30/06 & --   & 2541 & 2356 & --   & 2211 & 0.05\\
SDSS J092600.41+442736.1  & 09$^h$26$^m$00.40$^s$ &  +44$\degr$27$\arcmin$36.1$\arcsec$ & 0.181 & 11/06/06 & --   & --   & 2340 & --   & 2274 & 0.02\\
SDSS J102613.97+484458.9  & 10$^h$26$^m$13.97$^s$ &  +48$\degr$44$\arcmin$58.9$\arcsec$ & 0.160 & 11/22/06 & --   & 2565 & 2354 & --   & 2289 & 0.01\\
SDSS J135355.90+664800.5  & 13$^h$53$^m$55.90$^s$ &  +66$\degr$48$\arcmin$00.5$\arcsec$ & 0.198 & 01/04/07 & --   & 2661 & 2468 & --   & 2334 & 0.02\\
SDSS J214500.25+011157.6  & 21$^h$45$^m$00.25$^s$ &  +01$\degr$11$\arcmin$57.3$\arcsec$ & 0.204 & 07/10/07 & 2514 & --   & --   & 3600 & --   & 0.06\\
\end{tabular} 
\tablenotetext{a}{Spectroscopic redshifts from SDSS.}
\tablenotetext{b}{Galactic extinction as given by \citet{schlegel98}.}
\end{table*}

\begin{table*}[t]
\caption{\label{tab:sfrs}Star formation rates and stellar masses.} 
\begin{tabular}{llcll}
\hline
\hline
\multicolumn{1}{c}{Galaxy}  &  \multicolumn{1}{c}{SFR$_{FUV}^a$}  &  \multicolumn{1}{c}{SFR$_{IR}^b$} &  \multicolumn{1}{c}{SFR$_{1.4 GHz}^a$} & \multicolumn{1}{c}{$M_*^c$} \\
 & \multicolumn{1}{c}{($M_\odot$ yr$^{-1}$)} & \multicolumn{1}{c}{($M_\odot$ yr$^{-1}$)} & \multicolumn{1}{c}{($M_\odot$ yr$^{-1}$)} & \multicolumn{1}{c}{(10$^{10}$ $M_\odot$)}\\
\hline
SDSS J005527.46--002148.7 &  $4.2\pm0.8$ & 28    &  14.9$\pm$2.0  & 1.17 \\
SDSS J032845.99+011150.8  &  $3.2\pm1.1$ & $<$3.4  &   $<$4.6       & 0.55 \\
SDSS J040208.86--050642.0 &  $3.7\pm0.6$ & $<$2.8  &   $<$4.1       & 0.25 \\
SDSS J080844.26+394852.4  &  $3.1\pm0.4$ & 7.3   &   $<$3.7       & 2.14 \\
SDSS J092600.41+442736.1  &  $6.8\pm0.4$ & 3.7    &   7.8$\pm$1.4  & 0.16 \\
SDSS J102613.97+484458.9  &  $3.4\pm0.1$ & 6.6    &   9.2$\pm$1.3  & 0.81 \\
SDSS J135355.90+664800.5  &  $7.1\pm0.9$ & 14.6    &  22.5$\pm$2.7  & 0.98 \\
SDSS J214500.25+011157.6  &  $3.4\pm0.8$ & 7.5    &  14.5$\pm$2.9  & 1.32 
\end{tabular}
\tablenotetext{a}{UV SFR derived from the FUV luminosity (uncorrected for dust). Adapted from \citet{basuzych07}.}
\tablenotetext{b}{IR SFR derived from the bolometric dust luminosity (given in Table \ref{tab:ir}), and calculated using the formula 
$\mathrm{log}_{10}L_{dust}=9.75+\mathrm{log}_{10}SFR_{IR}$ and assuming a 100 Myr old constant SFR model with Salpeter IMF as given in \citet{takeuchi05}.}
\tablenotetext{c}{Stellar mass based on SED model fitting. Adapted from H07 \citep[see also][]{salim07}.}
\end{table*}

\begin{table*}[h]
\caption{\label{tab:phot}Photometry and UV spectral slope.} 
\begin{tabular}{lrrrrrr}
\hline
\hline
\multicolumn{1}{c}{Galaxy}  &  \multicolumn{1}{c}{$m_{FUV}$} & \multicolumn{1}{c}{$m_{NUV}$}  & \multicolumn{1}{c}{$m_{330}$}  & \multicolumn{1}{c}{$m_{u}$} & \multicolumn{1}{c}{$m_{850}$} &  \multicolumn{1}{c}{$\beta_{FUV-NUV}$} \\
 & (mag) & (mag) & (mag)  & (mag) & (mag)    & \\
\hline
SDSS J005527.46--002148.7 & 19.23$^{+0.15}_{-0.17}$ & 18.85$^{+0.08}_{-0.08}$ & 18.78 & $18.65\pm0.02$ & 18.25 &  $-1.13\pm0.22$  \\
SDSS J032845.99+011150.8   & 19.42$^{+0.24}_{-0.31}$ & 19.23$^{+0.14}_{-0.16}$ & 18.97 & $18.68\pm0.04$ & 18.15 &  $-1.56\pm0.39$  \\
SDSS J040208.86--050642.0 & 18.90$^{+0.06}_{-0.06}$ & 18.70$^{+0.03}_{-0.03}$ & 18.81 & $18.67\pm0.03$ & 18.08 &  $-1.54\pm0.09$  \\
SDSS J080844.26+394852.4   & 18.22$^{+0.10}_{-0.11}$ & 17.59$^{+0.05}_{-0.05}$ & 17.62 & $17.48\pm0.01$ & 17.05 &  $-0.54\pm0.15$  \\
SDSS J092600.41+442736.1   & 18.83$^{+0.11}_{-0.13}$ & 18.89$^{+0.07}_{-0.08}$ & --$^a$& $19.04\pm0.03$ & 19.28 &  $-2.12\pm0.18$  \\
SDSS J102613.97+484458.9   & 19.27$^{+0.03}_{-0.03}$ & 19.00$^{+0.02}_{-0.02}$ & 18.86 & $18.85\pm0.02$ & 18.23 &  $-1.38\pm0.05$  \\
SDSS J135355.90+664800.5   & 18.99$^{+0.04}_{-0.04}$ & 18.57$^{+0.02}_{-0.02}$ & 18.42 & $18.35\pm0.02$ & 17.67 &  $-1.03\pm0.06$  \\
SDSS J214500.25+011157.6   & 19.94$^{+0.25}_{-0.32}$ & 19.22$^{+0.12}_{-0.13}$ &19.61$^b$&$19.24\pm0.04$& 18.62$^c$ &$-0.35\pm0.41$\\
\end{tabular}
\tablenotetext{a}{We did not obtain data in F330W.}
\tablenotetext{b}{These measurements correspond to the ACS/SBC F150LP data, instead of ACS/HRC F330W.}
\tablenotetext{c}{These measurements correspond to the WFPC2 F606W data, instead of ACS/WFC F850LP.}
\end{table*} 

\clearpage

\begin{table*}[t]
\scriptsize
\caption{\label{tab:sizes}Size measurements and colors} 
\begin{tabular}{lccccccccc}
\hline
\hline
\multicolumn{1}{c}{Galaxy}  & \multicolumn{1}{c}{$r_{50,u}$} & \multicolumn{1}{c}{$r_{50,330}$} & \multicolumn{1}{c}{$r_{90,330}$} & \multicolumn{1}{c}{$r_{50,850}$} & \multicolumn{1}{c}{$r_{90,850}$} & \multicolumn{1}{c}{$r_{50,H\alpha}$} & \multicolumn{1}{c}{$r_{90,H\alpha}$} & \multicolumn{1}{c}{(\up--\zp)$_\mathrm{inner}^e$} & \multicolumn{1}{c}{(\up--\zp)$_\mathrm{outer}^f$} \\
  & (kpc)  & (kpc) & (kpc) & (kpc)   & (kpc) & (kpc) & (kpc) & (mag)  & (mag)\\
\hline
SDSS J005527.46--002148.7 & 0.77  & 0.37    & 1.26    & 0.78    & 2.13     &  0.50  & 1.69   & 0.03 & 0.89  \\
SDSS J032845.99+011150.8  & 1.66  & 0.87    & 1.96    & 1.79    & 4.93     &  0.93  & 2.49   & 0.24 & 1.32 \\
SDSS J040208.86--050642.0 & 1.16  & 0.84    & 1.94    & 1.44    & 4.26     &  1.23  & 2.81   & 0.48 & 0.84  \\
SDSS J080844.26+394852.4  & 0.47  & 0.16    & 1.41    & 0.57    & 2.77     &   0.47    &  2.94& --0.58 & 1.52 \\
SDSS J092600.41+442736.1  & 0.97  & --$^a$  & --$^a$  & 1.10    & 3.33     &   0.87 & 2.67   & --$^a$ & --$^a$ \\
SDSS J102613.97+484458.9  & 2.07  & 1.89    & 3.86    & 1.99    & 4.48     &   2.13 & 4.22   & 0.53 & 0.67  \\
SDSS J135355.90+664800.5  & 1.81  & 1.46    & 3.19    & 3.56    & 9.06     &   2.74 & 7.23   & 0.48 & 0.81   \\
SDSS J214500.25+011157.6  & 0.74  & 1.10$^b$& 3.55$^b$& 1.30$^c$& 4.64$^c$  & --$^d$ & --$^d$ & 0.93 & 1.00 \\
\end{tabular}
\tablenotetext{a}{We did not obtain data in F330W.}
\tablenotetext{b}{These measurements correspond to the ACS/SBC F150LP data, instead of ACS/HRC F330W.}
\tablenotetext{c}{These measurements correspond to the WFPC2 F606W data, instead of ACS/WFC F850LP.}
\tablenotetext{d}{We did not obtain data in \ha.}
\tablenotetext{e}{Inner color measured within $r_{50,330}$ (not corrected for internal reddening).}
\tablenotetext{f}{Outer color measured between $r_{50,850}$ and $r_{90,850}$ (not corrected for internal reddening).}
\end{table*} 

\begin{table*}[t]
\scriptsize
\caption{\label{tab:ir}Infrared photometry and extinction.} 
\begin{tabular}{lrrrrrcccc}
\hline
\hline
\multicolumn{1}{c}{Galaxy}  & \multicolumn{1}{c}{$f_{3.6}$} & \multicolumn{1}{c}{$f_{8}$} & \multicolumn{1}{c}{$f_{24}$} & \multicolumn{1}{c}{$f_{70}$} & \multicolumn{1}{c}{$f_{160}^a$} &  \multicolumn{1}{c}{$L_{TIR}^b$} & \multicolumn{1}{c}{$L_{FUV}^c$} &  \multicolumn{1}{c}{$A_{FUV}^d$} &  \multicolumn{1}{c}{$E(B-V)_{int}^e$} \\
 & ($\mu$Jy)  & ($\mu$Jy) & (mJy)    & (mJy) & (mJy) & \multicolumn{1}{c}{($10^{10}$ $L_\odot$)}  & \multicolumn{1}{c}{($10^{10}$ $L_\odot$)} & \multicolumn{1}{c}{(mag)} &  \multicolumn{1}{c}{(mag)}\\
\hline
SDSS J005527.46--002148.7 &  725.8$\pm$33.2 &   4280.5$\pm$80.8 & 33.2$\pm$0.4 & 163.4$\pm$6.4 & 115.4 & 16   & 3.2  & 1.48  & 0.14\\
SDSS J032845.99+011150.8  &  218.3$\pm$39.0 &    972.0$\pm$71.9 &  3.4$\pm$0.4 & $<$23.4$^f$    &  25.9 & $<$1.9  & 1.8  & 0.56  & 0.05\\
SDSS J040208.86--050642.0 &  131.6$\pm$34.8 &    415.3$\pm$66.6 &  1.1$\pm$0.1 & $<$19.8$^f$    &  25.5 & $<$1.6  & 2.8  & 0.31  & 0.03\\
SDSS J080844.26+394852.4  & 1048.6$\pm$39.8 &   5786.5$\pm$66.2 & 39.9$\pm$0.6 & 156.0$\pm$6.8 & 117.0 & 4.1  & 2.0  & 0.91  & 0.08\\
SDSS J092600.41+442736.1  &   90.7$\pm$4.2  &    229.4$\pm$24.0 &  3.4$\pm$0.4 &  20.7$\pm$3.7 &  10.2 & 2.1  & 5.5  & 0.16  & 0.01\\
SDSS J102613.97+484458.9  &  176.6$\pm$36.2 &    614.5$\pm$56.8 &  4.2$\pm$0.3 &  42.5$\pm$4.2 &  31.6 & 3.7  & 2.7  & 0.70  & 0.06\\
SDSS J135355.90+664800.5  &  369.2$\pm$32.4 &   1660.3$\pm$53.1 & 44.4$\pm$0.4 &  70.1$\pm$4.3 &  27.2 & 8.2  & 6.0  & 0.71  & 0.07\\
SDSS J214500.25+011157.6  &  200.4$\pm$31.5 &   1617.5$\pm$64.2 &  5.0$\pm$0.4 &  23.0$\pm$5.5 &  27.7 & 4.2  & 2.6  & 0.79  & 0.07\\
\end{tabular}
\tablenotetext{a}{Estimated using the observed flux densities at 3.6, 8.0, 24 and 70$\mu$m and the dust models of \citet{dale02}.}
\tablenotetext{b}{Bolometric dust luminosity.}
\tablenotetext{c}{Bolometric FUV luminosity.}
\tablenotetext{d}{FUV attentuation calculated from $L_{TIR}/L_{FUV}$ and using the fitting formula of \citet{burgarella05}.}
\tablenotetext{e}{Internal (stellar) extinction $E(B-V)_{int}=A_{FUV}/k(\lambda)$ \citep{calzetti01}.}
\tablenotetext{f}{Indicates a 3$\sigma$ upper limit.}
\end{table*} 

\begin{table*}[t]
\caption{\label{tab:morphologies}Morphological parameters} 
\begin{tabular}{lcccc|cccc}
\hline
\hline
\multicolumn{1}{c}{Galaxy}  & $r_{P,330}$ & $G_{330}$ & $M_{20,330}$ & $C_{330}$ & $r_{P,850}$ & $G_{850}$ & $M_{20,850}$ & $C_{850}$\\
 & (\arcsec) & & & & (\arcsec) & & &\\
\hline
SDSS J005527.46--002148.7 &  0.55    & 0.63 & -2.41 & 4.73  & 0.83 & 0.58 & -1.98 & 3.53\\
SDSS J032845.99+011150.8  &  1.66    & 0.66 & -1.48 & 3.97  & 2.62 & 0.58 & -1.55 & 3.34\\
SDSS J040208.86--050642.0 &  1.02    & 0.56 & -1.60 & 3.29  & 1.92 & 0.60 & -1.98 & 3.92\\
SDSS J080844.26+394852.4  &  0.13    & $>$0.57 & $<$-2.55 & $>$3.01 & 2.97 & 0.66 & -3.23 & 7.11\\
SDSS J092600.41+442736.1  &  --$^a$  & --$^a$   & --$^a$    & --$^a$    & 1.29 & 0.73 & -1.60 & 3.27\\
SDSS J102613.97+484458.9  &  2.21    & 0.60 & -1.31 & 3.28 & 1.92 & 0.58 & -1.48 & 2.93\\
SDSS J135355.90+664800.5  &  1.26    & 0.60 & -1.27 & 2.88 & 1.29 & 0.53 & -1.57 & 2.88\\
SDSS J214500.25+011157.6  &  0.75$^b$ & 0.56$^b$ & -1.47$^b$ & 2.95$^b$ & 1.40$^c$ & 0.61$^c$ & -1.80$^c$ & 3.56$^c$
\end{tabular}
\tablenotetext{a}{We did not obtain data in F330W.}
\tablenotetext{b}{These measurements correspond to the ACS/SBC F150LP data, instead of ACS/HRC F330W.}
\tablenotetext{c}{These measurements correspond to the WFPC2 F606W data, instead of ACS/WFC F850LP.}
\end{table*}

\clearpage

\begin{figure*}
\includegraphics[width=\textwidth]{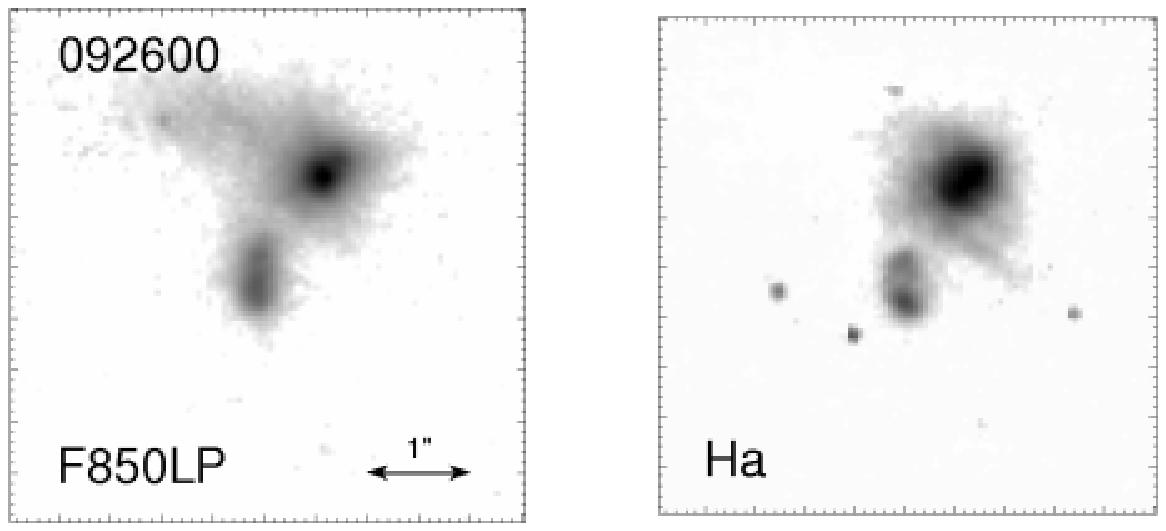}
\includegraphics[width=\textwidth]{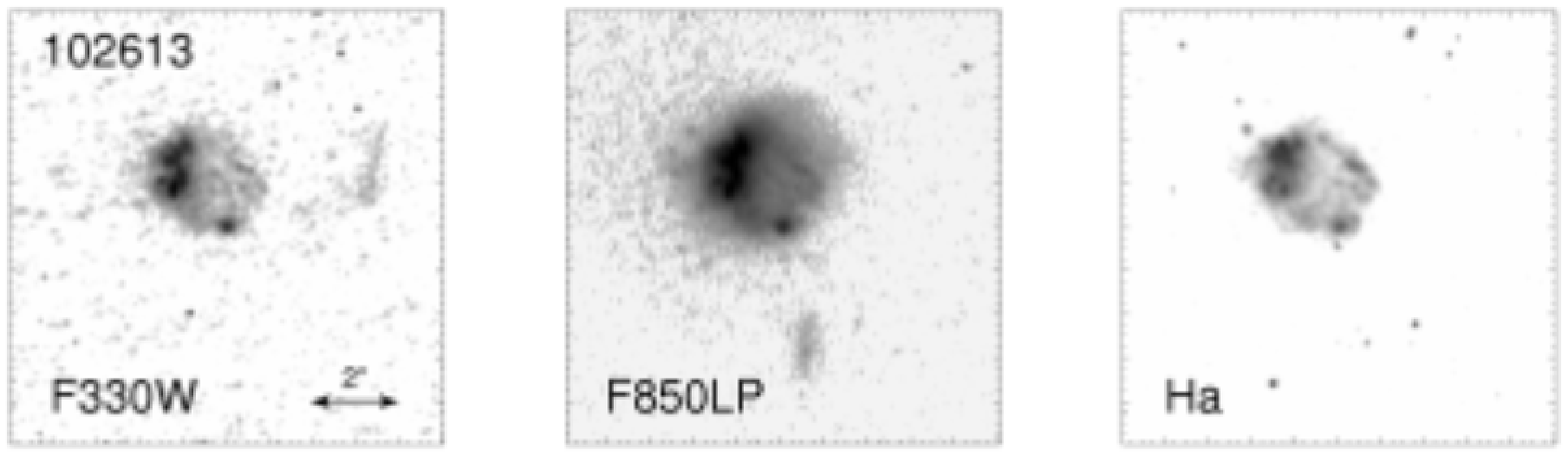}
\includegraphics[width=\textwidth]{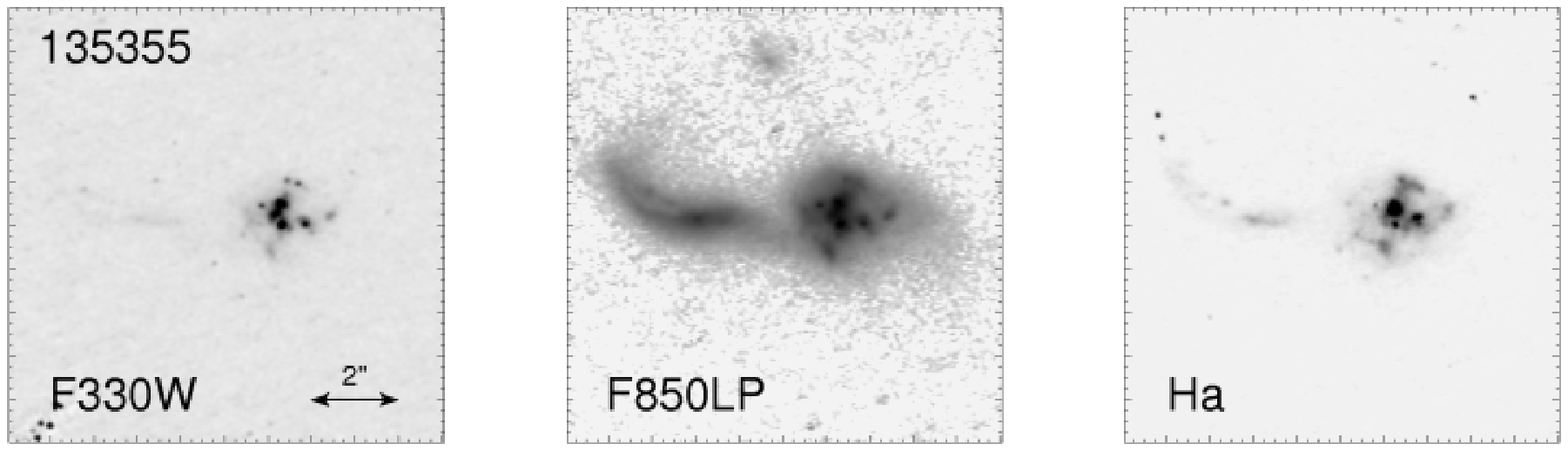}
\includegraphics[width=\textwidth]{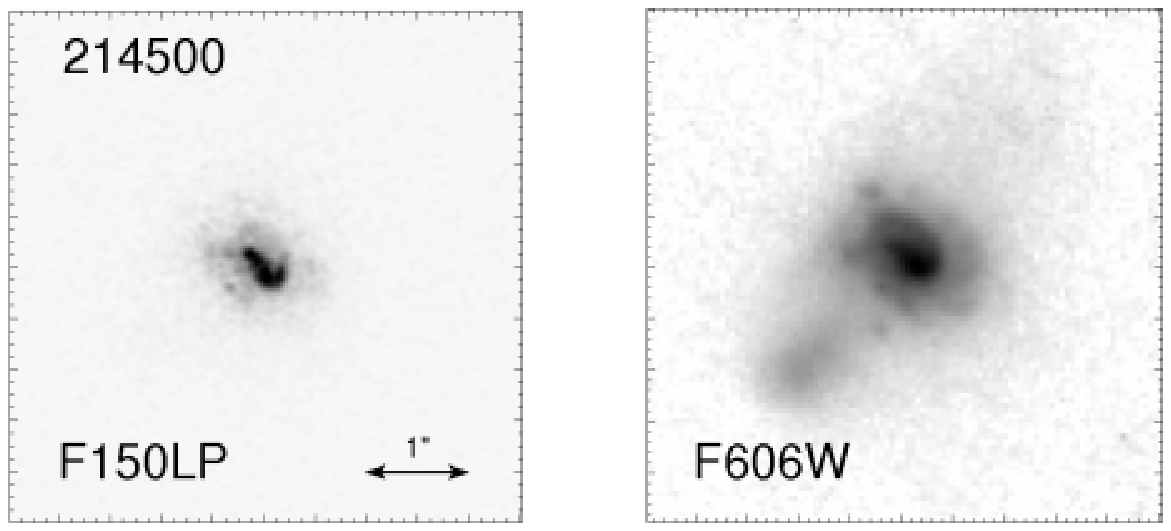}
Fig. 1.-- Cont.
\end{figure*}

\begin{figure*}[t]
\begin{center}
\mbox{
\includegraphics[width=0.3\textwidth]{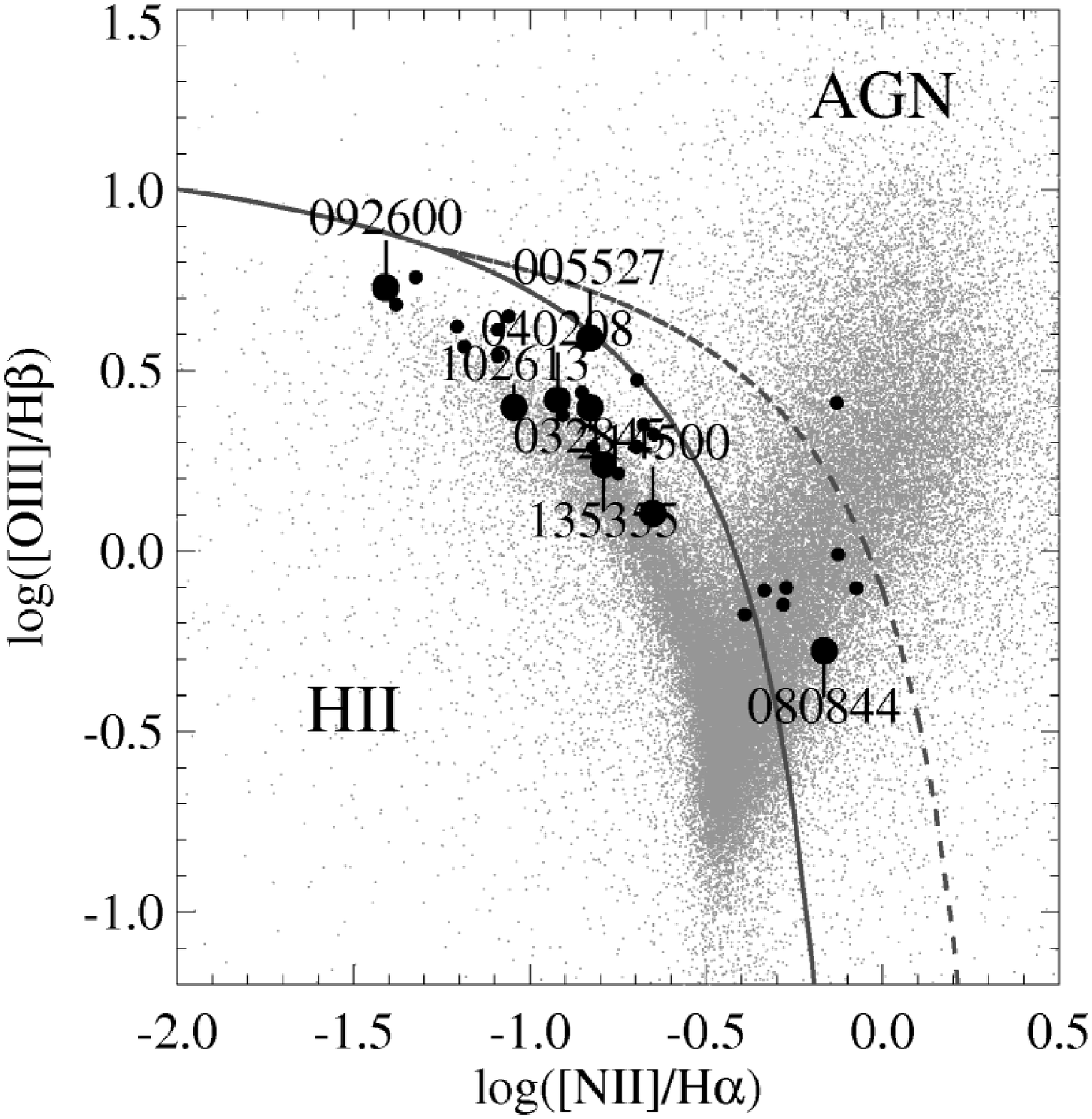}
\includegraphics[width=0.3\textwidth]{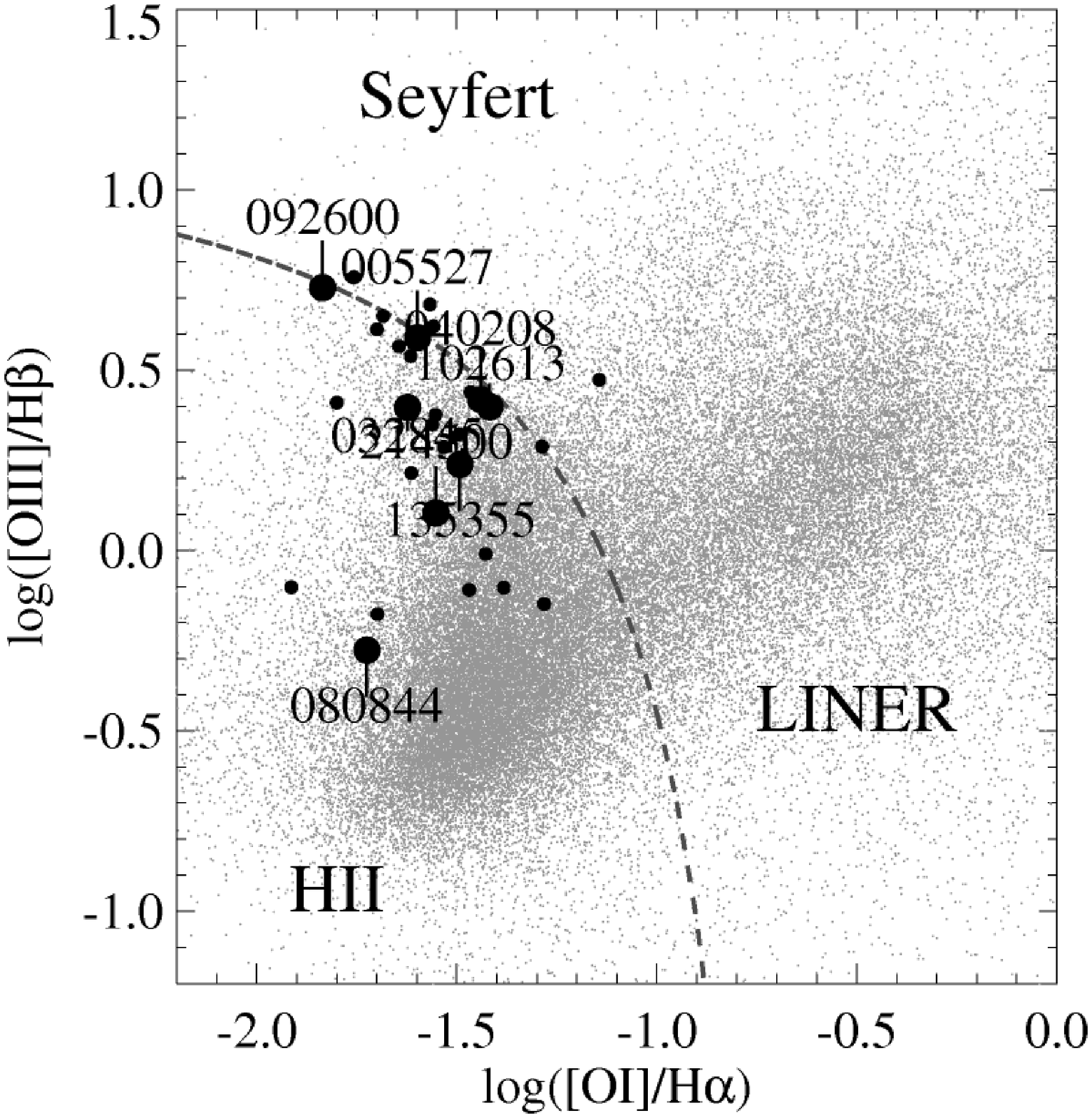}
\includegraphics[width=0.3\textwidth]{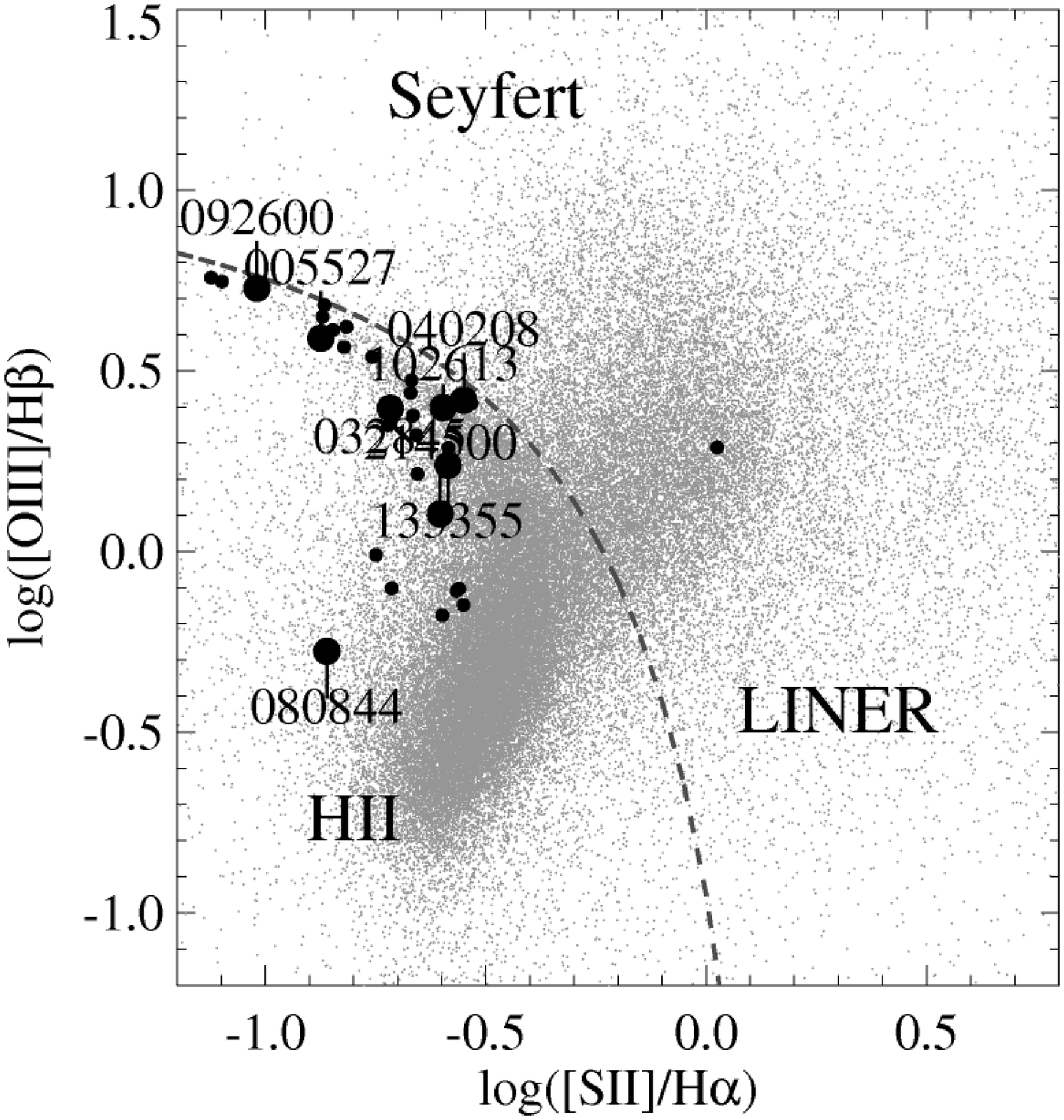}}
\end{center}
\caption{\label{fig:bpt}Optical emission line diagnostic diagrams
  log([OIII]/H$\beta$) vs. log([NII]/H$\alpha$) ({\it left panel}),
  log([OI]/H$\alpha$) ({\it middle panel}), and log([SII]/H$\alpha$)
  ({\it right panel}). Points indicate all objects in the GALEX Medium
  Imaging Survey (MIS) Galex Release 2 (GR2) sample cross-matched with
  the SDSS DR4 sample. The objects that are the subject of the current
  paper are indicated by the large filled circles.  Small filled
  circles indicates the other objects in our main sample of compact
  UVLGs that have HST, Spitzer and radio data either taken or
  scheduled. The solid line indicates the boundary between regions
  typically populated by starbursts and Seyferts from \citet{kauffmann03}.
  Dashed lines indicates the boundary between objects of starburst
  spectral type and pure AGN from \citet{kewley06}. Objects between
  the solid and dashed lines in the left panel have typically
  composite spectra consisting of a metal rich stellar population and
  an AGN. Note that, based on these diagnostics, object
  J080844.26+394852.3 is peculiar. The standard diagnostics
  diagram (left panel) indicates it could be of composite type, while
  it lies well on the starburst side in the other diagrams (middle and
  right panels). This behavior is typical for galaxies with very high ionization parameters \citep{kewley01}. 
The remainder of our HST objects lies in the starburst dominated region, albeit offset towards relatively high values of log([OIII]/H$\beta$. A similar offset 
has been found for LBGs at high redshift \citep{shapley05,erb06a}.}
\end{figure*}

\begin{figure}[t]
\begin{center}
\includegraphics[width=\columnwidth]{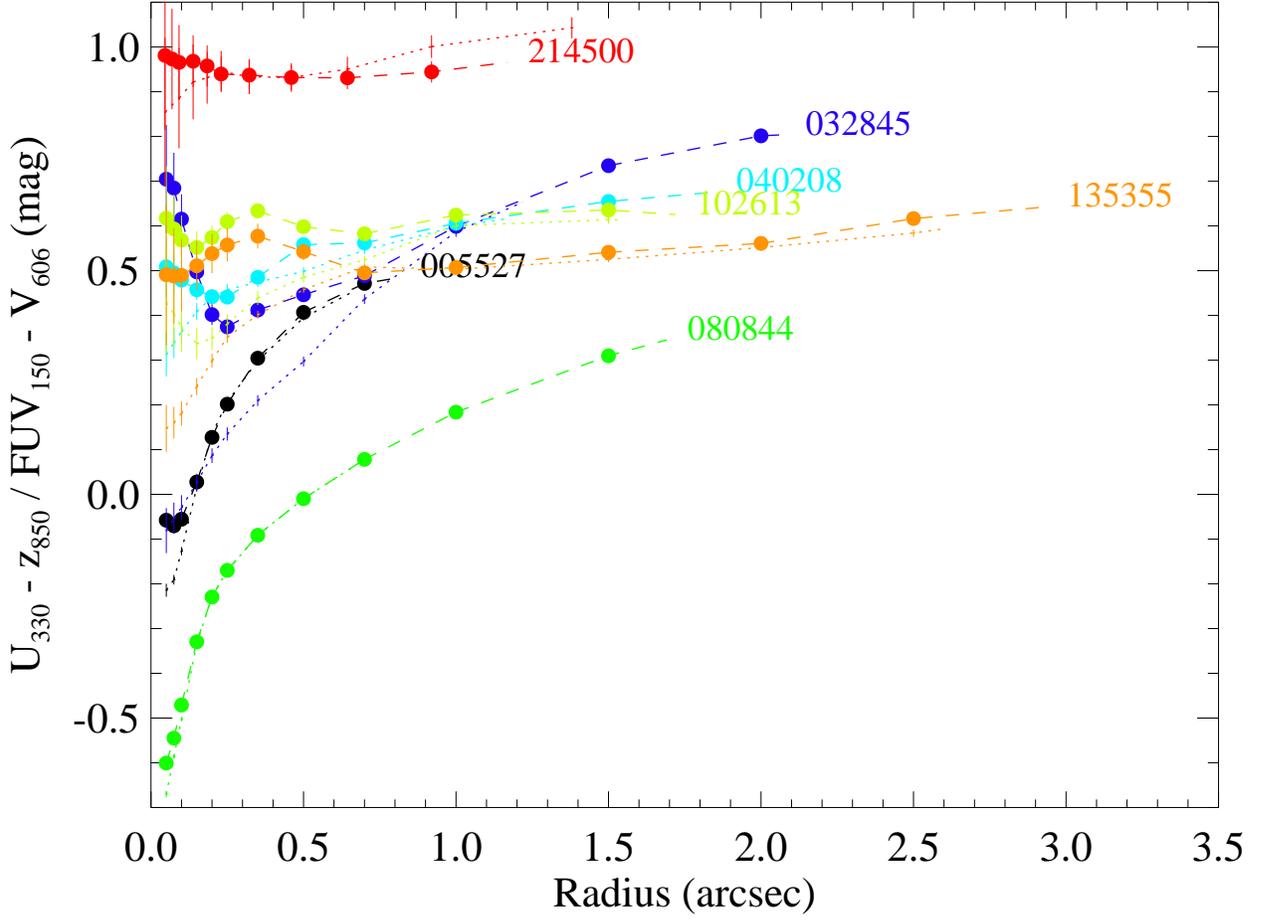}
\end{center}
\caption{\label{fig:radial}UV-optical radial color profiles of the UVLGs shown in Fig. \ref{fig:stamps}. 
Lines indicate the (cumulative) color determined in circular apertures measured from the 
centroids of the object detected in the optical image out to the 90\% flux radius (dashed colored lines),  
and with respect to the object centroids defined in the UV image (dotted colored lines).  
The UVLGs are generally very blue within approximately the half light radius, and become  
redder at larger radii. The objects often have steeper inner color gradients when then UV image is used for object detection (dotted lines), 
compared to when the optical image is used (dashed lines) due to the 
different positions of the object centroids in the UV and optical. 
For 214500, the UV-optical color corresponds to $FUV_{150}$--\vp, 
whereas for the other objects the colors measured are \up--\zp. Errors are 3$\sigma$.}  
\end{figure}

\begin{figure*}[t]
\begin{center}
\includegraphics[width=0.5\textwidth]{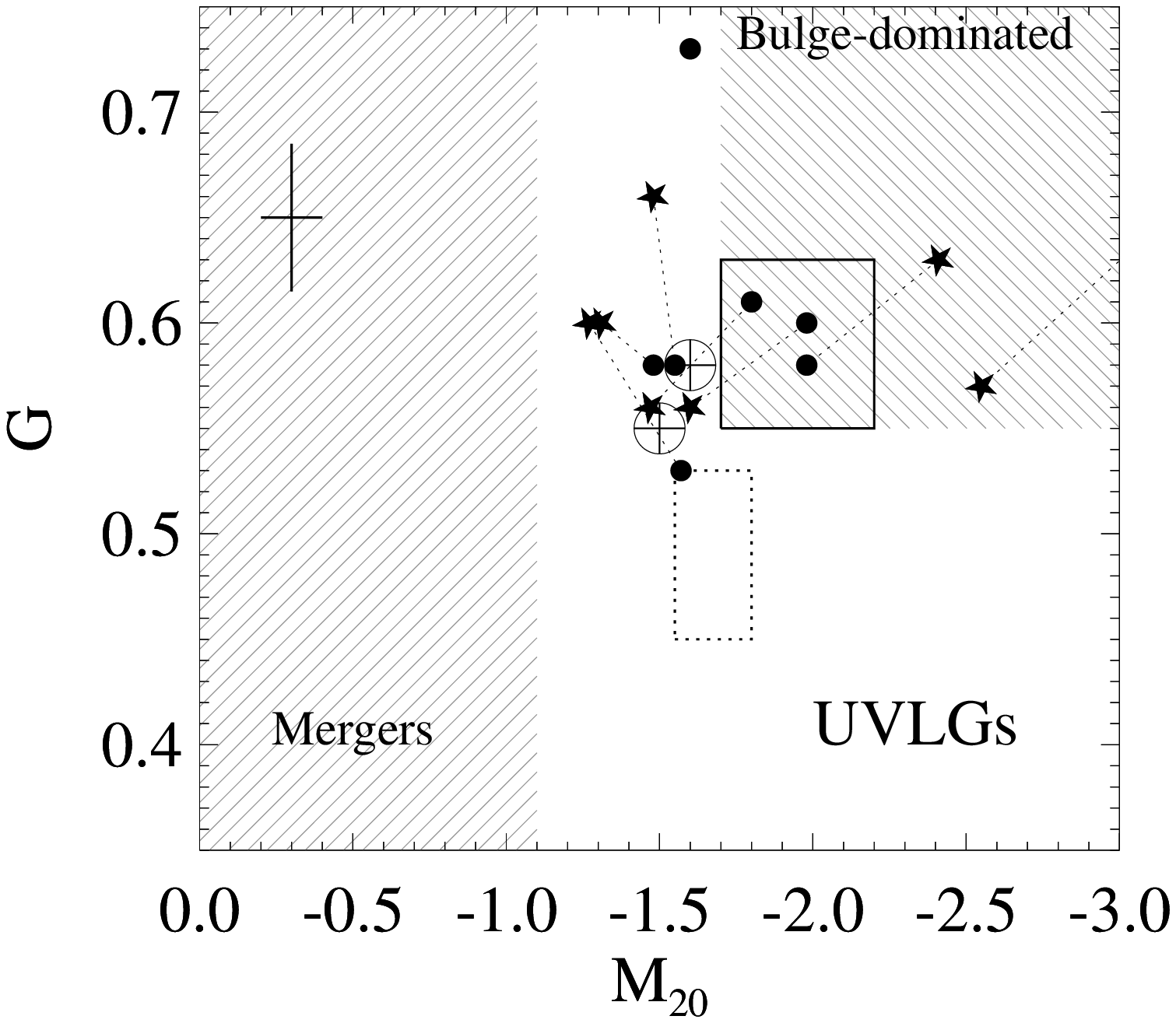}\includegraphics[width=0.5\textwidth]{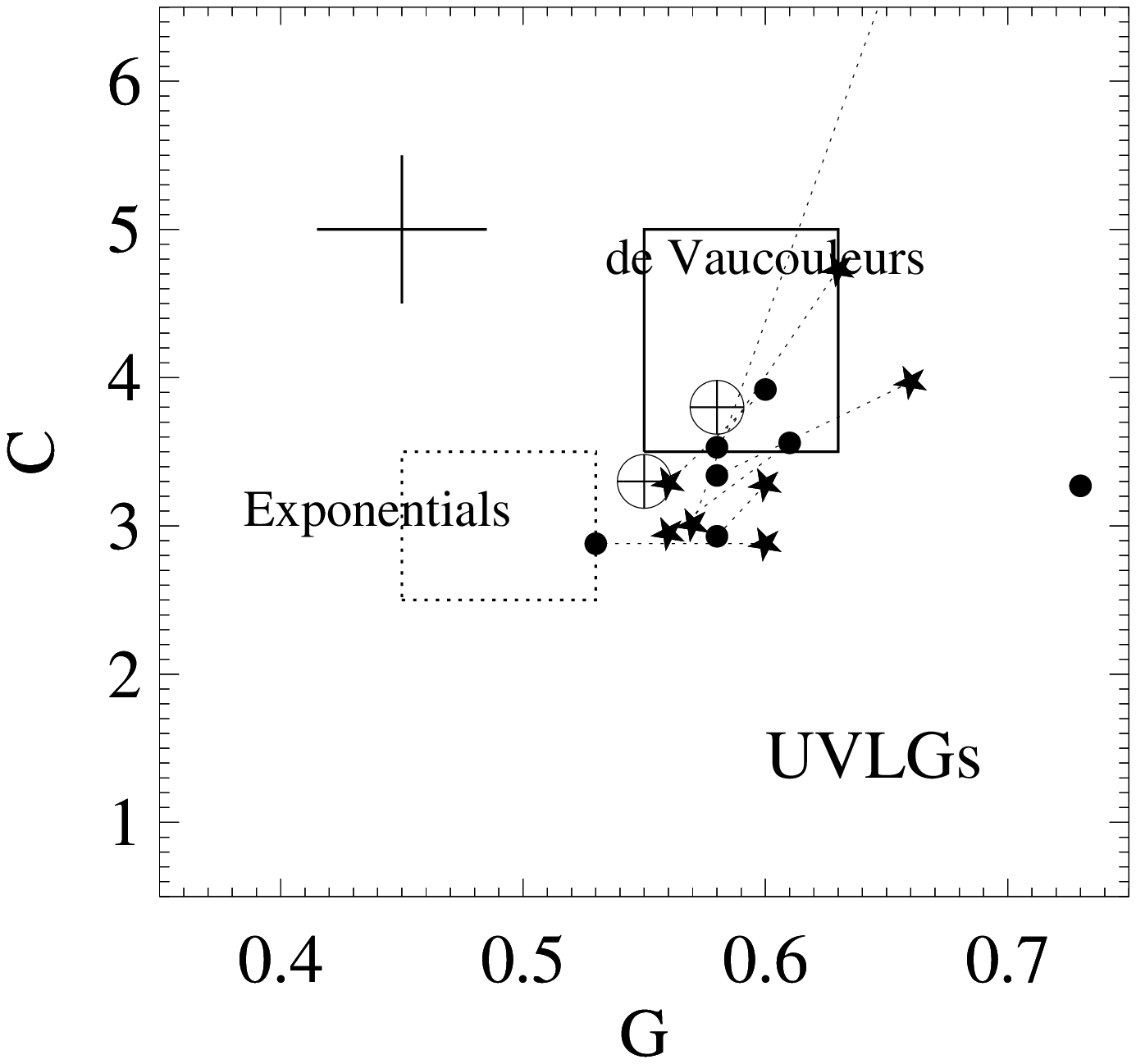}
\end{center}
\caption{\label{fig:morphologies}Distribution of morphological parameters $M_{20}$ vs. $G$ (left panel) and $G$ vs. $C$ (right panel) for the sample of $z\sim0.15$ UVLGs. 
Morphologies measured from the UV images are indicated by stars, morphologies measured from the optical images by filled circles. Dotted lines connect the UV and optical morphologies measured for each source. For comparison, we have indicated the following (all adapted from \citet{lotz06}): 
regions populated by likely mergers having multiple nuclei (left hatched area) and by bulge-dominated objects (right hatched area), regions occupied by simulated de Vaucouleurs and exponential profiles (top and bottom boxes, resp.), the median values found for star-forming galaxies at $z\sim1.5$ and LBGs at $z\sim4$ in GOODS (encircled crosses), and 
the typical errors for the high redshift samples (large crosses).}
\end{figure*}

\begin{figure}[t]
\begin{center}
\includegraphics[width=0.4\columnwidth]{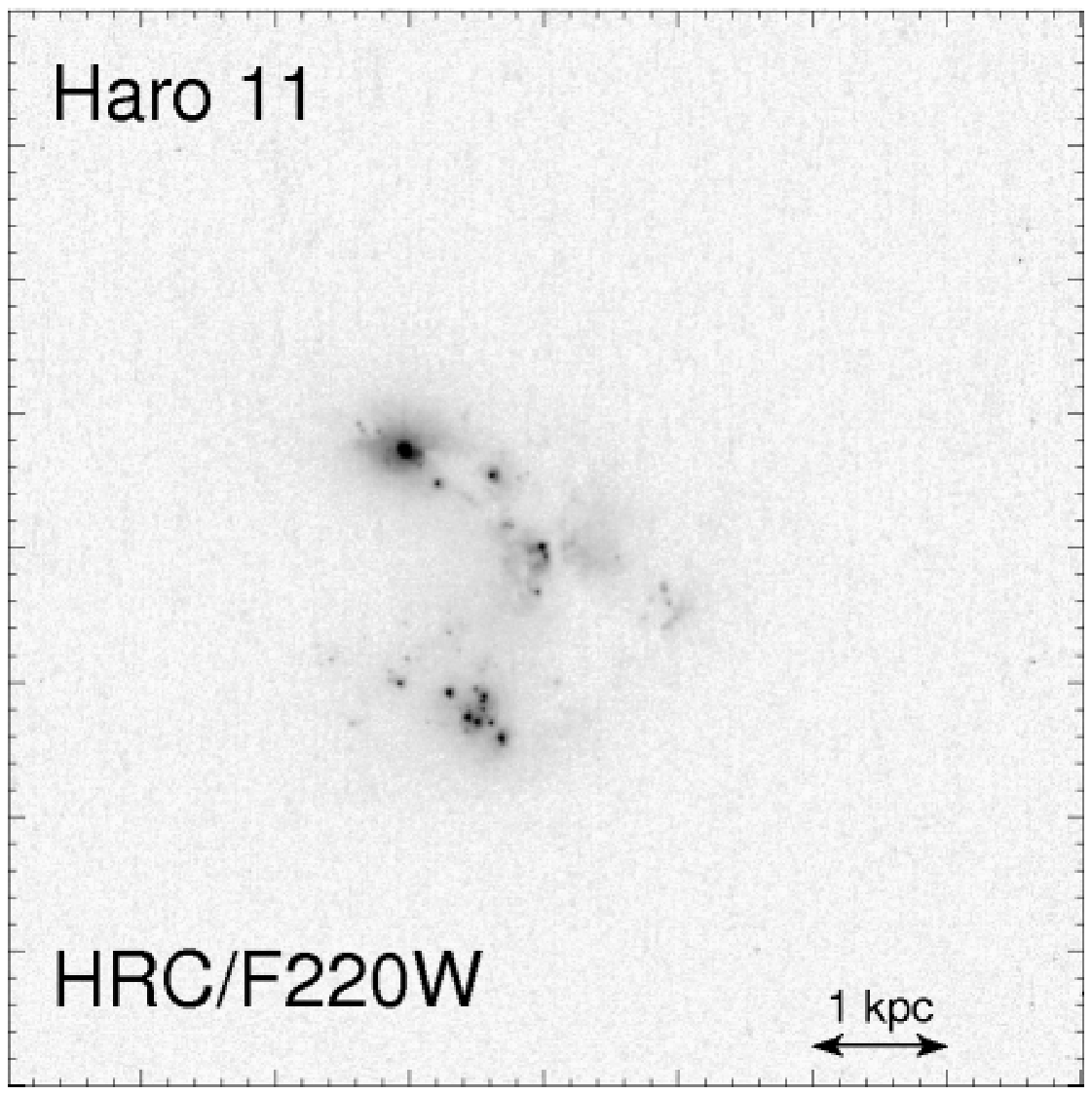}\hspace{5mm}\includegraphics[width=0.4\columnwidth]{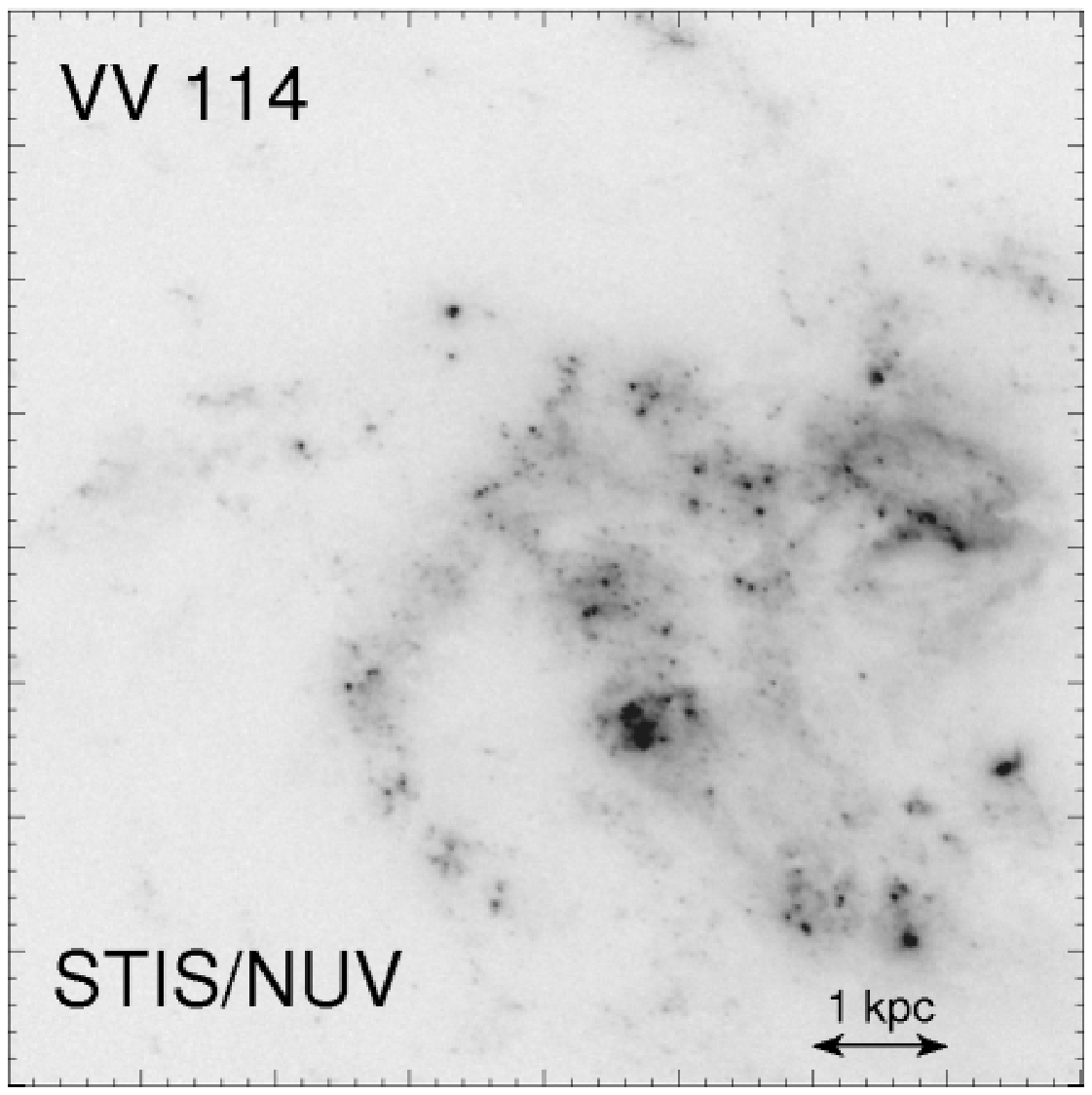}\\
\vspace{5mm}
\includegraphics[width=0.4\columnwidth]{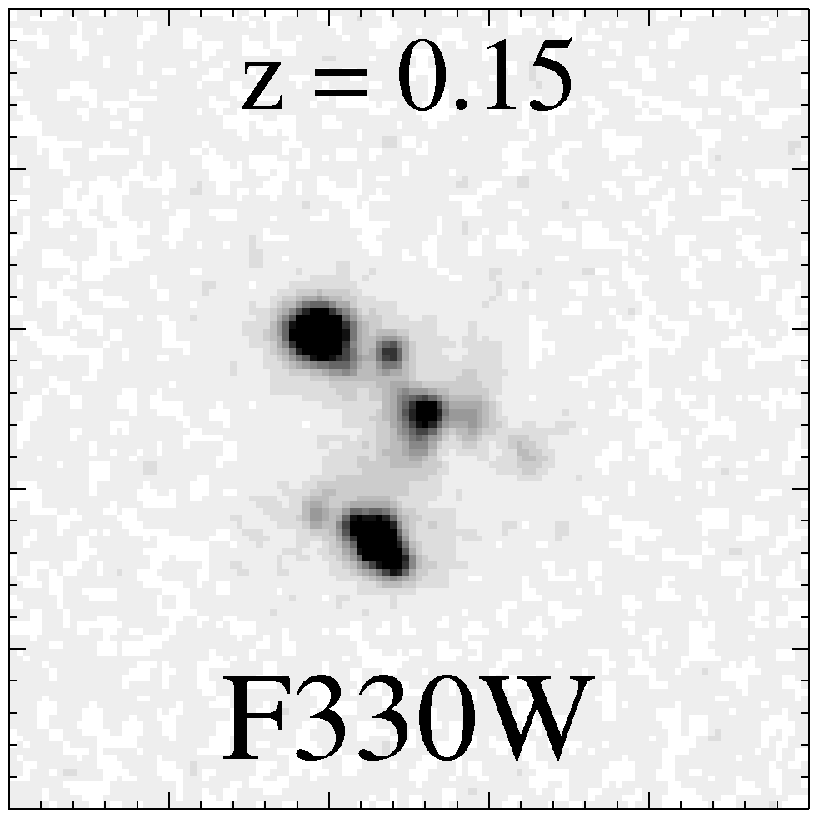}\hspace{5mm}\includegraphics[width=0.4\columnwidth]{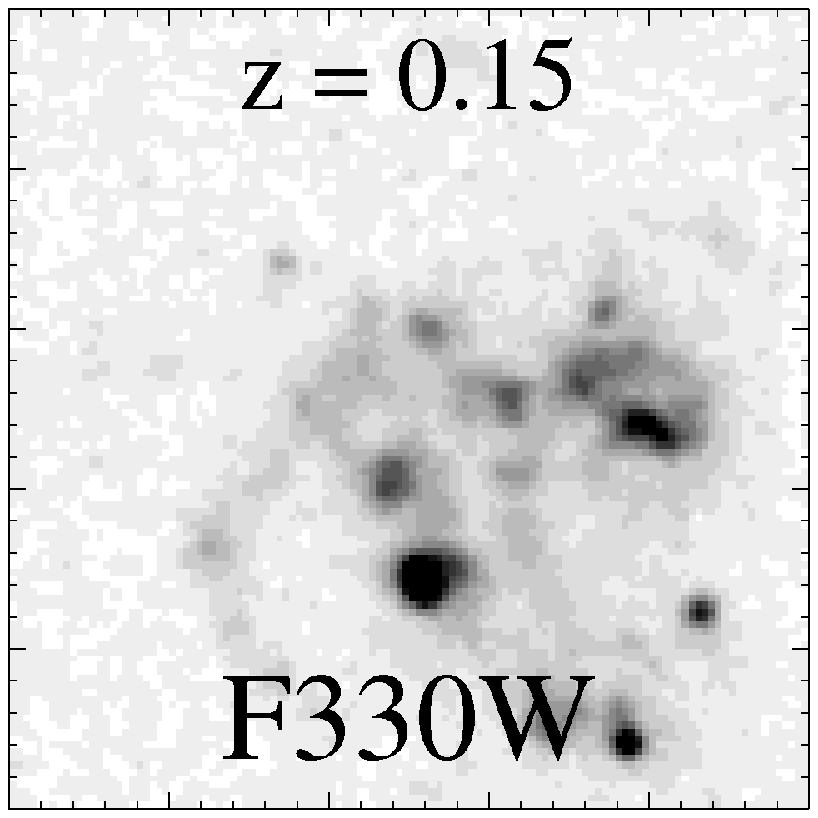}
\end{center}
\caption{\label{fig:haro11vv114sims}{\it Top panels:} NUV archival images of the local starburst galaxies Haro 11 and VV 114, both at $z=0.02$. 
The image of Haro 11 was taken with the ACS/HRC (Program 10575, PI: G\"oran \"Ostlin). 
The image of VV 114 was taken with STIS (Program 8201, PI: Gerhardt Meurer). See Sect. \ref{sec:archival} for details. 
{\it Bottom panels:} Simulated F330W images of Haro 11 (left) and VV 114 (right) at $z=0.15$ 
for comparison with our UVLG sample. The images measure 3\arcsec$\times$3\arcsec. See Sect. \ref{sec:sims} for details.}  
\end{figure}

\begin{figure}[t]
\begin{center}
\includegraphics[width=0.5\columnwidth]{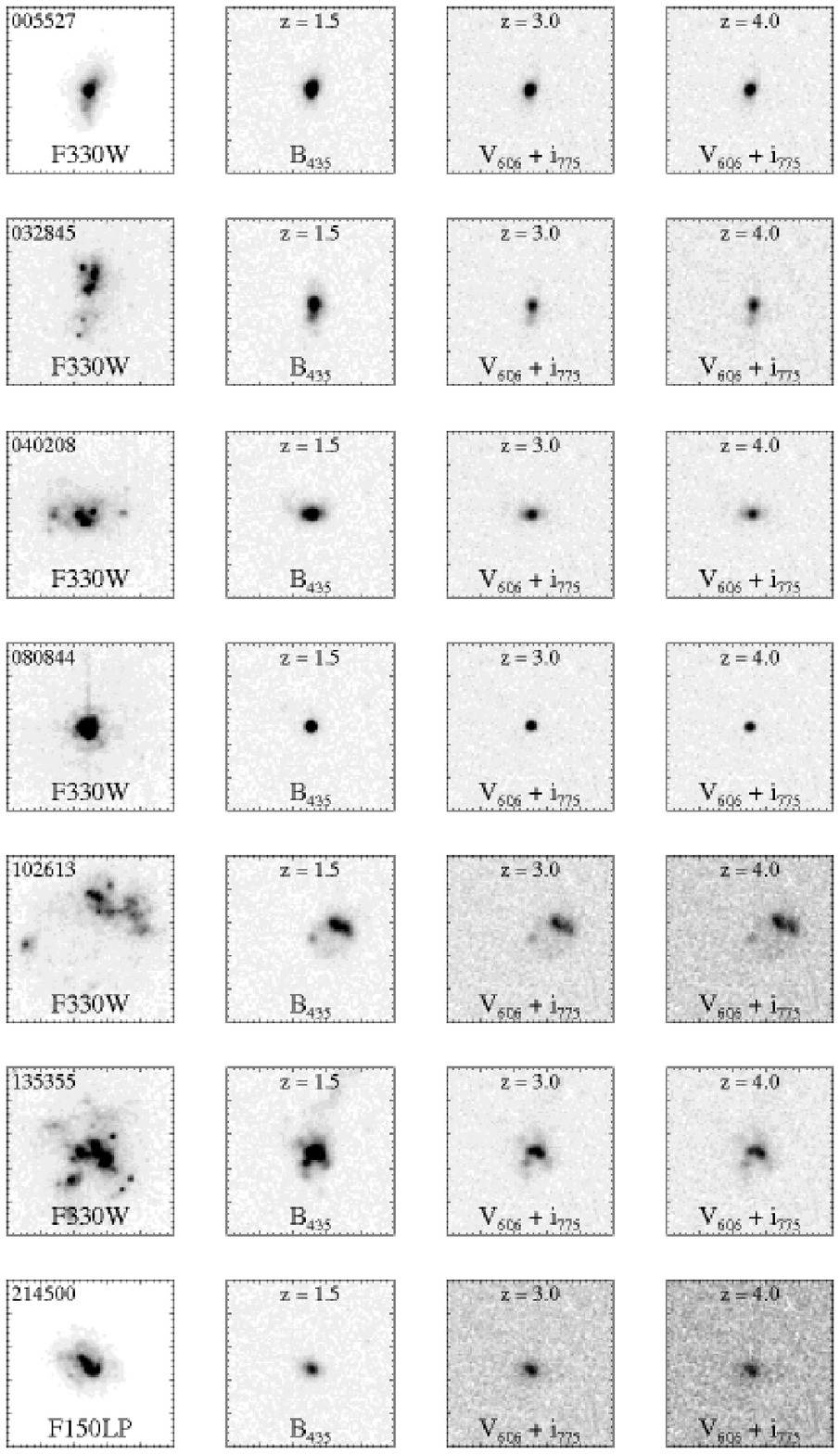}\\
\vspace{10mm}
\includegraphics[width=0.5\columnwidth]{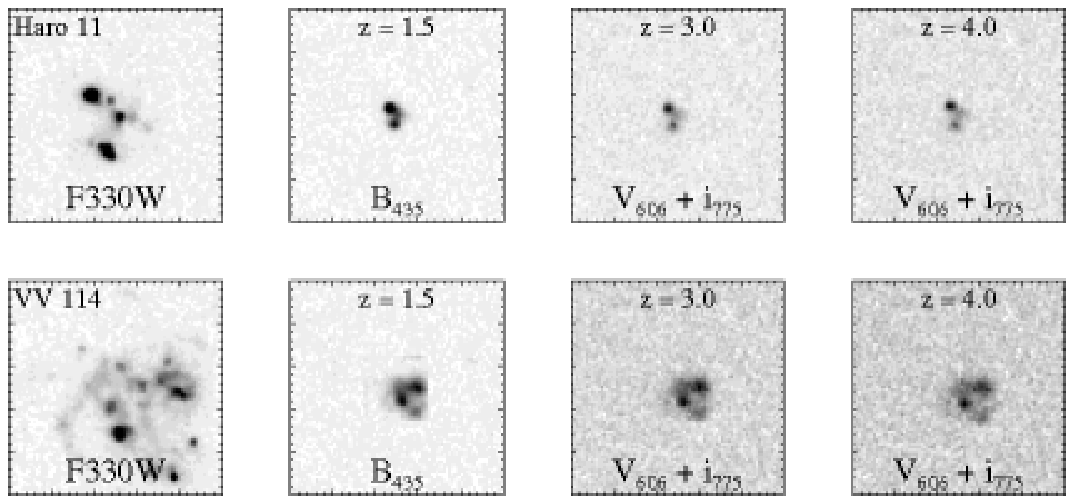}
\end{center}
\caption{\label{fig:sims_goods_uvlgs}Redshift simulations of the UVLG sample at the depth of GOODS (Sect. \ref{sec:sims}). 
Panels ($3\arcsec\times3\arcsec$) show the original, unredshifted HRC/F330W or SBC/F150LP images (left), and GOODS \bp\ or \vp+\ip\ 
simulations at $z=1.5$, $z=3.0$ and $z=4.0$. The last two rows show the redshift simulations of Haro 11 and VV 114 (Sect. \ref{sec:sims}) for comparison.}  
\end{figure}

\begin{figure}[t]
\begin{center}
\includegraphics[width=0.5\columnwidth]{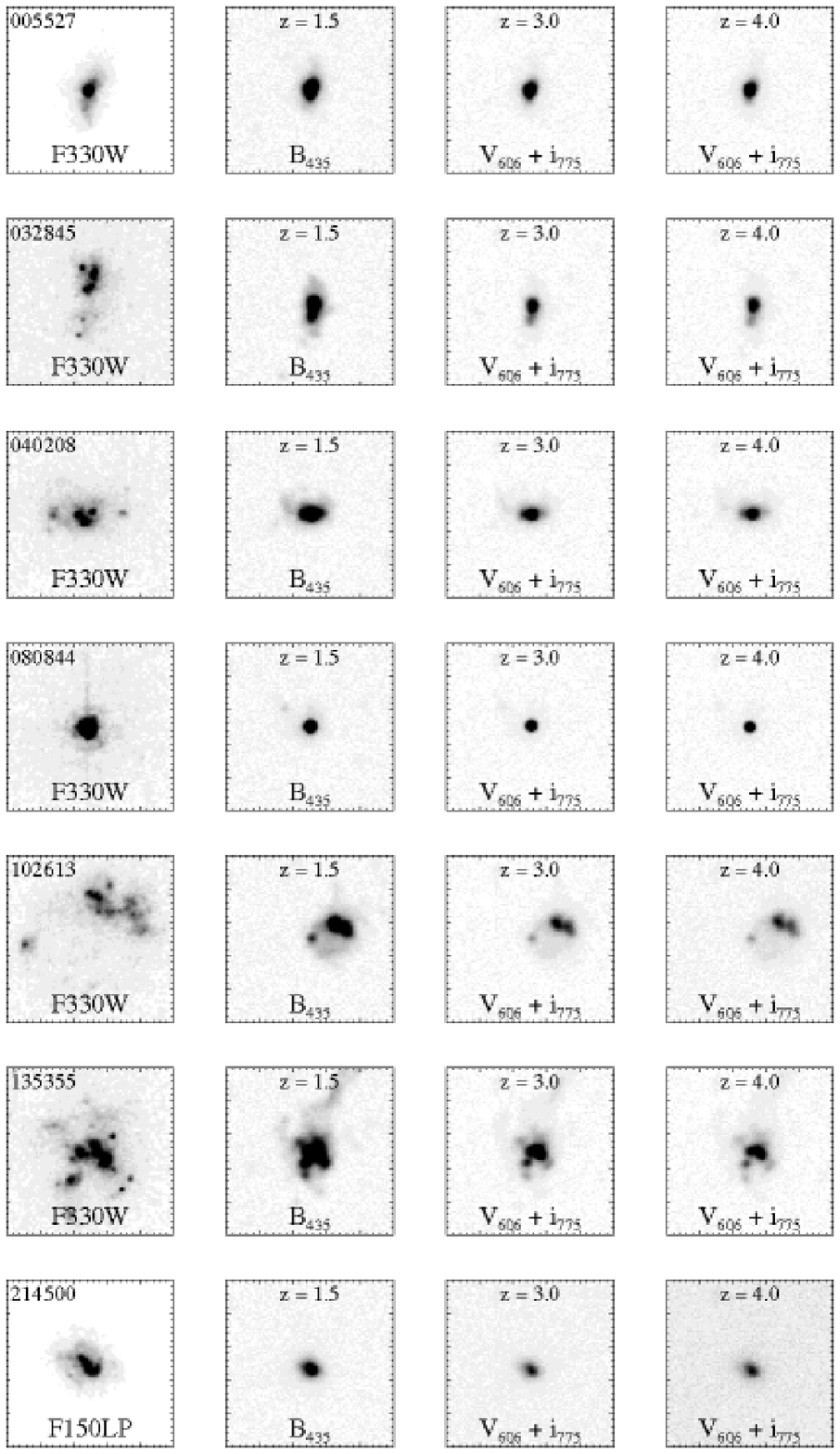}\\
\vspace{10mm}
\includegraphics[width=0.5\columnwidth]{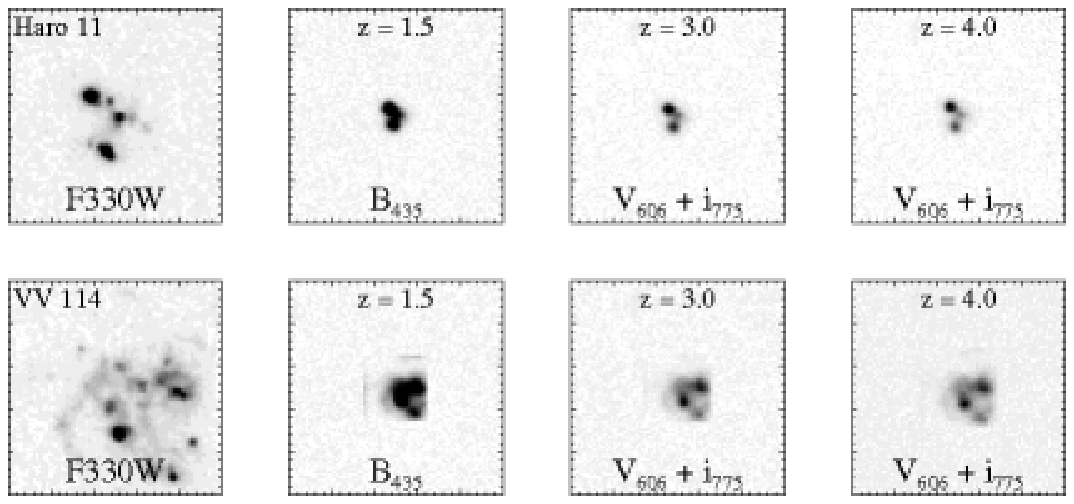}
\end{center}
\caption{\label{fig:sims_udf_uvlgs}Redshift simulations at the depth of the UDF. See Sect. \ref{sec:sims} and the caption of Fig. \ref{fig:sims_goods_uvlgs} for details.}
\end{figure}

\begin{figure}[t]
\begin{center}
\includegraphics[width=0.5\columnwidth]{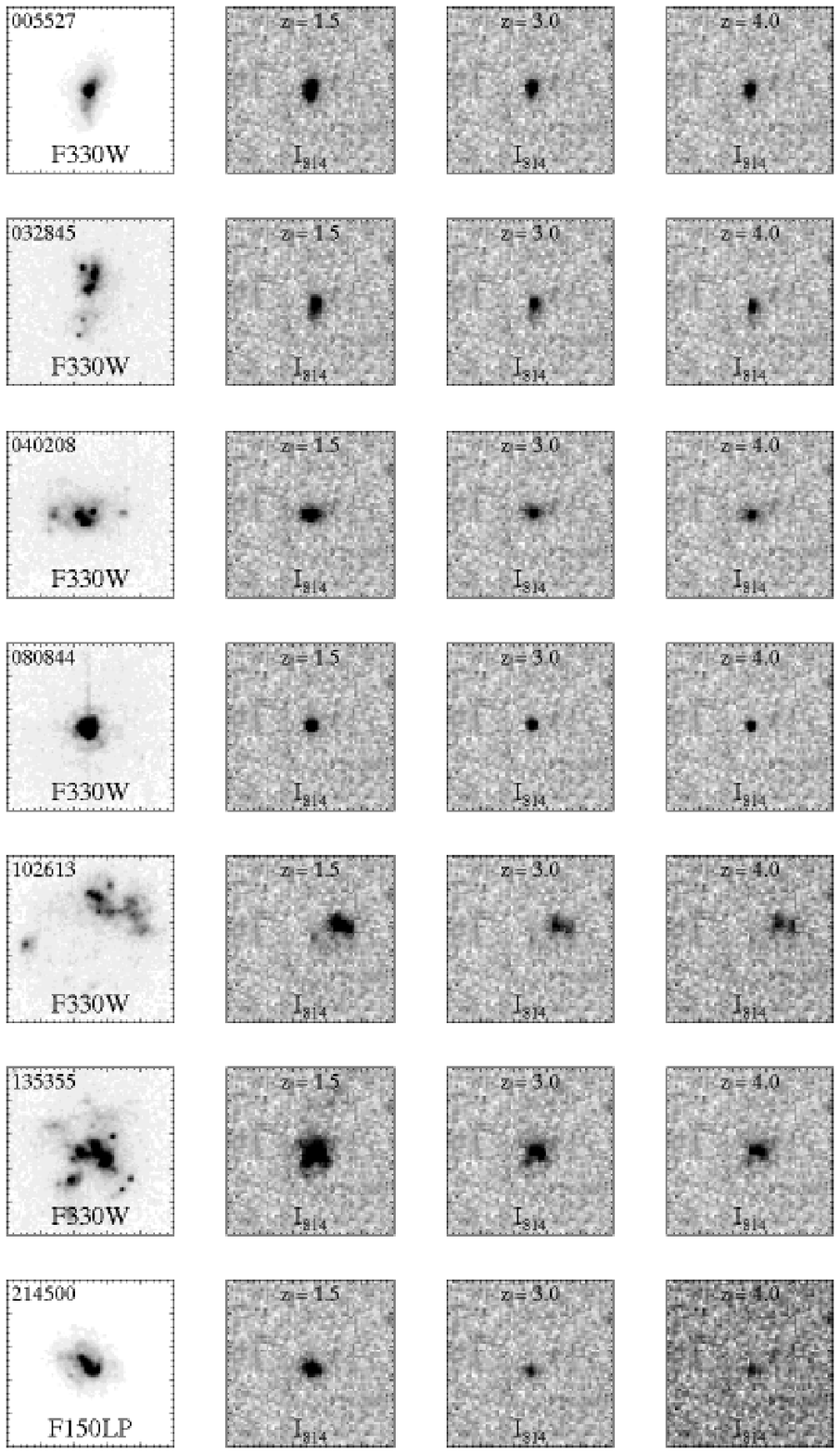}\\
\vspace{10mm}
\includegraphics[width=0.5\columnwidth]{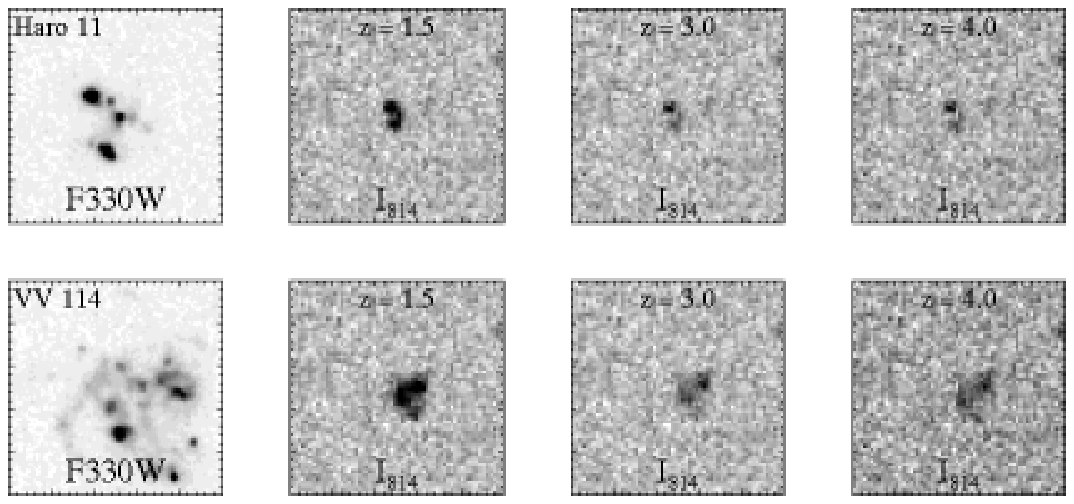}
\end{center}
\caption{\label{fig:sims_cosmos_uvlgs}Redshift simulations at the depth of COSMOS. See Sect. \ref{sec:sims} and the caption of Fig. \ref{fig:sims_goods_uvlgs} for details.}
\end{figure}

\begin{figure}[t]
\begin{center}
\includegraphics[width=0.7\textwidth]{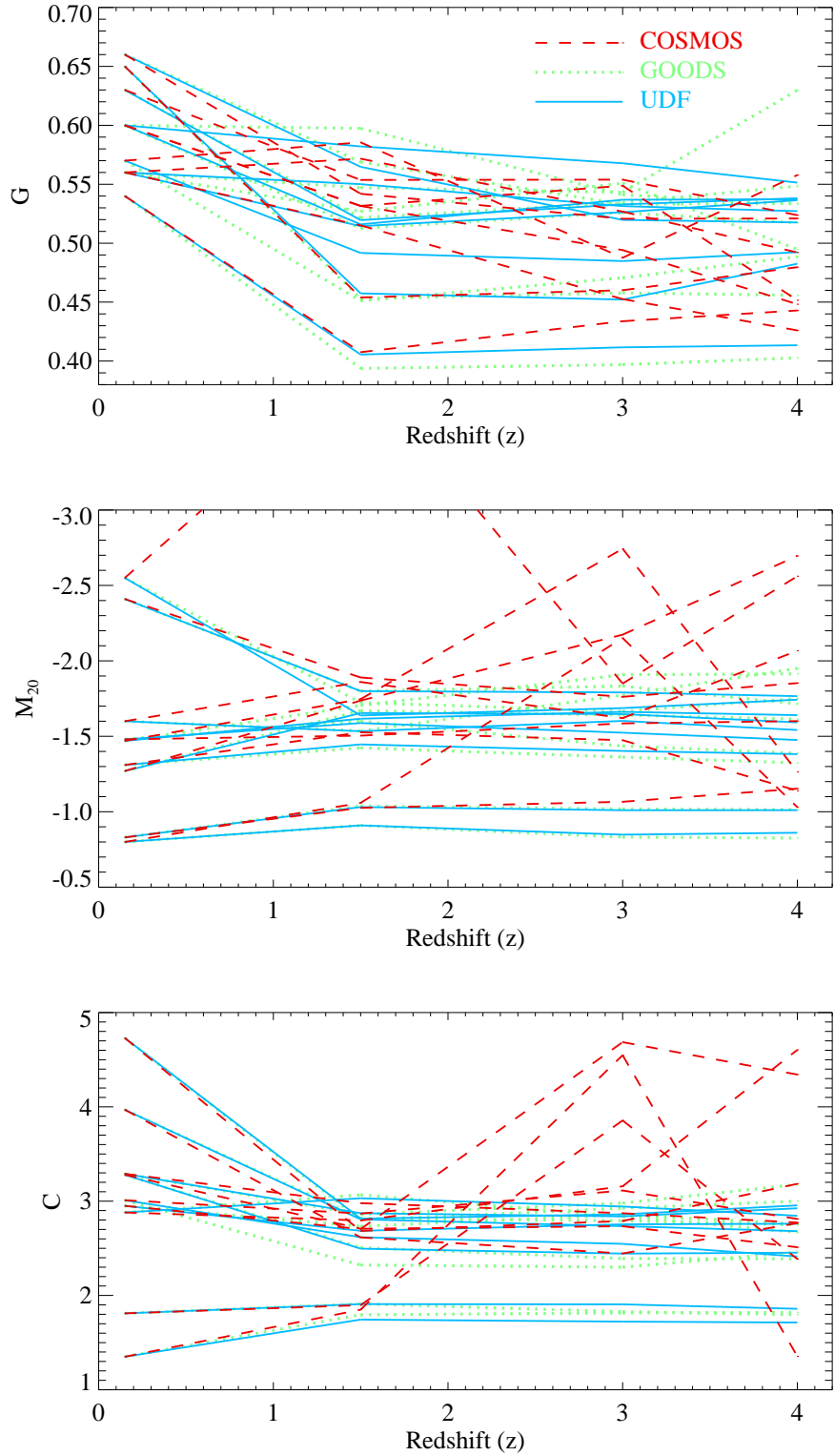}
\end{center}
\caption{\label{fig:morphologies_redshift}Morphological parameters of the UVLGs measured from the simulated images at different redshift. 
Objects were redshifted to $z=1.5$, $z=3$ and $z=4$, and simulated at the depths of COSMOS (red dashed lines), GOODS (green dotted lines) and the UDF (blue solid lines).}
\end{figure}

\begin{figure}[t]
\begin{center}
\includegraphics[width=0.7\textwidth]{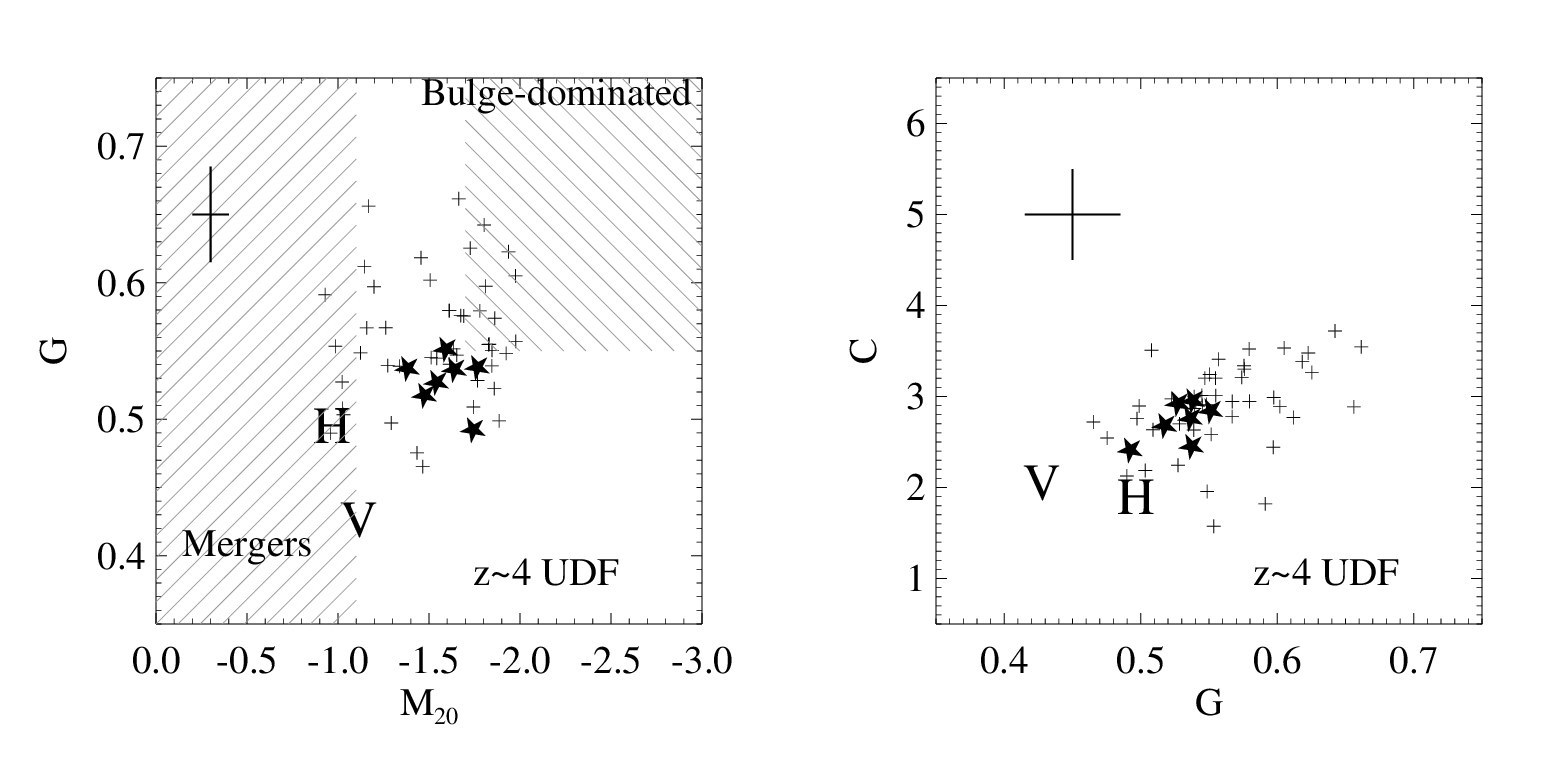}\\
\includegraphics[width=0.7\textwidth]{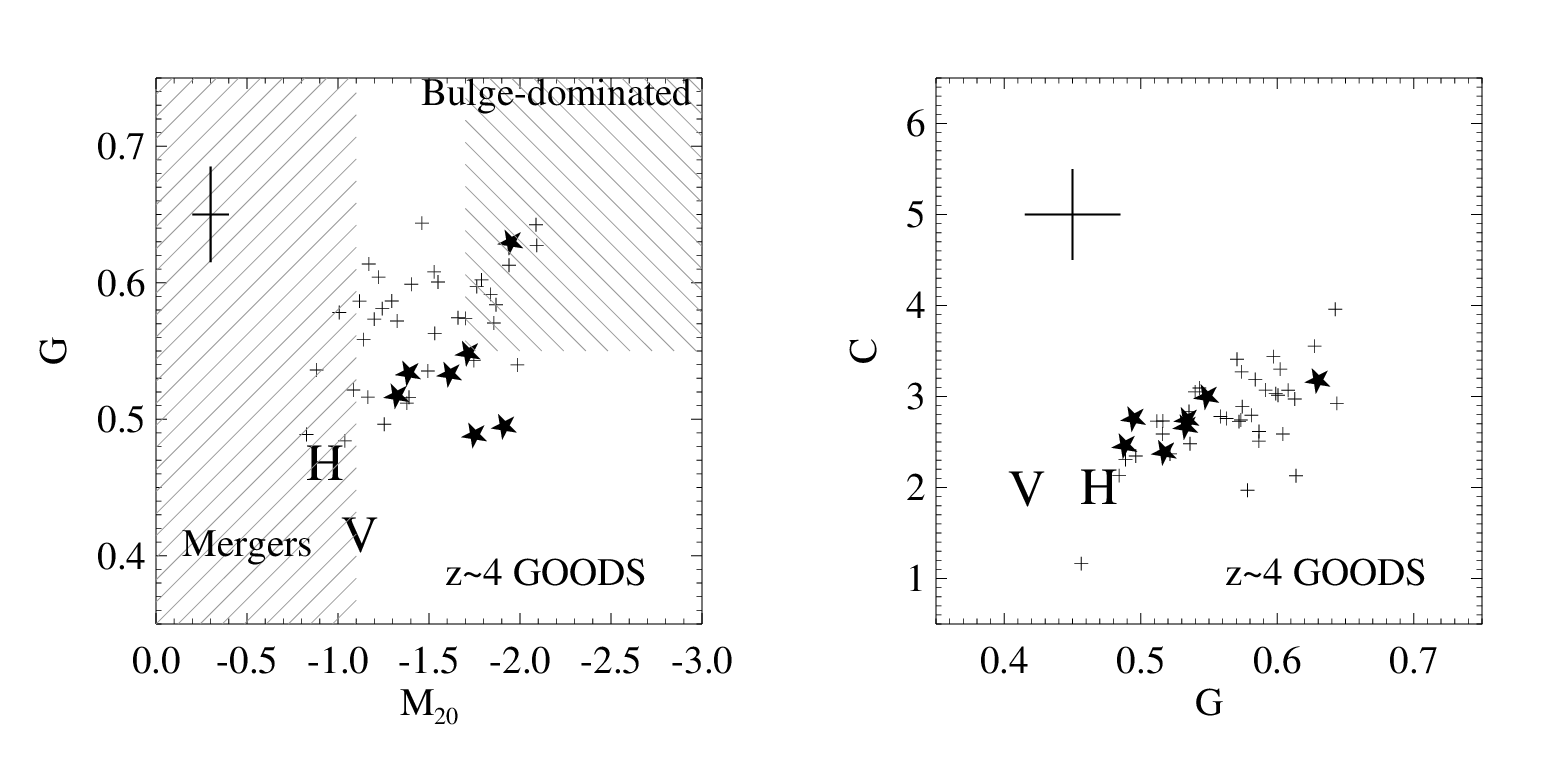}\\
\includegraphics[width=0.7\textwidth]{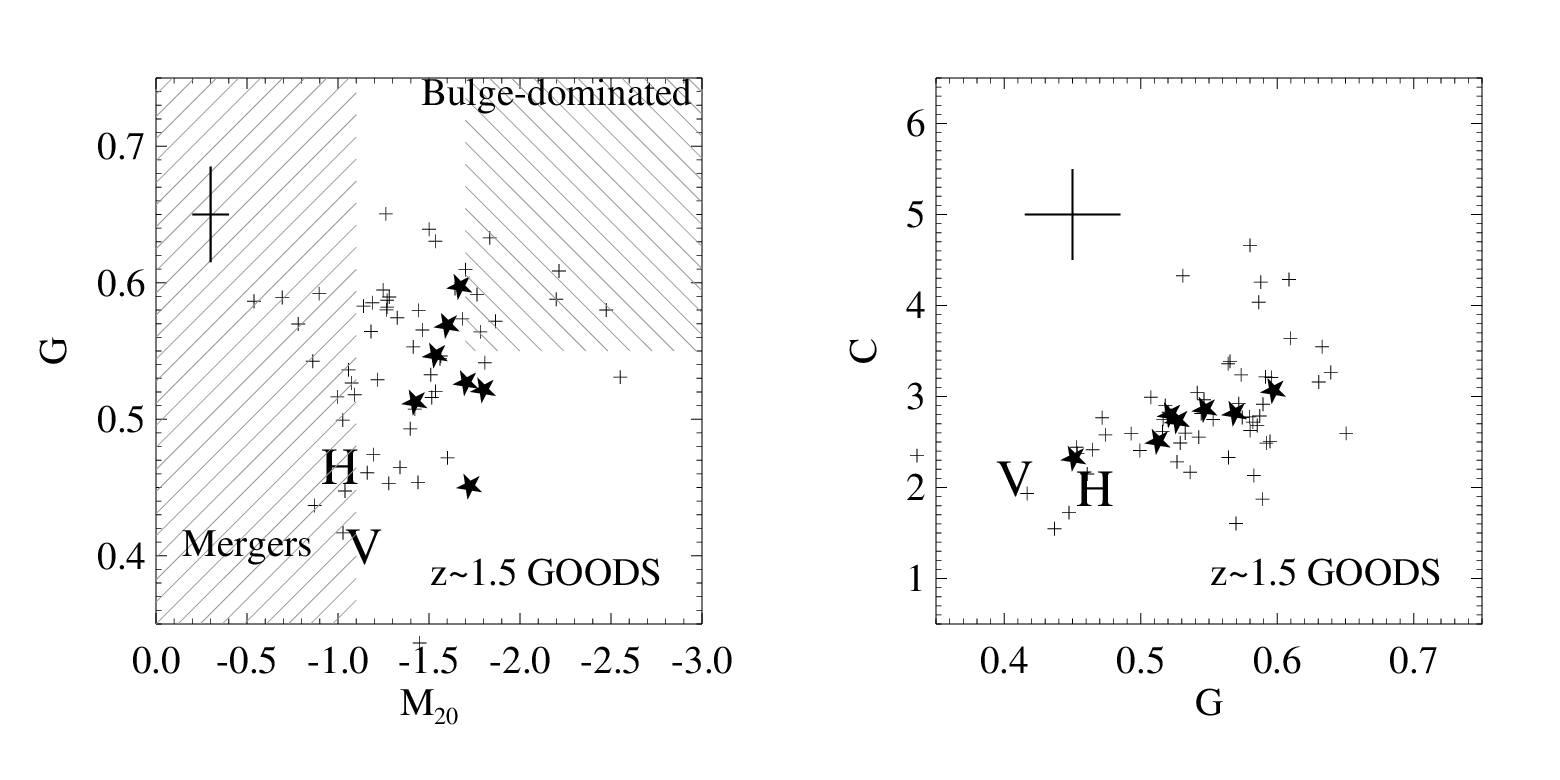}
\end{center}
\caption{\label{fig:morphologies_lotz}Comparison of the morphologies
  of Lyman break galaxies at $\sim4$ and starburst galaxies at
  $z\sim1.5$ from the samples of \citet{lotz06} (plusses) with the
  morphologies of the UVLGs (stars). The UVLGs were artificially
  redshifted to the mean redshift of the comparison samples, and
  simulated at a similar depth. Top panels show the results for LBGs
  and UVLGs at the UDF depth at $z\sim4$, middle panels show LBGs and
  UVLGs at the GOODS depth at $z\sim4$, and bottom panels show star
  forming galaxies and UVLGs at the GOODS depth at $z\sim1.5$. Haro 11 ('H') 
  and VV 114 ('V') were simulated at the same redshift and depth as the
  UVLGs.  For comparison, we have indicated the
  following (all adapted from \citet{lotz06}): regions populated by
  likely mergers having multiple nuclei (left hatched area) and by
  bulge-dominated objects (right hatched area), and the typical errors
  for the high redshift samples (large crosses). In all panels, the
  UVLGs span a very similar range in parameter space compared to the
  comparison samples, indicating that UVLGs and LBGs have very similar
  quantitative morphologies.}
\end{figure}

\begin{figure*}[t]
\begin{center}
\includegraphics[width=\textwidth]{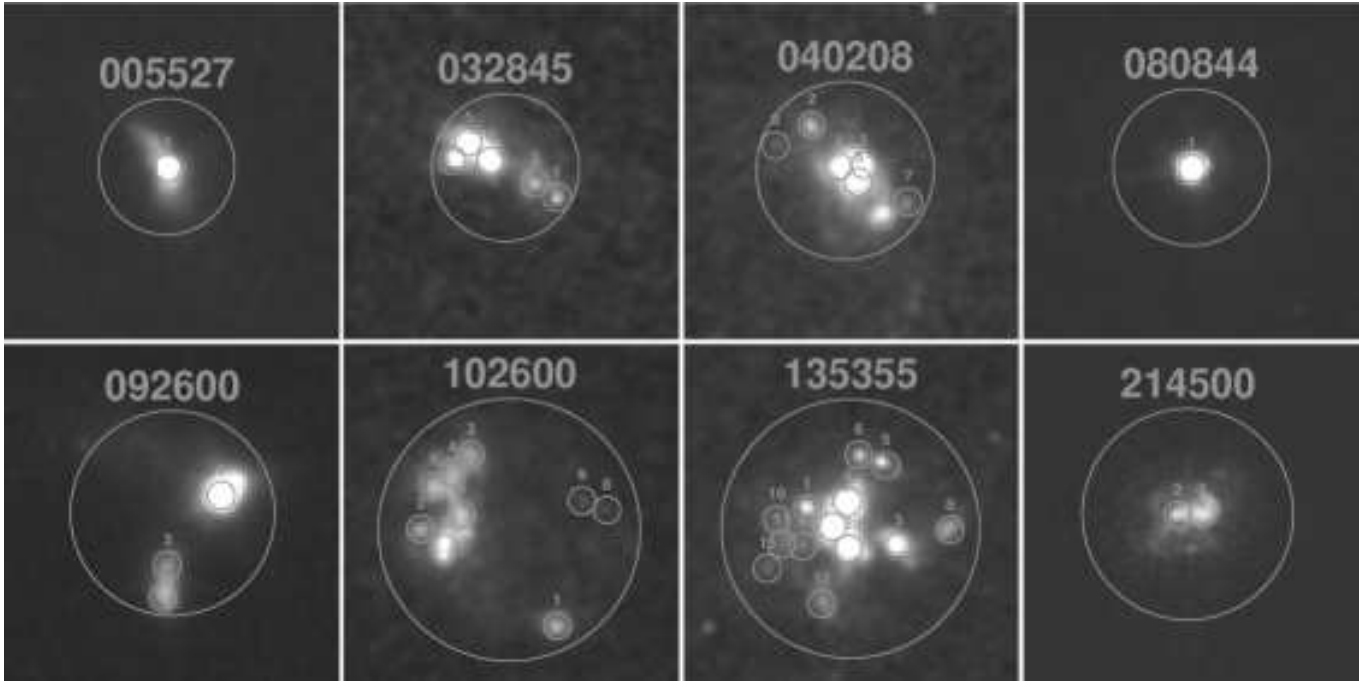}
\end{center}
\caption{\label{fig:ssc_apertures}Panels show the \up\ images indicating the apertures placed around the unresolved starburst regions identified by eye 
in each of the sources (small circles). Apertures measure 0\farcs3 in diameter, and are numbered for identification of each region in Fig. \ref{fig:ssc_cm}.  
Larger apertures were chosen to measure the average color of the regions in which the unresolved starbursts are situated (large circles). 
For object 214500 the image shown is $FUV_{150}$. For completeness, we show the \zp\ image for 092600 as no \up\ image is available.}
\end{figure*}

\begin{figure}[t]
\begin{center}
\mbox{
\includegraphics[width=0.49\columnwidth]{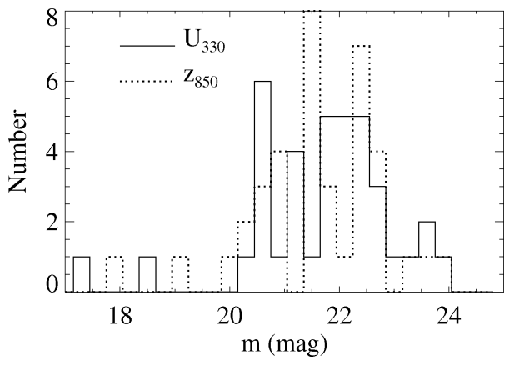}\hspace{0.02\columnwidth}
\includegraphics[width=0.49\columnwidth]{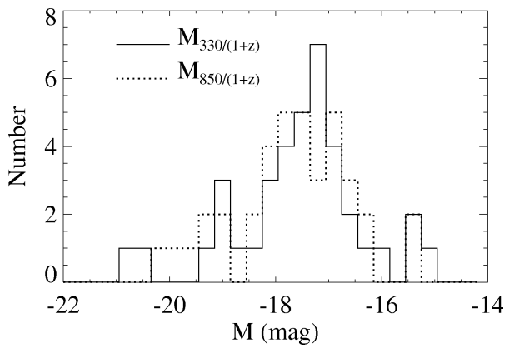}}
\end{center}
\caption{\label{fig:ssc_maghist}Apparent ({\it left panel}) and absolute ({\it right panel}) magnitude distributions in \up\ (solid lines) and \zp\ (dotted lines) 
of the unresolved starburst regions identified in Fig. \ref{fig:ssc_apertures}. The magnitudes were extinction-corrected using a global reddening value 
derived from the bolometric dust to FUV luminosity ratio  (see Table \ref{tab:phot}).}
\end{figure}

\begin{figure*}[t]
\begin{center}
\includegraphics[width=0.45\textwidth]{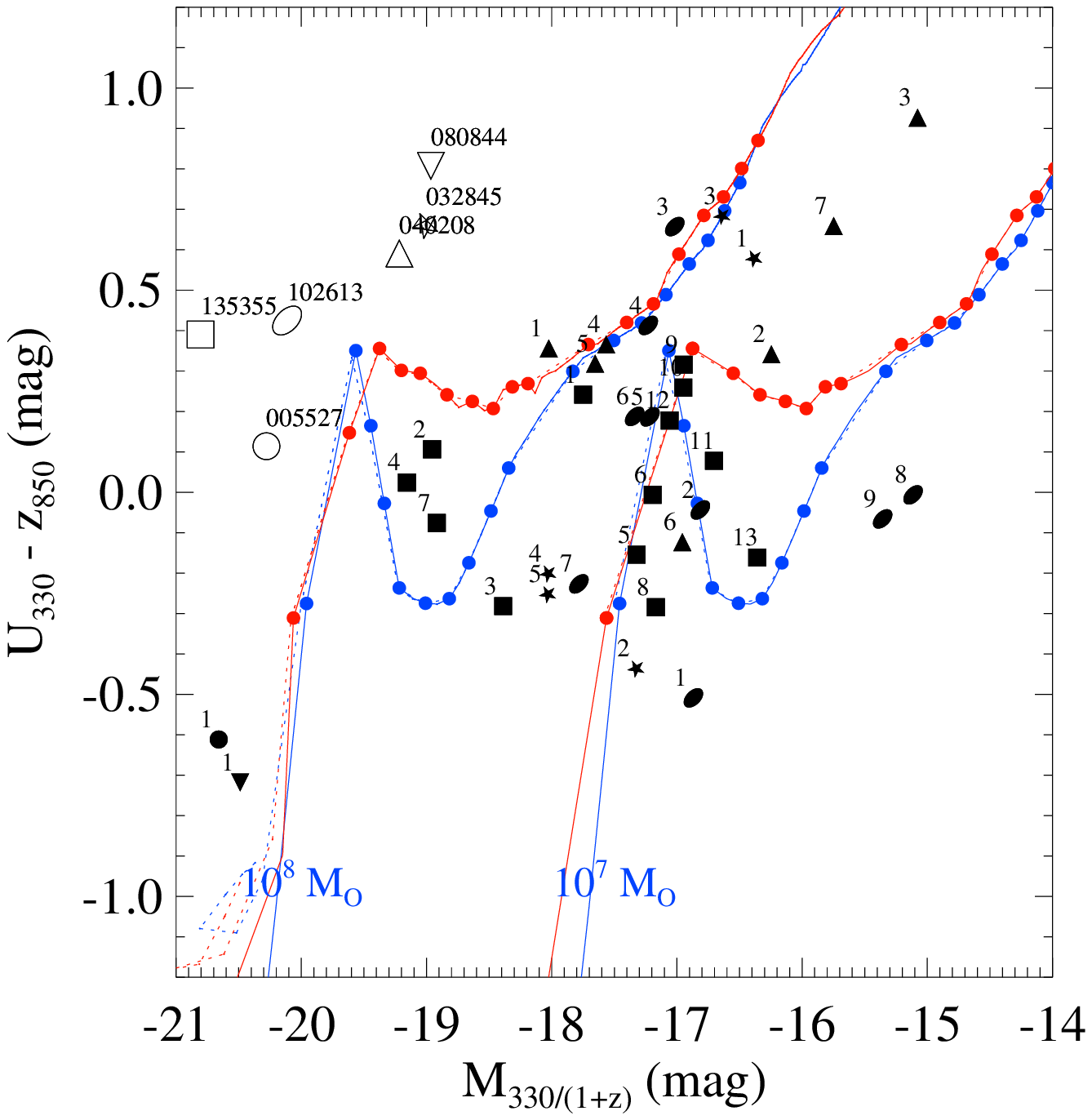}
\includegraphics[width=0.45\textwidth]{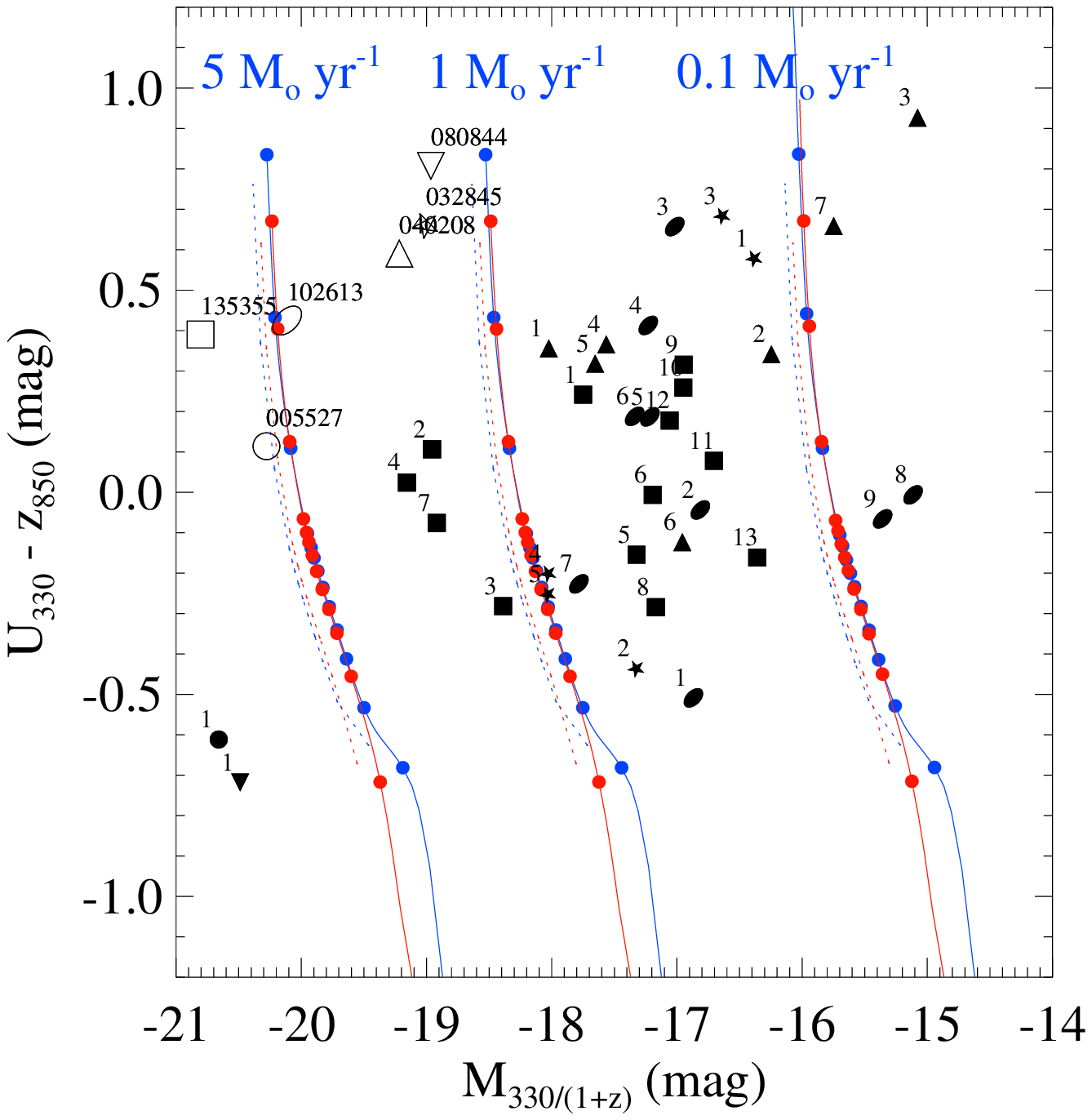}
\end{center}
\caption{\label{fig:ssc_cm}Color magnitude diagram of the individual
  starburst regions identified in
  Fig. \ref{fig:ssc_apertures}. Filled, numbered symbols correspond to
  the unresolved regions identified in 005527 (circle), 032845 (star),
  040208 (upward triangle), 080844 (downward triangle), 102613
  (ellipse), and 135355 (square). The large, open symbols correspond
  to the large apertures in Fig. \ref{fig:ssc_apertures} with the
  point sources removed. Tracks indicate the color and magnitude
  evolution of an instantaneous burst model ({\it left panel}) and a
  continuous star formation model ({\it right panel}) simulated using
  STARBURST99. Blue tracks are for LMC metallicity ($Z=0.008$), red
  tracks for solar metallicity ($Z=0.02$). For instantaneous bursts we
  plot tracks having a burst mass of $10^8$ $M_\odot$ and $10^7$
  $M_\odot$. For continuous models we plot tracks having SFRs of 5, 1,
  and 0.1 $M_\odot$ yr$^{-1}$.  Ages in Myr have been indicated along
  the tracks (marked at
  6,7,8,9,10,12,14,16,18,20,30,40,50,60,70,80,90,100 Myr for the burst
  models, and at 10,20,30,40,50,60,70,80,90,100,200,500,1000 Myr for
  the continuous models). Dotted lines indicate the colors if we
  include the contribution from the nebular continuum. The model
  tracks are reddening-free, but the observed colors and magnitudes
  were dereddened using a global reddening value derived from the
  bolometric dust to FUV luminosity ratio (see Table \ref{tab:phot}).
  The starburst regions of 214500 were omitted because of the
  different filterset used during those observations. Object 092600
  was omitted because no \up\ is available.}
\end{figure*}

\begin{figure}[t]
\begin{center}
\includegraphics[width=0.5\columnwidth]{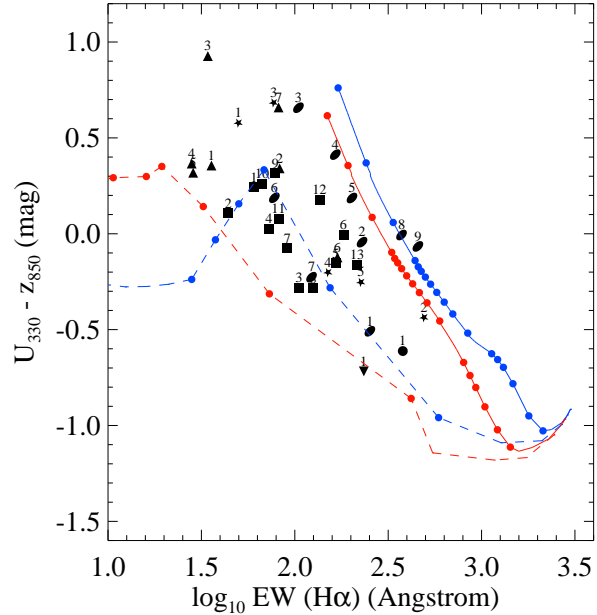}
\end{center}
\caption{\label{fig:ew}Rest-frame equivalent width \ha\ vs. \up--\zp\ color for the super starburst regions. Tracks indicate the 
age evolution of the \ha\ EW for instantaneous burst ({\it dashed lines}, with ages marked at 5,6,7,8,9,10 Myr) and continuous star formation ({\it solid lines}, with ages marged at 5,6,7,8,9,10,20,30,40,50,60,70,80,90,100,200,500,1000 Myr) models from STARBURST99. 
See text and the caption of Fig. \ref{fig:ssc_cm} for further details.}
\end{figure}

\clearpage

\appendix

\section{Scarpa et al. (2007) revisited}
\label{sec:scarpa}

\noindent
In a recent paper, \citet{scarpa07} have tried to argue that the
sample of compact UVLGs selected by H05 and H07 are not true analogs
of LBGs. \citet{scarpa07} claim that the sizes estimated from the SDSS
{\it u}-band images grossly underestimate the actual size in the UV,
and that the UVLGs therefore are much larger and lower in surface
brightness than LBGs.  Here, we will briefly show that this claim is
based on the combination of a faulty analysis and incorrect
assumptions,
that are easily refuted by the HST data presented in this paper.\\

\noindent
$\bullet$ \citet{scarpa07} claim that it is only possible to detect
central bulges in the SDSS {\it u}-band images, and that fainter,
extended disks would be missed.  They reach this conclusion by fitting
the SDSS {\it r}-band images with a two component model (point source
plus disk) and then using the half-light radius of {\it only the disk}
component to recompute the size of the UVLGs. In doing this, the
authors made a serious mistake, because they
ignore the fact that on average, half the light in their model fits comes from the central {\it ``point source''} (see their Table 1).\\

\noindent
$\bullet$ The claim in the \citet{scarpa07} paper that most of the FUV
flux in compact UVLGs should come from a large disk component relies
on the observation that there exists a population of galaxies with
relatively red bulges not visible in the FUV and very blue outer rings
or disks that are only apparent in the UV
\citep[e.g.][]{thilker05}. It should be clear both from the results
presented in H05 and H07, and from the analysis of HST data presented
here, that UVLGs are an entirely different class of objects. In fact,
the UVLGs are clearly more extended in the optical than in the UV due
to an underlying (older) component. The UVLGs have very blue inner
colors, and become redder outwards (see Fig. \ref{fig:radial} and
Table \ref{tab:sizes}).
This is opposite to the assumption of \citet{scarpa07}.\\

\noindent
$\bullet$ The HST \up\ images confirm that the UVLGs are indeed very
compact objects. In fact, in most cases, the half light radius is even
smaller than the seeing deconvolved SDSS {\it u}-band size (Table
\ref{tab:sizes}). In the case of both J005527.46--002148.7 and
J080844.26+394852.3 the UV flux is dominated by an unresolved
component even at the HST resolution. Furthermore, when we extrapolate
the slope of the UV continuum measured from the FUV--NUV color and
predict the \up\ flux, we find good agreement with the measured \up\
flux (Sect. \ref{sec:sizes}), indicating that the \up\ (and SDSS {\it
  u}-band) image is a good tracer of the FUV morphology. In the case
of J214500.25+011157.3 we have an ACS/SBC image taken through F150LP,
a filter that is almost identical to the GALEX FUV filter
\citep[see][]{teplitz06}.  The corresponding FUV surface brightness
calculated solely from the ACS data,
$\textrm{log}_{10}I_{FUV,ACS}=9.53$ $L_\odot$ kpc$^{-2}$, is slightly
lower than the value of H07 based on GALEX and SDSS
($\textrm{log}_{10}I_{FUV}=9.82$ $L_\odot$ kpc$^{-2}$), but still well
above our LBG selection threshold
of $\textrm{log}_{10}I_{FUV}=9$ $L_\odot$ kpc$^{-2}$.\\

\noindent
$\bullet$ As we have discussed above, the galaxy sizes defined by
\citet{scarpa07} pertain only to the disk component and thus cannot be
compared with the half-light radii quoted by H05 and
H07. \citet{scarpa07} also did not apply their profile fitting
methodology to the full sample in a uniform way.  They only applied
their analysis to the objects that were marked as ``supercompact'' by
H07 and not to the other UVLGs. As a result, their concluding Figure 5
is misleading. It appears to show that the LBG analogs have a similar
FUV surface brightness as the main sample. Had they applied their
technique to all the objects in the sample, the compact UVLGs would
still form an upper envelope to the $I_{FUV}$ distribution, with the
overall $I_{FUV}$ distribution shifted downwards toward lower surface
brightness due to the fact that the ``point source'' flux in their fits was ignored.\\

\noindent
$\bullet$ Finally, we would like to remind the reader that the real
test of whether or not a given sample of nearby galaxies are good
analogs of LBGs is not the UV selection per s\'e (note that the LBG
luminosity function extends over many orders of magnitude
\citep[e.g.][]{bouwens07}), but whether their main physical properties
are indeed similar. We firmly believe that our sample of LBG analogs
have passed this test.

\end{document}